\documentclass[preprint,11pt]{elsarticle} 

\usepackage{amsmath}

\usepackage{amssymb}
\usepackage{amsfonts}
\usepackage{bm}
\usepackage{mathabx}
\usepackage{rotating}
\usepackage{dcolumn}
\usepackage{booktabs}
\usepackage[version-1-compatibility]{siunitx}
\usepackage{caption}
\usepackage{subfig}
\usepackage{multirow}
\usepackage{esint}
\usepackage{supertabular}
\usepackage{xcolor}

\allowdisplaybreaks

\DeclareMathOperator{\cint}{ci}
\DeclareMathOperator{\sint}{si}

\newcommand{\apriori}{\mbox{\it a priori}}
\newcommand{\ADC}{\mbox{ADC}}

\newcommand{\cD}{{\mathcal D}}

\newcommand{\degreeC}{\kern-.2em\r{}\kern-.3em\hbox{C}}
\newcommand{\degr}{\kern-.2em\r{}}

\newcommand{\vphi}{\varphi}

\newcommand{\lb}{\left}
\newcommand{\rb}{\right}

\newcommand{\zi}{\zeta}

\newcommand{\re}{\mbox{Re}}
\newcommand{\im}{\mbox{Im}}

\newcommand{\erfc}{\mbox{erfc}}

\newcommand{\aSk}{a_{\mbox{\tiny $S$}k}}
\newcommand{\aCk}{a_{\mbox{\tiny $C$}k}}
\newcommand{\Ci}{\mbox{Ci}}
\newcommand{\Cismall}{\mbox{\scriptsize Ci}}
\newcommand{\Si}{\mbox{Si}}
\newcommand{\Sismall}{\mbox{\scriptsize Si}}
\newcommand{\cinf}{c_{\mbox{\tiny$\infty$}}}
\newcommand{\czero}{c_{\mbox{\tiny$0$}}}
\newcommand{\Cinf}{C^{\mbox{\tiny$\infty$}}}
\newcommand{\Czero}{C^{\mbox{\tiny$0$}}}
\newcommand{\Czerol}{C_{\mbox{\tiny$0$}}}
\newcommand{\Dinf}{D_{\mbox{\tiny$\infty$}}}
\newcommand{\Dzero}{D_{\mbox{\tiny$0$}}}

\begin{document}

\noindent

\vspace{5mm}
\begin{center}
{\bf\Large
Oscillating-gradient spin-echo diffusion-weighted imaging (OGSE-DWI) 
with a limited number of oscillations: \\ II. Asymptotics \\
}
\vspace{5mm}
{\large
Jeff Kershaw$^{a,*}$ and
Takayuki Obata$^a$ \\
}

\vspace{5mm}
$^a$ Applied MRI Research,  
National Institute of Radiological Sciences, QST, Chiba, Japan \\[1mm]

$^*${\it Corresponding author:} len@qst.go.jp 

\end{center}

\noindent\rule{\textwidth}{0.3pt}
{\bf Abstract} \\[1mm]
Oscillating-gradient spin-echo diffusion-weighted magnetic resonance imaging
(OGSE-DWI) has been promoted as a promising technique for studying the
microstructure of complex hydrated matter in the frequency domain.
The target of the OGSE-DWI technique is the spectral density of molecular diffusion,
$u_2(\omega)$, which is predicted to obey a set of asymptotic universality relations
that are linked to the global organisation of the sample.
So, in principle the complex microstructure of a medium can be classified by
measuring the spectral density in its low- and high-frequency limits.
However, due to practical limitations on the spectral resolution and range of the
technique, it is not possible to directly sample the spectral density with OGSE-DWI.
Rather, information about the spectral density can be obtained only indirectly
through the quantities $U_{kk}$ \& $U_{k0}$,
which are filtered representations of $u_2(\omega)$.
The purpose of this study is to investigate how the universal behaviour of
$u_2(\omega)$ emerges in the asymptotic behaviour of OGSE-DWI signal. \\[2mm]

\noindent\rule{\textwidth}{0.3pt}



\section{Introduction}
\label{intro}

Ever since it was first outlined by Stepisnik \cite{Stepisnik1981},
oscillating-gradient spin-echo diffusion-weighted magnetic resonance imaging
(OGSE-DWI) has been spoken of as a promising technique for studying
the microstructure of complex hydrated matter. Although there are some variants
(e.g.~\cite{Callaghan1995,Callaghan1996, Parsons2003}), in its purest state
the OGSE-DWI technique is distinguised by motion-probing gradients (MPGs) of the form
\begin{equation}
   g(t) = G\cos(\omega_k t-\phi),
 \label{eqn:intro:g(t)}
\end{equation}
with amplitude $G$, frequency $\omega_k=2\pi k/T$, arbitrary phase $\phi$,
duration $T$, time $0\leqslant t\leqslant T$,
and $k>0$ an integer representing the number of oscillations.
Under the assumption that
$k$ is very large,
the conventional description for OGSE-DWI signal claimed that the spectral
density of molecular diffusion, $u_2(\omega)$
(i.e.~the Fourier transform of the velocity-autocorrelation function),
can be directly measured at the selected frequency
(e.g.~\cite{Callaghan1995,Parsons2003,Does2003,Novikov2011}).
More recently, it has been pointed out that practical limitations on the duration
and number of oscillations of a MPG restrict the spectral resolution and
range of the technique \cite{Kershaw2021}.
Given that $T$ and $k$ must be finite,
it was shown that the signal equation for an oscillating single-harmonic MPG is
\begin{equation}
   \ln s_k(T) = -\frac{\gamma^2 G^2 T}{4\omega_k^2}
      \lb[U_{kk} + 2 U_{k0}\sin^2\phi\rb],
 \label{eqn:intro:lnSk}
\end{equation}
where $U_{kk}$ \& $U_{k0}$ are to be the measured quantities (note also that this
equation is equivalent to Eq.~(B5) in \cite{Novikov2019}).
$U_{kk}$ \& $U_{k0}$ are related to the spectral density via the
relationship
\begin{equation}
   U_{kl} = \int_0^\infty\!\!d\omega\,u_2(\omega) H_{kl}(\omega; T)
 \label{eqn:intro:Ukl}
\end{equation}
where
\begin{equation}
   H_{kl}(\omega; T) = \frac{B_{kl}\,T (1-\cos\omega T)}
             {\lb[(\omega T)^2-{\varpi_k}^2\rb]\lb[(\omega T)^2-{\varpi_l}^2\rb]},
 \label{eqn:intro:Hkl}
\end{equation}
$B_{kl}=2(3\delta_{k,l}-1){\varpi_k}^2/\pi$, $\varpi_k=2\pi k$, $\delta_{k,l}$
is the Kronecker delta, and $l$ equals either 0 or $k$.
In short, $U_{kk}$ \& $U_{k0}$ are representations of $u_2(\omega)$ filtered by
the functions $H_{kl}(\omega;T)$.

%
%

%

Rather than directly sampling the spectral density, Eqs.~(\ref{eqn:intro:lnSk}) \&
(\ref{eqn:intro:Ukl}) imply that information about $u_2(\omega)$ can be obtained
only indirectly through measurements of $U_{kk}$ \& $U_{k0}$.
After applying Eq.~(\ref{eqn:intro:Ukl}) to three examples of $u_2(\omega)$ taken
from the literature, it was suggested that $U_{kk}$ provides a reasonable qualitative
understanding of the spectral density for a particular system, while $U_{k0}$
reflects the cumulative diffusion coefficient $D(T)$ of the system \cite{Kershaw2021}.
However, it remains to be demonstrated how such measurements can be used
to quantify the microstructure of complex media.
If $u_2(\omega)$ were a well-established function then the exact theoretical forms of
$U_{kk}$ \& $U_{k0}$ could be evaluated and
fitted to the data to obtain estimates of the important parameters.
Unfortunately, when making observations of real/natural systems it is unlikely
that $u_2(\omega)$ will be known \apriori.
It is fortunate then that certain global features of a system might be
quantified by observing the asymptotic behaviour of $u_2(\omega)$.
In the low and high frequency limits it has been predicted that $u_2(\omega)$ behaves
universally as \cite{Novikov2011,Novikov2014}
\footnote{It is common in the literature to associate the symbols $\Dinf$ \& $\Dzero$
with the low- \& high-frequency limits, respectively, of a frequency-dependent
diffusion metric. The notation originates from the association of the same symbols
with the long- and short-time limits of various time-dependent diffusion metrics.
For consistency, this work shall adopt the same convention of labeling coefficients
involved in the low- \& high-frequency limits with an index of $\infty$ \& 0,
respectively.
The index will be a subscript if there are no other indices, otherwise the index
will be a superscript.}
\begin{equation}
   u_2(\omega) \sim \lb\{
   \begin{array}{ll}
     2 \Dinf + \cinf\,\omega^\vartheta, & \quad\omega\rightarrow0 \vspace{2mm}\\
     2 \Dzero - \czero\,\omega^{-1/2},  & \quad  \omega\rightarrow\infty
   \end{array}\rb.
 \label{eqn:intro:u2}
\end{equation}
where $c_{\mbox{\tiny$\infty$}}$ is a constant that may depend on microscopic
details of the system and
$c_{\mbox{\tiny$0$}}=\sqrt{2}(S/V)$ ${\Dzero}^{3/2}/d$.
For the low-frequency limit the global behaviour is captured in the dynamical
exponent $\vartheta=(p+d)/2$, with $p$
being a structural exponent characterising the
global organisation of the microstructure in a $d$-dimensional medium.
In the high-frequency limit, the inverse-square-root dependence on frequency
is a feature that is universally true for complex media, which
means that the surface-to-volume ratio, $S/V$, is a quantity that can be used
to characterise the global properties of a medium.
It is the purpose of this study to investigate how these global characteristics appear
in the asymptotic behaviour of $U_{k0}$ \& $U_{kk}$.

The paper is organised so that the method used to derive asymptotic expansions for
$U_{k0}$ \& $U_{kk}$ is introduced first (Sec.~\ref{sec:asymptotics}),
after which the method is demonstrated by applying it to several examples
(Sec.~\ref{sec:apply}), before it is applied to achieve the intended goal
(Sec.~\ref{sec:Universality}).
Some discussion will then follow.

\section{Asymptotics of $U_{k0}$ \& $U_{kk}$}
\label{sec:asymptotics}

\subsection{Asymptotic variables}
\label{sub:asymptotics:meth}

To obtain the asymptotic behaviour of $u_2(\omega)$
first recall that there is a time-scale characterising the response of every system.
In fact, the response of a system may depend on several time scales, in which case
the smallest and largest will determine the behaviour in the asymptotic limits.
For simplicity, and
without loss of generality, it shall be assumed here that the important time-scale
in either limit is always represented by $\tau$.
The asymptotic behaviour of the system is obtained by expanding $u_2(\omega)$ in
the limits $\omega\tau\rightarrow0$ and $\omega\tau\rightarrow\infty$,
corresponding to the low and high frequency regimes, respectively.
Note that in this case $\omega$ is a continuous quantity that can be arbitrarily
increased or decreased relative to $\tau$ to meet the requirement for each limit.
For that reason the asymptotic tendency is often presented as $\omega\rightarrow0$ or
$\omega\rightarrow\infty$
when writing down the asymptotic behaviour of $u_2(\omega)$,
leaving out the dependence of the asymptotic variable on $\tau$.



The asymptotic behaviour of $U_{k0}$ \& $U_{kk}$ can be similarly obtained by
expanding with respect to $\omega_k\tau$.
The observation frequency $\omega_k$ can be manipulated by varying either
$k$ or $T$. Unfortunately, it is not possible to unboundedly increase or decrease
either of those parameters, which means that
$\omega_k$ cannot be arbitrarily set to meet the requirement for observations in either
of the asymptotic regimes. Therefore the value of $\tau$ relative to the accessible
range of $\omega_k$ becomes important.
As noted in \cite{Kershaw2021}, it is advantageous to choose $T$ to be as long as
possible in practice because it allows the widest range of accessible frequencies.
Nevertheless, the range of $k$ is limited, so it must actually be the value of $T$
relative to $\tau$
that determines whether $\omega_k\tau$ matches the condition for either the low- or
high-frequency regimes. 
With that understanding,
from here on $\omega_k\tau$ and $\beta=T/\tau$ will be used interchangeably as
the asymptotic variables
of $U_{k0}$ \& $U_{kk}$, and it should also be understood that
requiring the limits $\omega_k\tau\rightarrow0$ and $\omega_k\tau\rightarrow\infty$
is equivalent to taking $\beta\rightarrow\infty$ and $\beta\rightarrow0$, respectively.

%

\subsection{The Mellin transform method}
\label{sub:asymptotics:MTM}


Returning to Eq.~(\ref{eqn:intro:Ukl}) now, to obtain the asymptotic
behaviour of $U_{kl}$ the obvious course of
action is to evaluate the integral and then expand with respect to $\omega_k\tau$
(or $\beta$). However, it may be that the integral requires a lot of tedious work
or is simply just too difficult to evaluate exactly.
In that case a more direct
alternative is to apply the Mellin transform method (MTM) for the asymptotic
expansion of integrals \cite{BleisteinHandelsmanBook}.
Summarising its use for Eq.~(\ref{eqn:intro:Ukl}),
after the variable change $v=\omega\tau$ 
the integral can be written in the dimensionless form
\begin{equation}
   I_{kl}(\beta) = \frac{U_{kl}}{2\Dzero B_{kl} \beta}
      = \int_0^\infty\!\!dv\,f(v) h_{kl}(\beta v)
 \label{eqn:asymptotics:MTM:Ikl}
\end{equation}
where $f(v)=u_2(v/\tau)/2\Dzero$ and
\begin{equation*}
   h_{kl}(v) = \frac{1-\cos v}{(v^2-{\varpi_k}^2)(v^2-{\varpi_l}^2)}.
\end{equation*}

For $s\in\mathbb{C}$ the Mellin transform of $h_{kl}(v)$, $M[h_{kl};s]$, is well
defined in the ordinary sense (Supporting Material, Sec.~\ref{sub:supp:Model3}).
In contrast, other than the assumption that it is continuous, the properties of
$u_2(\omega)$ are usually unknown {\it a~priori}, which means it is not generally
true that the Mellin transform of $f(v)$, $M[f;s]$, exists in the ordinary sense.
Choose instead an arbitrary real number $v_0>0$ so that
$f(v)=f_1(v)+f_2(v)$ with
\begin{align}
   f_1(v) &= \lb\{\begin{array}{lll} f(v), & & 0\leqslant v < v_0 \\
                        0, & & v_0\leqslant v <\infty \end{array}\rb. 
 \label{eqn:asymptotics:MTM:f1}
\end{align}
and
\begin{align}
   f_2(v) &= \lb\{\begin{array}{lll} 0, & & 0\leqslant v < v_0 \\
                        f(v), & & v_0\leqslant v <\infty. \end{array}\rb. 
 \label{eqn:asymptotics:MTM:f2}
\end{align}
Under these conditions the Mellin transforms of
$f_1(v)$ \& $f_2(v)$ do exist, and therefore the quantity
$M[f_1;s]+M[f_2;s]$ is understood as the Mellin transform of $f(v)$ in the
generalised sense. 
Now let $G_{lj}(s)=M[f_j;1-s] M[h_{kl};s]$ ($j=1,2$) where both $M[h_{kl};s]$ and
$M[f_j;1-s]$ are analytic within the strip $a_{lj}<\re\,s<b_{lj}$
($a_{lj}$, $b_{lj}\in\mathbb{R}$) of the complex plane.
The poles $s_n$ of $G_{lj}(s)$ lie outside of the analytic strip and may be organised
into two sets,
$\{s_n\leqslant a_{lj}\}$ and $\{s_n\geqslant b_{lj}\}$,
according to whether $\re\,s_n$ lies to the left or right of the complex plane from the
analytic strip. Then the series
\begin{equation}
   I_{kl}^j(\beta) = \int_0^\infty\!\!dv\,f_j(v) h_{kl}(\beta v)
     \sim \lb\{\begin{aligned}
        & \sum_{\{s_n\leqslant a_{lj}\}} 
          \mbox{Res}\lb\{\beta^{-s} G_{lj}(s)\rb\}, & \quad & \beta\rightarrow0 \\
       & -\!\! \sum_{\{s_n\geqslant b_{lj}\}} 
                                \mbox{Res}\lb\{\beta^{-s} G_{lj}(s)\rb\}, 
                                            & & \beta\rightarrow\infty
     \end{aligned} \rb.
 \label{eqn:asymptotics:MTM:IklSeries}
\end{equation}
provide asymptotic expansions for $I_{kl}(\beta)=I_{kl}^1(\beta)+I_{kl}^2(\beta)$
as sums over the residues
of $G_{lj}(s)$ at its poles.
Terms in the series associated with poles having real parts closer to the analytic strip
will in general dominate terms corresponding to poles sitting further away.

More detail on the Mellin transform and MTM has been included in the Supporting
Material (Secs.~\ref{sub:supp:MTM} \& \ref{sub:supp:genMTM}).


\subsection{The poles of $M[h_{kl};s]$, $M[f_j;1-s]$ and $G_{lj}(s)$}
\label{sub:asymptotics:poles}

Evaluating the terms in the series of Eq.~(\ref{eqn:asymptotics:MTM:IklSeries})
is aided by the fact that the local behaviour of $M[h_{kl};s]$ and $M[f_j;1-s]$
near a pole is directly linked
to the powers of $v$ in expansions of $h_{kl}(v)$ and $f(v)$ about 0 and $\infty$.
For example, the function $h_{kl}(v)$ has the asymptotic expansions
\begin{equation}
   h_{kl}(v) \sim \lb\{ \begin{aligned}
        & \sum_{n=0}^\infty h^{0}_{ln}\,v^{2(n+\delta_{k,l})}, & & v\rightarrow 0 \\
        & (1-\cos v)\sum_{n=0}^\infty h^{\infty}_{ln}\,v^{-2(n+2)}, & \quad &
                                            v\rightarrow\infty
    \end{aligned} \rb.
 \label{eqn:asymptotics:poles:hkl}
\end{equation}
where
\begin{align*}
   h^0_{ln} &=\frac{(-)^{n+1-\delta_{k,l}}}{{\varpi_k}^{2(1+\delta_{k,l})}}
      \sum_{m=0}^n \frac{(-)^m (1+m\,\delta_{k,l})}{{\varpi_k}^{2m}(2n-2m+2)!}
 \\[2mm]
   h^\infty_{ln} &=(1+n\,\delta_{k,l}){\varpi_k}^{2n}.
\end{align*}
It follows from the properties of the Mellin transform that
\begin{equation}
   M[h_{kl};s] \sim \lb\{ \begin{aligned}
      &\frac{h^{0}_{ln}}{s+2(n+\delta_{k,l})}, & &  s\rightarrow -2(n+\delta_{k,l}) \\
     -&\frac{h^{\infty}_{ln}}{s-2(n+2)}, & \quad &  s\rightarrow 2(n+2)
    \end{aligned} \rb. 
\end{equation}
with $n=0,1,2,\cdots$.
Notice that the poles of $M[h_{kl};s]$ are all simple and equal to minus the powers of
$v$ in the asymptotic expansions.

For the case of $f(v)$, it is assumed that it may be expanded in power series
about 0 and $\infty$ like so
\begin{equation}
   f(v) \sim \lb\{ \begin{aligned}
        & \sum_{n=0}^\infty f^{0}_{n}\,v^{\theta_{1n}}, & & v\rightarrow 0 \\
        & \sum_{n=0}^\infty f^{\infty}_{n}\,v^{-\theta_{2n}}, & \quad &
                                            v\rightarrow\infty.
    \end{aligned} \rb.
 \label{eqn:asymptotics:poles:f}
\end{equation}
Here $\{\theta_{1n}\}$ and $\{\theta_{2n}\}$ are strictly monotonic increasing sequences
of real numbers.
In accordance with the definitions of $f_1(v)$ and $f_2(v)$ in
Eqs.~(\ref{eqn:asymptotics:MTM:f1}) \& (\ref{eqn:asymptotics:MTM:f2}) above,
it follows that
\begin{align}
   & M[f_1;1-s] \sim -\frac{f_n^0}{s-\theta_{1n}-1}, & & s\rightarrow \theta_{1n}+1
 \label{eqn:asymptotics:poles:M[f1;1-s]} \\[2mm]
   & M[f_2;1-s] \sim \frac{f_n^\infty}{s+\theta_{2n}-1}, & & s\rightarrow -\theta_{2n}+1
 \label{eqn:asymptotics:poles:M[f2;1-s]}
\end{align}
for $n=0,1,2,\cdots$.
The poles of the $M[f_j;1-s]$ are also simple and, due to the shift in argument from
$s$ to $1-s$, are equal to 1 plus the powers of $v$ in the expansions of $f(v)$.

The pole sets of $G_{lj}(s)$ are composed from the poles of $M[f_j;1-s]$ and
$M[h_{kl};s]$.
For $j=1$, the analytic strips of $M[f_1;1-s]$ and $M[h_{kl};s]$ intersect for
$-2\delta_{k,l}<\re\,s<1+\theta_{10}$ so
$a_{l1}=-2\delta_{k,l}$, $b_{l1}=1+\theta_{10}$ and the poles of $G_{l1}(s)$
are
\begin{align*}
   \{s_n\leqslant a_{l1}\}&=\{-2(n+\delta_{k,l})\} \\
   \{s_n\geqslant b_{l1}\}&=\{2(n+2)\}\cup\{\theta_{1n}+1\}.
\end{align*}
Likewise for $j=2$, the analytic strips of $M[f_2;1-s]$ and $M[h_{kl};s]$ intersect
when $1-\theta_{20}<\re\,s<4$ so the poles of $G_{l2}(s)$ are
\begin{align*}
   \{s_n\leqslant a_{l2}\}&=\{-2(n+\delta_{k,l})\}\cup\{-\theta_{2n}+1\} \\
   \{s_n\geqslant b_{l2}\}&=\{2(n+2)\} 
\end{align*}
with $a_{l2}=1-\theta_{20}=1$ and $b_{l2}=4$.
Realising that there may be some overlap between the poles of the $M[f_j;1-s]$ and
$M[h_{kl};s]$, let $Q_{l1}$ be a set containing all pairs of non-negative integers
$(p,r)$ such that $\theta_{1p}+1=2(r+2)$, and let $P_{l1}$ \&
$R_{l1}$ be sets containing all values of $p$ \& $r$, respectively, that 
are not included as part of one of the pairs in $Q_{l1}$.
Similarly, let $Q_{l2}$ contain non-negative integer pairs $(p,r)$ such that 
$-\theta_{2p}+1=-2(r+\delta_{k,l})$, and let $P_{l2}$ \& $R_{l2}$ be sets
respectively consisting of the $p$ \& $r$ that are not already elements
of $Q_{l2}$. These definitions allow the poles in $\{s_n\geqslant b_{l1}\}$ and
$\{s_n\leqslant a_{l2}\}$ to be organised into three disjoint sets. Moreover, as the
poles of $M[h_{kl};s]$ and the
$M[f_j;1-s]$ are all simple, it is easy to calculate the residues in
Eq.~(\ref{eqn:asymptotics:MTM:IklSeries}) and use them to write
$I_{kl}(\beta)=I_{kl}^{\mbox{\tiny $P$}}(\beta)+I_{kl}^{\mbox{\tiny $Q$}}(\beta)+
I_{kl}^{\mbox{\tiny $R$}}(\beta)$ where
\begin{equation}
   I_{kl}^{\mbox{\tiny $P$}}(\beta) \sim \lb\{ \begin{aligned}
       & \sum_{p\in P_{l2}} \beta^{\theta_{2p}-1} f_{p}^\infty 
                     M[h_{kl};-\theta_{2p}+1], & &   \beta\rightarrow0 \\[1mm]
       & \sum_{p\in P_{l1}} \beta^{-\theta_{1p}-1} 
            f_{p}^0 M[h_{kl};\theta_{1p}+1], & &    \beta\rightarrow\infty
     \end{aligned} \rb.
 \label{eqn:asymptotics:poles:IklP}
\end{equation}
\begin{equation}
   I_{kl}^{\mbox{\tiny $R$}}(\beta) \sim \lb\{ \begin{aligned}
       & \sum_{r\in R_{l2}} \beta^{2(r+\delta_{k,l})}
            h_{lr}^0 M[f;2r+2\delta_{k,l}+1], & &   \beta\rightarrow0 \\[1mm]
       & \sum_{r\in R_{l1}}  \beta^{-2(r+2)}
            h_{lr}^\infty M[f;-2r-3], & &    \beta\rightarrow\infty
     \end{aligned} \rb.
 \label{eqn:asymptotics:poles:IklR}
\end{equation}
\begin{equation}
   I_{kl}^{\mbox{\tiny $Q$}}(\beta) \sim \lb\{ \begin{aligned}
       & \sum_{(p,r)\in Q_{l2}}\!\!\!\!\beta^{\theta_{2p}-1} K_{lpr}^2 
          -\ln\beta\!\!\!\!\sum_{(p,r)\in Q_{l2}}\!\!\!\!\beta^{\theta_{2p}-1}
                     h_{lr}^0 f_p^\infty, & & \beta\rightarrow0 \\[1mm]
       & \sum_{(p,r)\in Q_{l1}}\!\!\!\!\beta^{-\theta_{1p}-1} K_{lpr}^1
          +\ln\beta\!\!\!\!\sum_{(p,r)\in Q_{l1}}\!\!\!\!\beta^{-\theta_{1p}-1}
                     h_{lr}^\infty f_p^0, & \quad &    \beta\rightarrow\infty
     \end{aligned} \rb.
 \label{eqn:asymptotics:poles:IklQ}
\end{equation}
\begin{align*}
   K_{lpr}^1 &= h_{lr}^\infty M[f_2;-\theta_{1p}] - \lim_{s\rightarrow\theta_{1p}+1}
     \frac{d}{ds}\lb\{(s-\theta_{1p}-1)^2 G_{l1}(s)\rb\} \\
   K_{lpr}^2 &= h_{lr}^0 M[f_1;\theta_{2p}] + \lim_{s\rightarrow-\theta_{2p}+1}
     \frac{d}{ds}\lb\{(s+\theta_{2p}-1)^2 G_{l2}(s)\rb\} 
\end{align*}
Note that the Mellin transform of $f(v)$ in $I_{kl}^{\mbox{\tiny $R$}}(\beta)$ is
to be understood in the generalised sense.


The series in $I_{kl}^{\mbox{\tiny $P$}}(\beta)$ have terms proportional to
$\beta^{\theta_{1p}-1}$ and $\beta^{-\theta_{2p}-1}$. Therefore, after multiplying by the
factor of $\beta$ from the central term in Eq.~(\ref{eqn:asymptotics:MTM:Ikl}) and
recalling that $\beta=\varpi_k/\omega_k\tau$, the effect of
$I_{kl}^{\mbox{\tiny $P$}}(\beta)$
on the asymptotic behaviour of $U_{kl}$ is to supply terms with the same powers of
$\omega_k\tau$ that appear in an expansion of $u_2(\omega_k)$.
In contrast, by the construction of $R_{l1}$ and $R_{l2}$,
$I_{kl}^{\mbox{\tiny $R$}}(\beta)$
provides terms to the asymptotic behaviour of $U_{kl}$
that are qualitatively different to those in an expansion of $u_2(\omega_k)$.
In some circumstances the qualitatively different behaviour of
$I_{kl}^{\mbox{\tiny $R$}}(\beta)$ may dominate the qualitatively similar behaviour of
$I_{kl}^{\mbox{\tiny $P$}}(\beta)$ so that the asymptotic response of the system
is masked by the behaviour of the filters $H_{kl}(\omega;T)$.


Since the poles in $Q_{l1}$ and $Q_{l2}$ are second order,
evaluation of the residues is a little more complicated, with the result that
$I_{kl}^{\mbox{\tiny $Q$}}(\beta)$ can be organised into two separate sums
(re Eq.~(\ref{eqn:asymptotics:poles:IklQ})).
The first of these has terms proportional to $\beta^{\theta_{1p}-1}$ and
$\beta^{-\theta_{2p}-1}$ so its effect on the asymptotic behaviour of $U_{kl}$ is similar
to that of $I_{kl}^{\mbox{\tiny $P$}}(\beta)$.
On the other hand, the terms in the second sum are multiplied by a factor of
$\ln\beta$ so, like the case for $I_{kl}^{\mbox{\tiny $R$}}(\beta)$, the effect on the
asymptotic behaviour of $U_{kl}$ is qualitatively different to that of $u_2(\omega_k)$.

If the $M[f_j;1-s]$ have poles of 2nd or higher order, evaluation of the
residues in $I_{kl}^{\mbox{\tiny $P$}}(\beta)$ and $I_{kl}^{\mbox{\tiny $Q$}}(\beta)$
will be more complicated and integer powers of $\ln\beta$ will appear.
In contrast, evaluation of the terms in $I_{kl}^{\mbox{\tiny $R$}}(\beta)$ will
not be altered because the elements of $R_{l1}$ and $R_{l2}$ will always be simple poles.

\section{Application to three models}
\label{sec:apply}



The method described in Sec.~\ref{sec:asymptotics} is now applied to derive the
asymptotic behaviour of $U_{k0}$ \& $U_{kk}$ for three examples taken from the literature.
A brief description for each model can be found in \cite{Kershaw2021}, as can the
exact forms of $U_{k0}$ \& $U_{kk}$ for the first two models.
Details of the MTM applied to each model have been included in the Supporting Material
(Secs.~\ref{sub:supp:Models1&2} \& \ref{sub:supp:Model3}).
The sets $Q_{l1}$ \& $Q_{l2}$ are empty for all three models so only
$I_{kl}^{\mbox{\tiny $P$}}(\beta)$ \& $I_{kl}^{\mbox{\tiny $R$}}(\beta)$ contribute
to the asymptotic behaviour.
Each result contains the two most dominant terms from $I_{kl}^{\mbox{\tiny $P$}}(\beta)$
\& $I_{kl}^{\mbox{\tiny $R$}}(\beta)$ together with the order of the next most
significant term.



\vspace{2mm}
{\it Model~1}. Expanding $u_2(\omega)$ in the appropriate
limits
\begin{equation}
   \frac{u_2(\omega)}{2\Dzero} \sim \lb\{
   \begin{aligned}
    & (1-\eta)+\eta\,(\omega\tau)^2 + O(\omega\tau)^4,
                                       & & \omega\tau\rightarrow 0 \\[2mm]
    & 1-\eta\,(\omega\tau)^{-2} + O(\omega\tau)^{-4}, & & \omega\tau\rightarrow\infty.
   \end{aligned}\rb.
 \label{eqn:asymptotics:apply:Model1:u2}
\end{equation}
%
%
%
The corresponding expansions for the measured quantities are
\begin{align}
   \frac{U_{k0}}{2\Dzero} &\sim \lb\{ \begin{aligned}
         & (1-\eta) + \frac{\eta}{\varpi_k} (\omega_k\tau)^3 + O(\omega_k\tau)^5,
                                         &  & \omega_k\tau\rightarrow 0 \\
         & 1-\frac{\eta\varpi_k}{2}(\omega_k\tau)^{-1} + O(\omega_k\tau)^{-2},
                                       &  & \omega_k\tau\rightarrow\infty
      \end{aligned} \rb. 
 \label{eqn:asymptotics:apply:Model1:Uk0} \\[2mm]
   \frac{U_{kk}}{2\Dzero} &\sim \lb\{ \begin{aligned}
         & (1-\eta) + \eta\,(\omega_k\tau)^2 + O(\omega_k\tau)^3,
                                          &  & \omega_k\tau\rightarrow 0 \\[2mm]
         & 1-3\eta\,(\omega_k\tau)^{-2} + O(\omega_k\tau)^{-3},
                                             &  & \omega_k\tau\rightarrow\infty.
      \end{aligned} \rb. 
 \label{eqn:asymptotics:apply:Model1:Ukk}
\end{align}

{\it Model~2}. Expansion of the spectral density for this model gives
\begin{equation}
   \frac{u_2(\omega)}{2\Dzero} \sim \lb\{
   \begin{aligned}
    & (1-\zi)+\frac{\zi}{\sqrt{2}}\,(\omega\tau)^{1/2} + O(\omega\tau)^{3/2},
                                       &  & \omega\tau\rightarrow 0 \\
    & 1-\frac{\zi}{\sqrt{2}}\,(\omega\tau)^{-1/2} + O(\omega\tau)^{-3/2},
                                                & & \omega\tau\rightarrow\infty.
   \end{aligned}\rb.
 \label{eqn:asymptotics:apply:Model2:u2}
\end{equation}
Direct expansion of the exact results for $U_{k0}$ \& $U_{kk}$ can be used to confirm
that
\begin{align}
   \frac{U_{k0}}{2\Dzero} &\sim \lb\{ \begin{aligned}
         & (1-\zi) + A_{01}^\infty
            \frac{\zi}{\sqrt{2}} (\omega_k\tau)^{1/2} + O(\omega_k\tau)^{3/2},
                                         &  & \omega_k\tau\rightarrow 0 \\
         & 1 - A_{01}^0 \frac{\zi}{\sqrt{2}} (\omega_k\tau)^{-1/2}
                + O(\omega_k\tau)^{-1}, &  & \omega_k\tau\rightarrow\infty
      \end{aligned} \rb. 
 \label{eqn:asymptotics:apply:Model2:Uk0} 
\end{align}
\begin{equation*}
   A_{01}^\infty = \frac{2}{\pi k^{1/2}}-\frac{\aCk}{\varpi_k}, \qquad
   A_{01}^0 =\frac{8k^{1/2}}{3}+\frac{\aSk}{\varpi_k}
\end{equation*}
\begin{align}
   \frac{U_{kk}}{2\Dzero} &\sim \lb\{ \begin{aligned}
         & (1-\zi) + A_{k1}^\infty
              \frac{\zi}{\sqrt{2}}(\omega_k\tau)^{1/2} + O(\omega_k\tau)^{3/2},
                                          &  & \omega_k\tau\rightarrow 0 \\
         & 1 - A_{k1}^0 \frac{\zi}{\sqrt{2}}(\omega_k\tau)^{-1/2}
                + O(\omega_k\tau)^{-3/2}, &  & \omega_k\tau\rightarrow\infty
      \end{aligned} \rb. 
 \label{eqn:asymptotics:apply:Model2:Ukk}
\end{align}
\begin{equation*}
   A_{k1}^\infty=\frac{1}{\pi k^{1/2}} -\frac{\aCk}{2\varpi_k}+\aSk, \qquad
   A_{k1}^0=\frac{3\aSk}{2\varpi_k}+\aCk.
\end{equation*}
Here $\aSk=2S(2k^{1/2})$ and $\aCk=2C(2k^{1/2})$, where $S(z)$ and $C(z)$ are the
Fresnel integrals.


{\it Model~3}. A description of how the following expansions of $u_2(\omega)$ were
obtained for this model
is contained in Sec.~\ref{sub:supp:Model3} of the Supporting Material.
\begin{equation}
   \frac{u_2(\omega)}{2\Dzero} \sim \lb\{ \begin{aligned}
     & \frac{1}{1+\zi} + \frac{\sqrt{2}[\sqrt{1+\zi}-1]}{(1+\zi)^2}
                                                (\omega\tau)^{\frac{1}{2}}
         + O(\omega\tau)^{\frac{3}{2}},  & \quad & \omega\tau\rightarrow 0 \\
     & 1 - \frac{\zi}{\sqrt{2}}(\omega\tau)^{-\frac{1}{2}}
         + O(\omega\tau)^{-\frac{3}{2}},  & & \omega\tau\rightarrow\infty.
    \end{aligned} \rb.
 \label{eqn:asymptotics:apply:Model3:u2}
\end{equation}
The leading terms in the final results are
\begin{equation}
   \frac{U_{k0}}{2\Dzero} \sim \lb\{ \begin{aligned}
     & \frac{1}{1+\zi} + A_{01}^\infty\frac{\sqrt{2}[\sqrt{1+\zi}-1]}{(1+\zi)^2}
         (\omega_k\tau)^{\frac{1}{2}}
       + O(\omega_k\tau)^{\frac{3}{2}},  & \quad & \omega_k\tau\rightarrow 0 \\
     & 1 -A_{01}^0\frac{\zi}{\sqrt{2}} (\omega_k\tau)^{-\frac{1}{2}}
        + O(\omega_k\tau)^{-1},  & & \omega_k\tau\rightarrow\infty
    \end{aligned} \rb.
 \label{eqn:asymptotics:apply:Model3:Uk0}
\end{equation}
\begin{equation}
   \frac{U_{kk}}{2\Dzero} \sim \lb\{ \begin{aligned}
     & \frac{1}{1+\zi} + A_{k1}^\infty \frac{\sqrt{2}[\sqrt{1+\zi}-1]}{(1+\zi)^2}
                                                  (\omega_k\tau)^{\frac{1}{2}}
         + O(\omega_k\tau)^{\frac{3}{2}}, & \quad & \omega_k\tau\rightarrow 0 \\
     & 1 -A_{k1}^0 \frac{\zi}{\sqrt{2}} (\omega_k\tau)^{-\frac{1}{2}}
        + O(\omega_k\tau)^{-\frac{3}{2}}, & & \omega_k\tau\rightarrow\infty
    \end{aligned} \rb.
 \label{eqn:asymptotics:apply:Model3:Ukk}
\end{equation}

\begin{figure}[tp]
 \centering
 \begin{tabular}{r}
  \subfloat[]{\setcounter{subfigure}{1}
                 \includegraphics[scale=0.72]{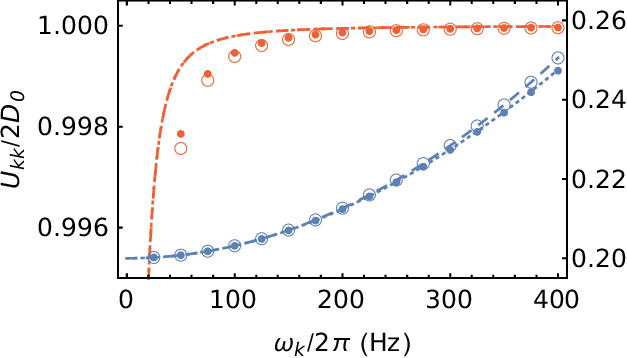}} \\[8mm]
  \subfloat[]{\setcounter{subfigure}{2}
                 \includegraphics[scale=0.72]{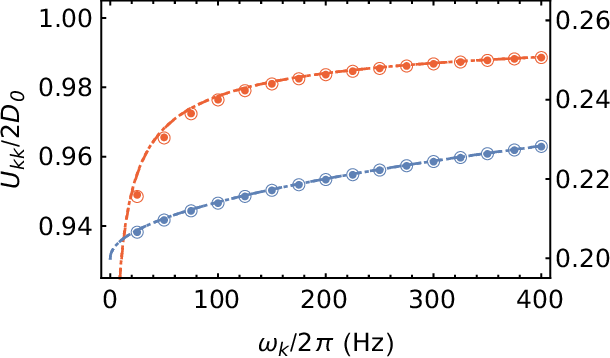}} \\[8mm]
  \subfloat[]{\setcounter{subfigure}{3}
                 \includegraphics[scale=0.72]{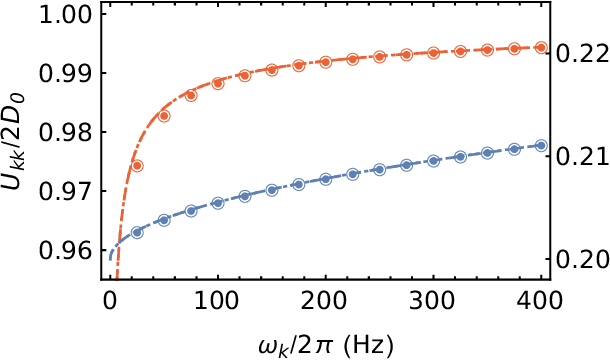}}
 \end{tabular}
 \caption{
Comparison of the exact and asymptotic behaviour of $U_{kk}$ with that of $u_2(\omega)$
on magnified scales for Models~1--3.
In all parts of the figure the dark-orange data correspond to the high-frequency regime
and the left vertical axis, while the blue data correspond to the low-frequency regime
and the right vertical axis.
Dotted lines denote the exact value of $u_2(\omega)$ for each model, while dashed lines
are the corresponding asymptotic approximations presented in
Eqs.~(\ref{eqn:asymptotics:apply:Model1:u2}), (\ref{eqn:asymptotics:apply:Model2:u2})
\& (\ref{eqn:asymptotics:apply:Model3:u2}).
In most cases the dotted and dashed lines overlap.
Filled circles represent the exact value of $U_{kk}$, with open circles
illustrating the asymptotic approximations of
Eqs.~(\ref{eqn:asymptotics:apply:Model1:Ukk}), (\ref{eqn:asymptotics:apply:Model2:Ukk})
\& (\ref{eqn:asymptotics:apply:Model3:Ukk}).
(a) Model~1 with $\eta=0.8$ and $T/\tau=0.4$ (dark-orange) and 400 (blue).
(b) Model~2 with $\zi=0.8$ and $T/\tau=0.04$ (dark-orange) and 40000 (blue).
(c) Model~3 with $\zi=4$ and $T/\tau=0.0004$ (dark-orange) and 400 (blue).
}
 \label{fig:asymptotics:apply:Model1-3:Ukk}
\end{figure}

\begin{figure}[t]
 \centering
 \begin{tabular}{cc}
  \subfloat[]{\setcounter{subfigure}{1}
                        \includegraphics[scale=0.72]{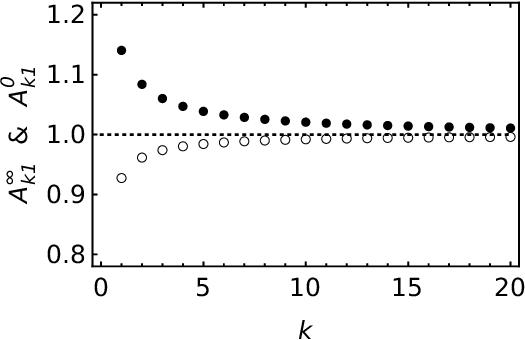}}
                                                             \quad  & \quad
  \subfloat[]{\setcounter{subfigure}{2}
                        \includegraphics[scale=0.72]{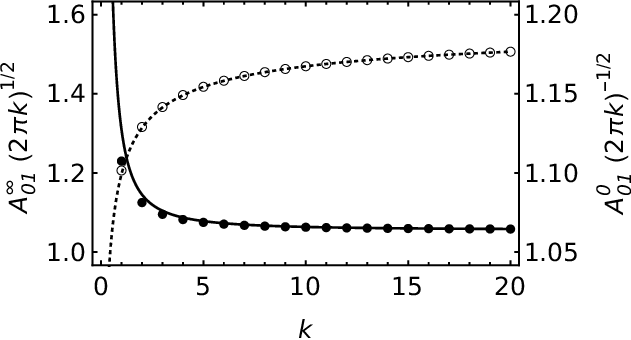}}
 \end{tabular}
 \caption{
(a) Plots of $A_{k1}^\infty$ (open circles) and $A_{k1}^0$ (closed circles) as a
function of $k$.
The largest deviation from unity occurs for $k=1$ ($A_{11}^\infty\approx0.93$
and $A_{11}^0\approx1.14$), and thereafter both coefficients approach 1 as
$k^{-1}$ for increasing $k$.
(b) Plots of $A_{01}^\infty{\varpi_k}^{1/2}$ (open circles) and
$A_{01}^0 {\varpi_k}^{-1/2}$ (closed circles) as a function of $k$.
The lines represent the asymptotic behaviour of the coefficients as
$k\rightarrow\infty$:
$A_{01}^\infty {\varpi_k}^{1/2} \sim 2\sqrt{2/\pi}-{\varpi_k}^{-1/2}$ (dotted) and
$A_{01}^0 {\varpi_k}^{-1/2} \sim (4/3)\sqrt{2/\pi}+{\varpi_k}^{-3/2}$ (solid).
}
 \label{fig:asymptotics:apply:A0k1etc}
\end{figure}

\begin{figure}[tp]
 \centering
 \begin{tabular}{c}
  \subfloat[]{\setcounter{subfigure}{1}
                 \includegraphics[scale=0.72]{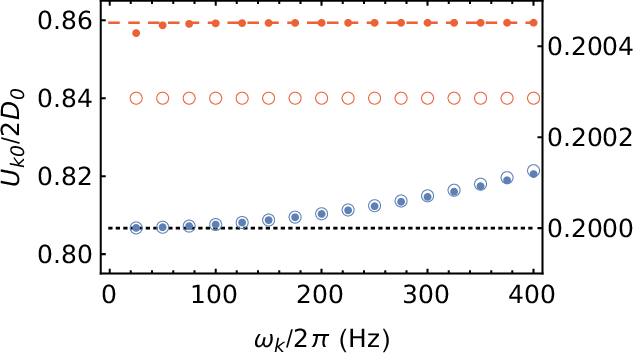}} \\[8mm]
  \subfloat[]{\setcounter{subfigure}{2}
                 \includegraphics[scale=0.72]{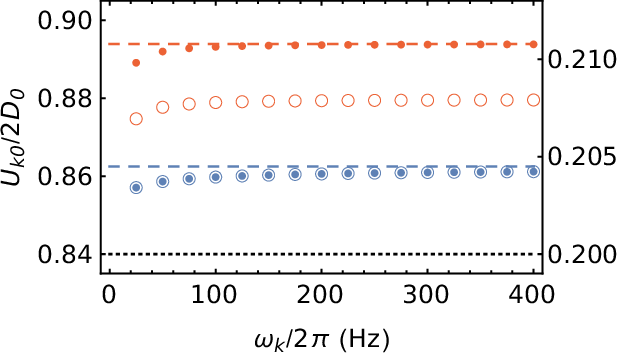}} \\[8mm]
  \subfloat[]{\setcounter{subfigure}{3}
                 \includegraphics[scale=0.72]{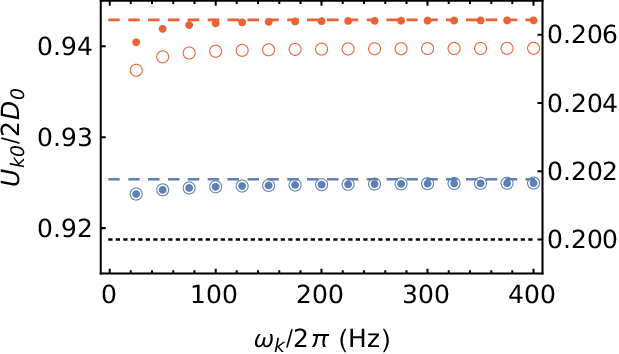}}
 \end{tabular}
 \caption{
Demonstration of the exact and asymptotic behaviour of $U_{k0}$ for Models~1--3.
In all parts of the figure the dark-orange data correspond to the high-frequency regime
and the left vertical axis, while the blue data correspond to the low-frequency regime
and the right vertical axis.
Filled circles represent the exact value of $U_{k0}$, with open circles
illustrating the asymptotic approximations of
Eqs.~(\ref{eqn:asymptotics:apply:Model1:Uk0}), (\ref{eqn:asymptotics:apply:Model2:Uk0})
\& (\ref{eqn:asymptotics:apply:Model3:Uk0}). 
The black dotted horizontal line is $\Dinf/\Dzero$, and the dashed lines are equal to
$D(T)/\Dzero$ for the corresponding value of $T/\tau$.
(a) Model~1 with $\eta=0.8$ and $T/\tau=0.4$ (dark-orange) and 400 (blue).
There is no dashed blue line for this part because $D(T)/\Dzero\approx 0.202$,
which is off the scale for the low-frequency regime data.
(b) Model~2 with $\zi=0.8$ and $T/\tau=0.04$ (dark-orange) and 40000 (blue).
(c) Model~3 with $\zi=4$ and $T/\tau=0.0004$ (dark-orange) and 400 (blue).
}
 \label{fig:asymptotics:apply:Model1-3:Uk0}
\end{figure}

Figure~\ref{fig:asymptotics:apply:Model1-3:Ukk} compares the asymptotic behaviour
of $U_{kk}$ with that of $u_2(\omega)$ for each of Models~1--3.
The upper (dark orange) data corresponds to the system being in the high-frequency
regime (i.e.~$\omega\tau$ or $\omega_k\tau\rightarrow\infty$),
while the lower (blue) data corresponds to the system being in the low-frequency
regime (i.e.~$\omega\tau$ or $\omega_k\tau\rightarrow0$).
For Model~1, comparison of Eqs.~(\ref{eqn:asymptotics:apply:Model1:u2}) \&
(\ref{eqn:asymptotics:apply:Model1:Ukk}) finds that in the low-frequency limit $U_{kk}$
has the same asymptotic behaviour as $u_2(\omega)$ up to $O(\omega_k\tau)^2$,
which is evident in Fig.~\ref{fig:asymptotics:apply:Model1-3:Ukk}a, but after
that there is an $O(\omega_k\tau)^{3}$ term that does not occur for $u_2(\omega)$.
In the high-frequency limit, the coefficient of the $O(\omega_k\tau)^{-2}$ term differs
from that of $u_2(\omega)$, and the high-frequency plots of $u_2(\omega)$ and $U_{kk}$
in Fig.~\ref{fig:asymptotics:apply:Model1-3:Ukk}a reflect this difference.
The $O(\omega_k\tau)^{\pm3}$ terms in Eq.~(\ref{eqn:asymptotics:apply:Model1:Ukk})
originate from the leading terms in $I_{kk}^{\mbox{\tiny $R$}}(\beta)$.

For Models~2 \& 3, the coefficients of powers of $\omega_k\tau$ in
Eqs.~(\ref{eqn:asymptotics:apply:Model2:Ukk}) \&
(\ref{eqn:asymptotics:apply:Model3:Ukk}) are presented in a way that allows differences
in the asymptotic behaviour of $U_{kk}$ and $u_2(\omega)$ to be quickly identified.
For both models the difference is characterised by the coefficients
$A_{k1}^\infty$ (in the low-frequency limit) and $A_{k1}^0$ (in the high-frequency limit).
Clearly, the further $A_{k1}^\infty$ and $A_{k1}^0$ are from unity the more the
asymptotic behaviour of $U_{kk}$ deviates from that of $u_2(\omega)$.
Figure~\ref{fig:asymptotics:apply:A0k1etc}a
shows that $A_{k1}^\infty$ and $A_{k1}^0$
differ from 1 most when $k=1$, but thereafter both
approach 1 as $1/k$, the former from below and the latter from above, for increasing $k$.
Neither Eq.~(\ref{eqn:asymptotics:apply:Model2:Ukk}) nor
Eq.~(\ref{eqn:asymptotics:apply:Model3:Ukk}) contains a term originating from
$I_{kk}^{\mbox{\tiny $R$}}(\beta)$.
Looking at the simulated data in Fig.~\ref{fig:asymptotics:apply:Model1-3:Ukk}b--c,
the asymptotic approximations to $u_2(\omega)$ and $U_{kk}$ appear indistinguishable
from the exact curves
even on the magnified scales used.


Figure~\ref{fig:asymptotics:apply:Model1-3:Uk0} illustrates the asymptotic behaviour
of $U_{k0}$ for all three models.
The asymptotic approximations in the low-frequency regime are almost
indistinguishable from the exact values of $U_{k0}$, while those in the high-frequency
regime have a clear offset, indicating that the next lowest order term in the
expansions is important at the selected scales.
Recalling that $U_{k0}$ approaches $2D(T)$ for large $k$ \cite{Kershaw2021},
those values have also been drawn on the figure as horizontal dashed lines.
Apart from one exception, $2D(T)$ provides a reasonable
approximation to the exact values of $U_{k0}$ on the scales shown. The exception
is the low-frequency data of Model~1, for which the value of $2D(T)$ is off the chosen
scale (Fig.~\ref{fig:asymptotics:apply:Model1-3:Uk0}a).

Consistent with the discussion on the qualitative behaviour of $U_{k0}$ in
Sec.~3.2 of \cite{Kershaw2021}, the leading terms of
Eqs.~(\ref{eqn:asymptotics:apply:Model1:Uk0}), (\ref{eqn:asymptotics:apply:Model2:Uk0})
\& (\ref{eqn:asymptotics:apply:Model3:Uk0}) are equal to $2\Dinf$ and $2\Dzero$ in
the low-frequency and high-frequency limits, respectively.
For Model~1, $M[h_{k0};s]=0$ for $s=3,5,7,\cdots$ so the asymptotic behaviour of $U_{k0}$
in the low-frequency regime stems from $I_{k0}^{\mbox{\tiny $R$}}(\beta)$ alone. In the
high-frequency regime the leading term of $I_{k0}^{\mbox{\tiny $R$}}(\beta)$ dominates
the second term in $I_{k0}^{\mbox{\tiny $P$}}(\beta)$ so $U_{k0}/2\Dzero$ approaches 1
as $(\omega_k\tau)^{-1}$ rather than $(\omega_k\tau)^{-2}$.
For Models 2 \& 3, the coefficients $A_{01}^\infty$ and $A_{01}^0$ have been used in
Eqs.~(\ref{eqn:asymptotics:apply:Model2:Uk0}) \&
(\ref{eqn:asymptotics:apply:Model3:Uk0}) to put
the asymptotic behaviour of $U_{k0}$ in a form similar to that of $u_2(\omega)$.
Even though $A_{01}^\infty$ diverges and $A_{01}^0$ tends to zero for
increasing $k$, the products $A_{01}^\infty {\varpi_k}^{1/2}$ and
$A_{01}^0 {\varpi_k}^{-1/2}$, where the factors of ${\varpi_k}^{\pm1/2}$ come from
$(\omega_k\tau)^{\pm 1/2}$, approach constant values so the
corresponding terms remain finite overall (Fig.~\ref{fig:asymptotics:apply:A0k1etc}b).
Only the $O(\omega_k\tau)^{-1}$ term in the high-frequency regime of
Eqs.~(\ref{eqn:asymptotics:apply:Model2:Uk0}) \& (\ref{eqn:asymptotics:apply:Model3:Uk0})
originates from $I_{kk}^{\mbox{\tiny $R$}}(\beta)$.

It is no coincidence that the coefficients $A_{k1}^\infty$, $A_{k1}^0$, $A_{01}^\infty$
and $A_{01}^0$ occur in the asymptotic expansions for both Models~2 \& 3.
The same thing will happen for any two models having the same powers
of $\omega\tau$ in the expansions of $u_2(\omega)$.

\section{Structural universality}
\label{sec:Universality}

The aim here is to derive the asymptotic behaviour of $U_{k0}$ \& $U_{kk}$ given only
the information in Eq.~(\ref{eqn:intro:u2}).
After first introducing an unspecified time-scale parameter $\tau$ so that the problem
may be written in the dimensionless form of Eq.~(\ref{eqn:asymptotics:MTM:Ikl}),
the MTM can be used to deduce the most sigificant terms in the asymptotic expansions
of $U_{k0}$ \& $U_{kk}$.
The details of this laborious process have been included in the Supporting Material
(Sec.~\ref{sub:supp:Universality}).

\subsection{High-frequency limit}
\label{sub:Universality:HiLim}

In this limit the known terms of $I_{kl}^{\mbox{\tiny $P$}}(\beta)$ dominate the
leading term of $I_{kl}^{\mbox{\tiny $R$}}(\beta)$ so that
the asymptotic behaviour may be summarised as
\begin{equation}
   U_{kl} \sim 
    2\Dzero - \Czero_{l1}(k)\,{\omega_k}^{-1/2}, \qquad \omega_k\tau\rightarrow\infty
 \label{eqn:Universality:HiLim:Ukl}
\end{equation}
where $\Czero_{l1}(k)/\czero=A_{l1}^0$.
The coefficients, $\Czero_{01}(k)$ \& $\Czero_{k1}(k)$, are respectively
proportional to $\czero$ via the same factors, $A_{01}^0$ \&
$A_{k1}^0$, that appeared in the results for Models~2 \& 3.
This was expected because those models were constructed in \cite{Novikov2011}
with the universal high-frequency behaviour built in.
For Models~2 \& 3 the constant $\czero$ equates to $\sqrt{2}\zi\tau^{-1/2} \Dzero$ so
that $S/V\equiv \zi d/\sqrt{\tau \Dzero}$.
The high-frequency behaviour of Model~1 is not consistent with the predicted
universality relation in  Eq.~(\ref{eqn:intro:u2}), so there is no $S/V$ equivalent
for that model.

Although Eq.~(\ref{eqn:Universality:HiLim:Ukl}) has been presented with the same
qualitative dependence on frequency
as $u_2(\omega)$, it should be remembered that the true asymptotic variable is $\beta$.
The $k$-dependence of $\Czero_{01}(k)$ \& $\Czero_{k1}(k)$ therefore complicates the
interpretation of $U_{kk}$ \& $U_{k0}$.
For example, even though $U_{kk}$ eventually approaches $2\Dzero$
as ${\omega_k}^{-1/2}$ because $A_{k1}^0\sim 1$ for large $k$
(re Fig.~\ref{fig:asymptotics:apply:A0k1etc}a), plots of $U_{kk}$ versus $\omega_k$
may show substantial deviation from that behaviour for the low values of $k$
accessible with OGSE-DWI.
On the other hand,
$A_{01}^0\,{\varpi_k}^{-1/2}\sim(4/3)\sqrt{2/\pi}+{\varpi_k}^{-3/2}$
for large $k$ (re Fig.~\ref{fig:asymptotics:apply:A0k1etc}b), so, rather than
varying as ${\omega_k}^{-1/2}$, plots of $U_{k0}$ against $\omega_k$ will quickly
approach a constant value that is consistent with the result
$\lim_{k\rightarrow\infty} U_{k0}=2D(T)$ proved in \cite{Kershaw2021}.

\subsection{Low-frequency limit}
\label{sub:Universality:LoLim}

The qualitative behaviour in the low-frequency limit depends on
the value of $\vartheta$.
For $\vartheta>0$ there are three separate cases:
\begin{enumerate}
\item[(a)]
$0<\vartheta<3$. 
In this case the leading $O(\beta^{-\vartheta-1})$ term of
$I_{kl}^{\mbox{\tiny $P$}}(\beta)$ dominates the $O(\beta^{-4})$
term of $I_{kl}^{\mbox{\tiny $R$}}(\beta)$.
Accordingly, the latter term is ignored and the asymptotic behaviour of $U_{kk}$ \&
$U_{k0}$ is
\begin{equation}
   U_{kl} \sim 2\Dinf + \Cinf_{l1}(k,\vartheta)\,{\omega_k}^\vartheta,
                                           \qquad  \omega_k\tau\rightarrow0
 \label{eqn:Universality:LoLim:Ukl1}
\end{equation}
with $\Cinf_{l1}(k,\vartheta)/\cinf=B_{kl}
M[h_{kl};\vartheta+1]/{\varpi_k}^{\vartheta}$.

\item[(b)]
$\vartheta=3$.
This is a special case where, since the pole at $\vartheta+1=4$ of $M[f_1;1-s]$ coincides
with the first pole of $M[h_{kl};s]$, $G_{l1}(s)$ has a second order pole.
As a consequence, the leading term of $I_{kl}^{\mbox{\tiny $Q$}}(\beta)$ dominates
everything except the leading $O(\beta^{-1})$ term of $I_{kl}^{\mbox{\tiny $P$}}(\beta)$.
The asymptotic behaviour of the observed quantities is therefore distinguished by a
logarithmic dependence on frequency
\begin{equation}
   U_{kl} \sim 2\Dinf - \Cinf_{l2}(k)\, {\omega_k}^3\ln\omega_k\tau,
                    \qquad \omega_k\tau\rightarrow0
 \label{eqn:Universality:LoLim:Ukl2}
\end{equation}
where $\Cinf_{l2}(k)/\cinf=B_{kl}/{\varpi_k}^3$.

\item[(c)]
$\vartheta>3$.
This case is the reverse of the $0<\vartheta<3$ case as the $O(\beta^{-\vartheta-1})$
term of $I_{kl}^{\mbox{\tiny $P$}}(\beta)$ is now subdominant to the $O(\beta^{-4})$
term of $I_{kl}^{\mbox{\tiny $R$}}(\beta)$. The qualitative behaviour of $U_{kk}$ \&
$U_{k0}$ therefore differs from that of $u_2(\omega)$:
\begin{equation}
   U_{kl} \sim 2\Dinf + \Cinf_{l3}(k)\,{\omega_k}^3, 
                                            \qquad \omega_k\tau\rightarrow0 \\
 \label{eqn:Universality:LoLim:Ukl3}
\end{equation}
where $\Cinf_{l3}(k)/2\Dzero=B_{kl}\tau^3 M[f;-3]/{\varpi_k}^3$.
Clearly, for systems of this type the global information associated with $\vartheta$
is obfuscated by an $O({\omega_k}^3)$ term.

\end{enumerate}

Details on how $M[h_{kl};\vartheta+1]$ may be evaluated
can be found in Sec.~\ref{sub:supp:Mhkl} of the Supporting Material,
while Sec.~\ref{sub:supp:Cl1inf} considers the behaviour of
$\Cinf_{l1}(k,\vartheta)/\cinf$ with respect to $\vartheta$ and $k$.
From that information it can be shown that for large $k$,
\begin{equation*}
   \Cinf_{k1}(k,\vartheta)\,{\omega_k}^\vartheta \sim O(k^\vartheta)
\end{equation*}
and
\begin{equation*}
   \Cinf_{01}(k,\vartheta)\,{\omega_k}^\vartheta \sim \lb\{ \begin{aligned}
     & O(1), & & 0<\vartheta<1 \\
     & O(\ln k), & \quad & \vartheta=1 \\
     & O(k^{\vartheta-1}), & & \vartheta>1. \\
   \end{aligned} \rb.
\end{equation*}
That is, the apparent frequency-dependence of $U_{kk}$ when plotted against
$\omega_k$ will approach the frequency-dependence of $u_2(\omega)$,
while the apparent frequency-dependence of $U_{k0}$ will never
match the frequency-dependence of the spectral density.

\section{Discussion}
\label{sec:Discussion}

The MTM has been introduced as a general technique that can be applied to derive
the asymptotic behaviour of $U_{kk}$ \& $U_{k0}$ in both the low and high frequency
limits.
Exact forms for $U_{kk}$ \& $U_{k0}$ for Models~1 \& 2 were calculated previously
in \cite{Kershaw2021}, and expansion of those results verifies the results of the MTM
when applied to those models.
The advantage of the MTM is that it can be used even when the exact forms of $U_{kk}$
\& $U_{k0}$ are difficult to evaluate, as was the case for Model~3, or when the
precise form of $u_2(\omega)$ is unknown, as was true in
Sec.~\ref{sec:Universality}.


\subsection{Apparent diffusion coefficient}
\label{sub:Discussion:ADC}

Given the popularity of the apparent diffusion coefficient (ADC) signal model for the
interpretation of DWI data, it is pertinent to also mention the 
asymptotic behaviour of the ADC.
A relationship between the ADC for a single-harmonic MPG
(i.e.~Eq.~(\ref{eqn:intro:g(t)})), and $U_{kk}$ \& $U_{k0}$ was presented as Eq.~(27)
in \cite{Kershaw2021}.
Inserting Eq.~(\ref{eqn:Universality:HiLim:Ukl}) into that relationship in the
appropriate way, the asymptotic behaviour
in the high-frequency limit is
\begin{equation}
   \ADC_k(\phi) \sim  \Dzero - \Czerol(k,\phi)\,{\omega_k}^{-1/2}, 
 \label{eqn:Discussion:ADC:HiLim}
\end{equation}
with $\Czerol(k,\phi)/\czero=(A_{k1}^0+2A_{01}^0\sin^2\phi)/(2+4\sin^2\phi)$.

An expression for the universal high-frequency behaviour of the ADC was previously
derived in \cite{Sukstanskii2013}.
Also, although the \ADC\ is not specifically mentioned, the high-frequency behaviour
of the signal observed using an oscillating MPG was discussed in an appendix of
\cite{Novikov2019}. After a suitable translation between the notations used in
\cite{Sukstanskii2013} \& \cite{Novikov2019} and that used here, it is found that
the results presented in those studies are equivalent to
Eq.~(\ref{eqn:Discussion:ADC:HiLim}).
In particular, the quantities $c'(\vphi,N)$ (in Eq.~(14) of \cite{Sukstanskii2013})
and $\tilde{c}(\vphi,N)/\sqrt{2}$ (in Eq.~(C6) of \cite{Novikov2019}) are equivalent to
$\sqrt{2} \Czerol(k,\phi)/\czero$.
More details of the comparison may be found in the Supporting Material
(Sec.~\ref{sub:supp:ADCcomments}).

An asymptotic expansion for the ADC in the low-frequency limit could be derived by
inserting
Eqs.~(\ref{eqn:Universality:LoLim:Ukl1})--(\ref{eqn:Universality:LoLim:Ukl3})
into the aforementioned relationship,
but that procedure is complicated by the dependence on $\vartheta$.
A more straightforward approach using the MTM is presented in
Sec.~\ref{sub:supp:ADCLoLim} of the Supporting Material.
For $\phi=0$ the resulting expansion is equivalent to that for $U_{kk}/2$
(re Sec.~\ref{sub:Universality:LoLim}) as expected.
For the other extreme when $\phi=\pi/2$, the result may be summarised as
\begin{equation}
   \ADC_k(\pi/2) \sim \lb\{ \begin{aligned}
   & \Dinf + \Cinf_1(k,\pi/2,\vartheta)\,{\omega_k}^{\vartheta},
                                                      & & 0<\vartheta<5 \\
   & \Dinf - \Cinf_2(k,\pi/2)\,{\omega_k}^{5} \ln\omega_k\tau, & \quad & \vartheta=5 \\
   & \Dinf + \Cinf_3(k,\pi/2)\,{\omega_k}^{5}, & & \vartheta>5
     \end{aligned} \rb.
 \label{eqn:Discussion:ADC:ADCk(pi/2)}
\end{equation}
where $\Cinf_1(k,\pi/2,\vartheta)/\cinf=2{\varpi_k}^{2-\vartheta}(M[h_{kk};\vartheta+1]
-M[h_{k0};\vartheta+1])/3\pi$, $\Cinf_2(k,\pi/2)/\cinf=1/3\pi^2 k$ and
$\Cinf_3(k,\pi/2)/\Dzero=M[f;-5] \tau^5/3\pi^2 k$.
Sec.~\ref{sub:supp:ADCLoLim} also considers the behaviour of
$\Cinf_1(k,\pi/2,\vartheta)$ with respect to $\vartheta$ and $k$.

\begin{figure}[t]
 \centering
 \begin{tabular}{c}
  \includegraphics[scale=0.72]{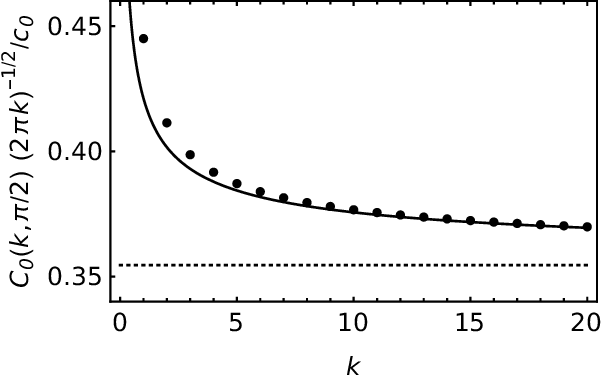}
 \end{tabular}
 \caption{
Plot of $\Czerol(k,\pi/2){\varpi_k}^{-1/2}/\czero$ as a function of $k$.
The solid line represents the asymptotic behaviour
$\Czerol(k,\pi/2)\,{\varpi_k}^{-1/2}/\czero \sim (4/9)\sqrt{2/\pi}+1/6{\varpi_k}^{1/2}$
for large $k$, while the dotted line is the limit $(4/9)\sqrt{2/\pi}$ as
$k\rightarrow\infty$.
}
 \label{fig:Discussion:practice:C0(k,pi/2)}
\end{figure}

\subsection{OGSE-DWI in practice}
\label{sub:Discussion:practice}

This study has concentrated on a form of OGSE-DWI where
the applied MPG is a pure single-harmonic waveform (i.e.~Eq.~(\ref{eqn:intro:g(t)})).
Unfortunately, the only way that a single-harmonic MPG can be implemented in
practice is when $\phi=\pi/2$,
which implies that precise independent measurements of $U_{kk}$ \& $U_{k0}$ are
impractical. The only quantity that can
actually be measured with {\it pure} OGSE-DWI is $\ADC_k(\pi/2)$.
Nevertheless, assuming that the range of accessible frequencies
($\approx10^1$--$10^3$~Hz) falls in either the high- or low-frequency regime for
a given system, measurements of $\ADC_k(\pi/2)$ can be used to extract structural
information in certain situations.
In the high-frequency limit, the $k$-dependence of $\Czerol(k,\pi/2)$ in 
Eq.~(\ref{eqn:Discussion:ADC:HiLim}) means that a slightly more subtle analysis is
required than simply fitting for a constant coefficient of ${\omega_k}^{-1/2}$.
$\Czerol(k,\pi/2)$ combines the
characteristics of both $A_{k1}^0$ and $A_{01}^0$
(re Fig.~\ref{fig:asymptotics:apply:A0k1etc}), so that
$\Czerol(k,\pi/2)\,{\varpi_k}^{-1/2}/\czero \sim (4/9)\sqrt{2/\pi}+1/6{\varpi_k}^{1/2}$
for large $k$ (Fig.~\ref{fig:Discussion:practice:C0(k,pi/2)}).
That is, measurements of the ADC will gradually approach a constant value as
${k}^{-1/2}$, and so independent estimates of $\Dzero$ and $\czero$ (and hence $S/V$)
may be possible if sufficient signal-to-noise is available.

The situation in the low-frequency limit is more complicated due to the added
dependence on $\vartheta$.
An analysis of the behaviour of the coefficient
$\Cinf_1(k,\pi/2,\vartheta)$ in Eq.~(\ref{eqn:Discussion:ADC:ADCk(pi/2)})
(see Supporting Material, Sec.~\ref{sub:supp:ADCLoLim}) finds that for large $k$
and $\vartheta>0$,
\begin{equation}
   \Cinf_1(k,\pi/2,\vartheta)\,{\omega_k}^\vartheta \sim O(k^\vartheta).
 \label{eqn:Discussion:practice:C1inf}
\end{equation}
So, in the low-frequency limit the apparent frequency-dependence of $\ADC_k(\pi/2)$
when plotted against $\omega_k$ will tend towards that of the spectral density when
$0<\vartheta<5$, and hence it is likely that structural information can be extracted
from the data for samples falling into that range of $\vartheta$.
On the other hand,
Eq.~(\ref{eqn:Discussion:ADC:ADCk(pi/2)})
indicates that the $O({\omega_k}^\vartheta)$ term will
always be subdominant to, and hence obfuscated by, an $O({\omega_k}^5)$ term when
$\vartheta>5$.
It is also unlikely that the logarithmic dependence on frequency corresponding to the
special case $\vartheta=5$ can be clearly identified within the
narrow range of frequencies accessible with OGSE-DWI.
Using pure OGSE-DWI to characterise the structural organisation of a
material with a dynamic exponent falling outside of the range $0<\vartheta<5$ may be
challenging.

It was suggested in \cite{Kershaw2021} that OGSE-DWI experiments be performed by
varying $k$ while keeping $T$ constant at its longest possible value, but such a
proctocol is not essential.
The same results would apply for the asymptotics of the ADC
if the experiments were performed after first setting $k$ and then varying $T$,
or even if some hybrid protocol were used where both $k$ and $T$ are systematically
varied.
Even though the limits on the spectral range will remain unaltered, using an
acquisition protocol where $T$ is varied has the potential benefit of improved
spectral resolution.
Making acquistions for multiple values of $T$ might also aid in the estimation of
$\vartheta$.
Rather than being a simple fit of $\ADC_k(\pi/2)$ against a single frequency variable,
the estimation problem requires that data is fitted against the two
independent variables $k$ and $T$.
Whatever protocol is adopted for the lab, it is important to remember the $k$ and $T$
dependence of the data in the analysis.

Instead of a pure single-harmonic MPG, OGSE-DWI is commonly performed using
some type of approximation to a cosinusoidal waveform
(e.g.~\cite{Parsons2003,Does2003}), or at least a waveform that has the symmetry
$g(t)=g(T-t)$ (e.g.~\cite{Callaghan1995,Callaghan1996}).
As previously noted in \cite{Kershaw2021}, any such waveform will contain
contributions from multiple harmonics so the precise signal equation will
be more complicated than the specialised case in Eq.~(\ref{eqn:intro:lnSk}).
More important to accurate quantification of complex microstructures,
it is unclear how well the expansions derived for $U_{kk}$
in this study will hold as approximations to the asymptotic behaviour of the
signal when observed with a multiharmonic MPG.
The effects and potential errors associated with multiharmonic
MPGs will be investigated in a forthcoming article.


\section{Conclusions}
\label{sec:Conclusion}

The goal of this study was to investigate how the global features associated with
the universality relations in Eq.~(\ref{eqn:intro:u2}) emerge in the asymptotic
behaviour of OGSE-DWI signal. The MTM was introduced as a technique to achieve
that goal without having to first evaluate exact forms for $U_{kk}$ \& $U_{k0}$.
Given that the limitations on $T$ and $k$ restrict the range of frequencies
accessible to OGSE-DWI, it is the value of $T$ relative to the time-scale $\tau$
characterising the response of the system that actually determines whether the
condition for observations in either asymptotic regime is met or not.
In the high-frequency limit, although
$U_{kk}$ \& $U_{k0}$ have the same qualitative dependence on frequency as
$u_2(\omega)$, the $k$-dependence of the coefficients complicates the interpretation.
In the low-frequency limit, the asymptotic behaviour of $U_{kk}$ \& $U_{k0}$ is
further complicated by the dependence on $\vartheta$.
In some circumstances the true asymptotic behaviour of the system
may be obscured by the characteristics of the filters $H_{kl}(\omega;T)$.

Asymptotic expansions were also derived for the ADC associated with a single-harmonic
MPG. The expression for the high-frequency limit was found to be equivalent to the
results presented in two earlier studies.
Unfortunately, it is not possible to make specific independent measurements of
$U_{kk}$ \& $U_{k0}$ in practice due to the difficulty associated with implementing
a pure cosinusoidal MPG. This means that the only quantity that can be measured
with a single-harmonic MPG is $\ADC_k(\pi/2)$.
Nevertheless, measurements of $\ADC_k(\pi/2)$ can be used to estimate $S/V$ when
the window of frequencies accessible with OGSE-DWI lies in the high-frequency
regime for a particular sample. Estimates of $\vartheta$ in the low-frequency limit
might also be possible for materials with $0<\vartheta<5$, but outside of that range
producing reliable estimates may be challenging.
In any case, the estimation procedure requires that the data is fitted against
both $k$ and $T$ rather than a single frequency variable.
Methods to analyse the effects of multiharmonic MPGs will be addressed in a future
article.

\newpage

\pagestyle{empty}

\begin{center}

%


\tablefirsthead{\multicolumn{3}{c}{ } \\
  \multicolumn{3}{c}{\bf List of Symbols} \\[2mm]
   \hline \\[-2mm] {\bf Symbol} & \multicolumn{1}{c}
    {\bf Definition} & {\bf First use} \\[2mm]  \hline}

\tablehead{\hline}
\tabletail{\hline}

\begin{supertabular}{lll}

$A_{01}^\infty$, $A_{01}^0$ & $k$-dependent coefficients present in the expansions
           of $U_{k0}$ & Eq.~(\ref{eqn:asymptotics:apply:Model2:Uk0}) \\
                    & for Models~2 \& 3 \\
$A_{k1}^\infty$, $A_{k1}^0$ & $k$-dependent coefficients present in the expansions
           of $U_{kk}$ & Eq.~(\ref{eqn:asymptotics:apply:Model2:Ukk}) \\
                    & for Models~2 \& 3 \\
$\ADC_k(\phi)$ & apparent diffusion coefficient for a single-harmonic MPG
                                           & Eq.~(\ref{eqn:Discussion:ADC:HiLim})\\
          &  of frequency $\omega_k$ and phase $\phi$ & \\
$a_{lj}$ & lower limit of the analytic strip of $G_{lj}(s)$ 
                                           & Sec.~\ref{sub:asymptotics:MTM} \\
$\aSk$, $\aCk$ & symbols used for convenience in Models~2 \& 3 &
                 Eq.~(\ref{eqn:asymptotics:apply:Model2:Uk0})  \\
$B_{kl}$ & constant used in the definition of $H_{kl}(\omega;T)$
                 & Eq.~(\ref{eqn:intro:Hkl}) \\
$b_{lj}$ & upper limit of the analytic strip of $G_{lj}(s)$
                                           & Sec.~\ref{sub:asymptotics:MTM} \\
$\beta$ & ratio of $T$ to $\tau$ & Sec.~\ref{sub:asymptotics:meth} \\
$C(z)$ & Fresnel cosine integral & Sec.~\ref{sec:apply} \\
$\Czero_{l1}(k)$ & coefficient in the high-frequency universal behaviour
        of $U_{kl}$ & Eq.~(\ref{eqn:Universality:HiLim:Ukl}) \\
$\lb. \!\! \begin{aligned}
  &\Cinf_{l1}(k,\vartheta) \\ &\Cinf_{l2}(k) \\ &\Cinf_{l3}(k)
\end{aligned} \rb\}$ & coefficients in the low-frequency universal behaviour
     of $U_{kl}$ & $\begin{aligned} & \mbox{Eq.~(\ref{eqn:Universality:LoLim:Ukl1})} \\
           & \mbox{Eq.~(\ref{eqn:Universality:LoLim:Ukl2})} \\ 
              & \mbox{Eq.~(\ref{eqn:Universality:LoLim:Ukl3})} \end{aligned}$ \\
$\Czerol(k,\phi)$ & coefficient in the
        high-frequency universal behaviour of $\ADC_k(\phi)$ 
                                         & Eq.~(\ref{eqn:Discussion:ADC:HiLim}) \\
$\lb. \!\! \begin{aligned}
  &\Cinf_{1}(k,\vartheta) \\ &\Cinf_{2}(k) \\ &\Cinf_{3}(k)
\end{aligned} \rb\}$ 
     & coefficients in the low-frequency universal behaviour of $\ADC_k(\phi)$
      & Eq.~(\ref{eqn:Discussion:ADC:ADCk(pi/2)})  \\
$\czero$, $\cinf$ & coefficients characterising the universal asymptotic behaviour 
       & Eq.~(\ref{eqn:intro:u2}) \\
        & of the spectral density \\
$\Dzero$, $\Dinf$ & diffusion coefficients associated with the asymptotic limits
       & Eq.~(\ref{eqn:intro:u2}) \\
        & of the spectral density \\
$D(T)$ & cumulative diffusion coefficient at time $T$ & Sec.~\ref{intro} \\
$d$ & spatial dimension & Eq.~(\ref{eqn:intro:u2}) \\
$\delta_{n,m}$ & Kronecker delta & Eq.~(\ref{intro}) \\
$\eta$ & parameter used in Model~1 to represent disorder strength
                                  & Eq.~(\ref{eqn:asymptotics:apply:Model1:u2}) \\
$G$    & amplitude of the MPG & Sec.~\ref{intro} \\
$G_{lj}(s)$ & product of $M[f_j;1-s]$ and $M[h_{kl};s]$ 
                                               & Sec.~\ref{sub:asymptotics:MTM} \\
$f(v)$ & dimensionless form of $u_2(\omega)$ & Eq.~(\ref{eqn:asymptotics:MTM:Ikl}) \\
$f_j(v)$ & partitions of $f(v)$, $j=1,2$ & Eq.~(\ref{eqn:asymptotics:MTM:f1}) \\
$f_{n}^0$, $f_{n}^\infty$ & $n$th coefficients in the expansions of $f(v)$
                  & Eq.~(\ref{eqn:asymptotics:poles:f}) \\
$g(t)$ & MPG waveform & Eq.~(\ref{eqn:intro:g(t)}) \\
$\gamma$ & gyromagnetic ratio of the proton & Eq.~(\ref{eqn:intro:lnSk}) \\
$H_{kl}(\omega;T)$ & kernel in the integral relationship between $U_{kl}$ and
                                 $u_2(\omega)$ & Eq.~(\ref{eqn:intro:Ukl}) \\
$h_{kl}(v)$ & dimensionless form of $H_{kl}(\omega;T)$ 
                                        & Eq.~(\ref{eqn:asymptotics:MTM:Ikl}) \\
$h_{ln}^0$, $h_{ln}^\infty$ & $n$th coefficients in the expansions of $h_{kl}(v)$
                  & Eq.~(\ref{eqn:asymptotics:poles:hkl}) \\
$I_{kl}(\beta)$ & dimensionless form of $U_{kl}$ 
                                      & Eq.~(\ref{eqn:asymptotics:MTM:Ikl}) \\
$I_{kl}^j(\beta)$ & partition of $I_{kl}(\beta)$ associated with $f_j(v)$ &
                                Eq.~(\ref{eqn:asymptotics:MTM:IklSeries}) \\
$I_{kl}^P(\beta)$ & partition of $I_{kl}(\beta)$ associated with the set $P_{lj}$ &
                                Eq.~(\ref{eqn:asymptotics:poles:IklP}) \\
$I_{kl}^Q(\beta)$ & partition of $I_{kl}(\beta)$ associated with the set $Q_{lj}$ &
                                Eq.~(\ref{eqn:asymptotics:poles:IklQ}) \\
$I_{kl}^R(\beta)$ & partition of $I_{kl}(\beta)$ associated with the set $R_{lj}$ &
                                Eq.~(\ref{eqn:asymptotics:poles:IklR}) \\
$K_{lpr}^j$ & quantity associated with the residue of the second order
      & Eq.~(\ref{eqn:asymptotics:poles:IklQ}) \\ 
          & poles in the sets $Q_{lj}$, $j=1,2$ \\
$k$    & number of oscillations in the selected MPG & Eq.~(\ref{eqn:intro:g(t)}) \\
$l$    & equals either $0$ or $k$ & Eq.~(\ref{eqn:intro:Ukl}) \\
$M[f;s]$ & Mellin transform of $f(v)$ & Sec.~\ref{sub:asymptotics:MTM} \\
$M[f_j;s]$ & Mellin transform of $f_j(v)$, $j=1,2$ & Sec.~\ref{sub:asymptotics:MTM} \\
$M[h_{kl};s]$ & Mellin transform of $h_{kl}(v)$ & Sec.~\ref{sub:asymptotics:MTM} \\
$\omega$ & angular frequency & Eq.~(\ref{eqn:intro:Ukl}) \\
$\omega_k$ & angular frequency of the selected MPG & Eq.~(\ref{eqn:intro:g(t)}) \\
$P_{lj}$ & sets containing non-negative integers $p$, $j=1,2$ 
                                          & Sec.~\ref{sub:asymptotics:poles} \\
$p$ & non-negative integer indexing the poles of $G_{lj}(s)$
                                          & Sec.~\ref{sub:asymptotics:poles} \\
$p$ & structural exponent & Eq.~(\ref{eqn:intro:u2}) \\
$\phi$ & phase of the selected MPG & Eq.~(\ref{eqn:intro:g(t)}) \\
$\varpi_k$ & $=2\pi k$ & Eq.~(\ref{eqn:intro:Hkl}) \\
$Q_{lj}$ & sets containing non-negative integer pairs $(p,q)$, $j=1,2$
                                          & Sec.~\ref{sub:asymptotics:poles} \\
$R_{lj}$ & sets containing non-negative integers $r$, $j=1,2$
                                          & Sec.~\ref{sub:asymptotics:poles} \\
$r$ & non-negative integer indexing the poles of $G_{lj}(s)$
                                          & Sec.~\ref{sub:asymptotics:poles} \\
$S/V$ & surface-to-volume ratio & Eq.~(\ref{eqn:intro:u2}) \\
$S(z)$ & Fresnel sine integral & Sec.~\ref{sec:apply} \\
$s$ & arbitrary complex variable & Sec.~\ref{sub:asymptotics:MTM} \\
$s_n$ & $n$th pole of $G_{lj}(s)$ & Sec.~\ref{sub:asymptotics:MTM} \\
$s_k(T)$ & signal for single-harmonic MPG of frequency $\omega_k$
                                 & Eq.~(\ref{eqn:intro:lnSk}) \\
$T$    & duration of a MPG & Eq.~(\ref{eqn:intro:g(t)}) \\
$t$ & time variable & Sec.~\ref{eqn:intro:g(t)} \\
$\tau$ & correlation time & Sec.~\ref{sub:asymptotics:meth} \\
$\vartheta$ & dynamical exponent & Eq.~(\ref{eqn:intro:u2}) \\
$\theta_{1n}$, $\theta_{2n}$ & $n$th powers of $v$ in the expansions of $f(v)$
                                  & Eq.~(\ref{eqn:asymptotics:poles:f}) \\
$U_{kl}$ & $l=0$ or $k$; quantities in the signal equation for a single
           harmonic MPG & Eq.~(\ref{eqn:intro:lnSk}) \\
$u_2(\omega)$ & spectral density of molecular diffusion
                                         & Sec.~\ref{intro}\\
$v$ & dimensionless variable equal to $\omega\tau$ 
                                    & Eq.~(\ref{eqn:asymptotics:MTM:Ikl}) \\
$v_0$ & arbitrary positive real number & Eq.~(\ref{eqn:asymptotics:MTM:f1}) \\
$\zi$  & parameter related to the volume fraction occupied by permeable
                                 & Eq.~(\ref{eqn:asymptotics:apply:Model2:Uk0})  \\
       &  membranes in Models~2 \& 3 &  \\

\end{supertabular}

\end{center}

\newpage


\addtocontents{toc}{\setcounter{tocdepth}{2}}

\section*{Supporting Material}

\setcounter{subsection}{0}
\renewcommand{\thesubsection}{S-\arabic{subsection}}
\setcounter{equation}{0}
\renewcommand{\theequation}{S.\arabic{equation}}
\setcounter{figure}{0}
\renewcommand{\thefigure}{S.\arabic{figure}}
\setcounter{table}{0}
\renewcommand{\thetable}{S.\arabic{table}}
\pagestyle{plain}
\setcounter{page}{1}
\setcounter{footnote}{0}

This document contains material supporting the results presented in the
manuscript ``Oscillating-gradient spin-echo diffusion-weighted imaging (OGSE-DWI)
with a limited number of oscillations: II. Asymptotics'' by Jeff Kershaw \&
Takayuki Obata.

Section, equation, figure and table numbers local to this supporting material
are prefaced by an ``S''. All other numbering refers to items in the main text.

\renewcommand\contentsname{\center\normalfont\bf Table of contents}
\begin{minipage}{6.0in}
{\it
\tableofcontents
}
\end{minipage}
\vspace{5mm}

\subsection{Asymptotic expansion of integrals using the Mellin transform method}
\label{sub:supp:MTM}

Let $a, b\in\mathbb{R}$ and
let $h(x)$ be a locally integrable function\footnote{A function $h(x)$ is locally
integrable on $(0,\infty)$ if $\int_{x_1}^{x_2}dx\,h(x)<\infty$ for all $x_1$,
$x_2$ that satisfy $0<x_1<x_2<\infty$.}
on $(0,\infty)$ such that
$h(x)\sim O(x^{-a})$ as $x\rightarrow0$ and $h(x)\sim O(x^{-b})$ as
$x\rightarrow\infty$.
Then, if $a<b$ and $s\in\mathbb{C}$, the {\em Mellin transform} of $h(x)$ is
defined as
\begin{equation}
   M\lb[h;s\rb] = \int_0^\infty\!\!dx\,x^{s-1}h(x),
 \label{eqn:supp:MTM:MT}
\end{equation}
which is analytic in the infinite strip $a<\re\,s<b$ parallel to the imaginary axis.
The corresponding inversion formula is
\begin{equation}
   h(x) =\frac{1}{2\pi i}\int_{r-i\infty}^{r+i\infty}\!\!ds\,x^{-s}M\lb[h;s\rb]
 \label{eqn:supp:MTM:IMT}
\end{equation}
with $r$ being a real number such that $a<r<b$.
Also, note that by making a simple change of variables in Eq.~(\ref{eqn:supp:MTM:MT})
it can be shown for real $\lambda>0$ that
\begin{equation}
   M[h(\lambda x);s]=\lambda^{-s} M[h;s].
 \label{eqn:supp:MTM:M[h(lambda x);s]}
\end{equation}

Even though Eq.~(\ref{eqn:supp:MTM:MT}) only defines $M[h;s]$ for the analytic
strip $a<\re\,s<b$, it is a useful property of the Mellin transform that it
can be analytically continued to a meromorphic function throughout the whole
complex plane \cite{BleisteinHandelsmanBook}.
Moreover, the behaviour of $M[h;s]$ near the poles can be predicted from the
asymptotic behaviour of $h(x)$. Specifically, if
\begin{equation}
   h(x) \sim \sum_{n=0}^\infty \sum_{m=0}^{M^0_{n}} h_{nm}^0
                        \lb(\ln x\rb)^m x^{a_n} , \qquad x\rightarrow0
 \label{eqn:supp:MTM:h->0}
\end{equation}
with $\re\,a_n$ being a strictly monotonic increasing sequence of numbers,
starting from $\re\,a_0=-a$,
and $M^0_n$ is a finite non-negative integer for each $n$,
then $M[h;s]$ can be continued analytically into the left-hand plane $\re\,s<a$ and
\begin{equation}
   M[h;s] \sim \sum_{m=0}^{M^0_n}\frac{(-)^m h_{mn}^0 m!}{(s+a_n)^{m+1}},
                                                 \qquad s\rightarrow -a_n.
 \label{eqn:supp:MTM:analytCont0}
\end{equation}
Similarly, if
\begin{equation}
   h(x) \sim \sum_{n=0}^\infty \sum_{m=0}^{M^\infty_n} h_{nm}^\infty
                        \lb(\ln x\rb)^m x^{-b_n}, \qquad x\rightarrow\infty,
 \label{eqn:supp:MTM:h->infty}
\end{equation}
where $\re\,b_n$ is a strictly monotonic increasing sequence of numbers,
beginning with $\re\,b_0=b$, and $M^\infty_n$ is a finite
non-negative integer for each $n$,
then $M[h;s]$ can be continued analytically into the right-hand plane $\re\,s>b$ with
\begin{equation}
   M[h;s] \sim -\sum_{m=0}^{M^\infty_n}\frac{(-)^{m} h_{nm}^\infty m!}
                                    {(s-b_n)^{m+1}}, \qquad s\rightarrow b_n.
 \label{eqn:supp:MTM:analytContInfty}
\end{equation}
Notice that the poles of $M[h;s]$ in the left-hand (or right-hand) complex plane from the
analytic strip correspond to
the powers of $x$ in the expansion about 0 (or $\infty$). Furthermore, the order of the
$n$th pole is determined by the highest power of $\ln x$ in the $n$th term of the
corresponding expansion. If $M^0_n=M^\infty_n=0$ $\forall\,n$ then the poles of $M[h;s]$
are all simple.

There is also an interesting relationship between the Mellin and Fourier transforms.
After making the change of variable $y=\ln x$ in
Eq.~(\ref{eqn:supp:MTM:MT}) it is easy to show that
\begin{equation}
   M[h;s] = \int_{-\infty}^\infty\!\!dy\,e^{\sigma y}
                                                       h(e^y) e^{i\nu y}
\end{equation}
where $\sigma=\re\,s$.
That is, the Mellin transform of $h(x)$ is equal to the Fourier transform of
the function $h_\sigma(y)=e^{\sigma y} h(e^y)$.
This is useful because if $h_\sigma(y)$ is integrable on $\mathbb{R}$, then
the Riemann-Lebesgue lemma implies that
\begin{equation}
   \lim_{|\nu|\rightarrow\infty}M[h;s]=0
 \label{eqn:supp:MTM:Riemann-Lebesgue}
\end{equation}
with $\nu=\im\,s$.
Note that $h_\sigma(y)$ will always be integrable for $\sigma$ within the
analytic strip.
Similar arguments as those used to
analytically continue the Mellin transform can then be used to show that the
limit is true $\forall\sigma\in\mathbb{R}$.

%

\begin{figure}[bht]
 \centering
  \includegraphics[scale=0.52]{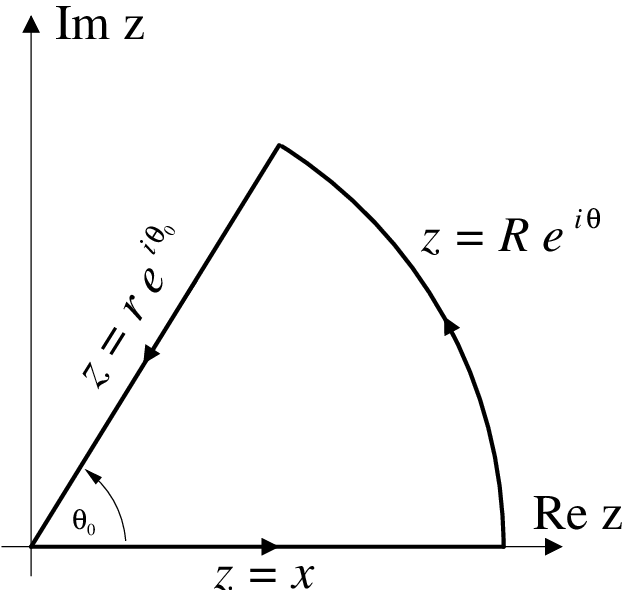}
 \caption{
Contour $S$ used to obtain Eq.~(\ref{eqn:supp:MTM:sector}). The contour
consists of three lines with the parametrisation for each written next to it.
When $\theta_0<0$ is chosen the contour is reflected about the $\re\,z$ axis
to lie in the lower-half plane.
}
 \label{fig:supp:MTM:Sector}
\end{figure}

Another useful identity for Mellin transforms can be obtained by integrating
$z^{s-1} h(z)$ around the sector S shown in Fig.~\ref{fig:supp:MTM:Sector}. Here
$\theta_0$ is chosen so that the poles and branch points of $h(z)$ are excluded
from being either on or inside the boundary of the sector.
$\theta_0$ is also selected as positive or negative according to whether
$\nu$ is positive or negative. 
When $\sigma$ is within the analytic strip it is straightforward to show using
the asymptotic behaviour of $h$ that
the integral along the arc approaches zero as $R\rightarrow\infty$.
The following relationship  is therefore obtained
\begin{equation}
   M[h(x);s] = e^{i\theta_0 s} \int_0^\infty\!\!dr\, r^{s-1} h(r e^{i\theta_0})
             = e^{i\theta_0 s} M[h(r e^{i\theta_0});s].
 \label{eqn:supp:MTM:sector}
\end{equation}
Since the asymptotic behaviour of $h(r e^{i\theta_0})$ wrt $r$ is similar to that
of $h(x)$ wrt $x$, all of the preceding results for Mellin transforms
can be shown to also apply for $M[h(r e^{i\theta_0});s]$. In particular, it
is analytic for $a<\sigma<b$, goes to zero as $\nu\rightarrow\infty$
(re Eq.~(\ref{eqn:supp:MTM:Riemann-Lebesgue})) and can be analytically continued
into the whole complex plane.
This means that
\begin{equation}
   \lb|M[h(x);s]\rb| \sim O(e^{-\theta_0\nu}), \qquad |\nu|\rightarrow\infty.
\end{equation}
Analytic continuation guarantees that this behaviour holds
$\forall\sigma\in\mathbb{R}$.

The Parseval formula for Mellin transforms is introduced next. Analogous to the
case for $h(x)$, let $f(x)$ be another locally integrable function such that
\begin{equation}
   f(x) \sim \lb\{ \begin{aligned}
       & \sum_{n=0}^\infty \sum_{m=0}^{M_n^0} f_{nm}^0 \lb(\ln x\rb)^m x^{c_n},
                                                & & x\rightarrow0 \\
       & \sum_{n=0}^\infty \sum_{m=0}^{M_n^\infty} f_{nm}^\infty \lb(\ln x\rb)^m
                                   x^{-d_n}, & & x\rightarrow\infty
    \end{aligned} \rb.
 \label{eqn:supp:MTM:f}
\end{equation}
with $\re\,c_0=-c$ and $\re\,d_0=d$. Assume also that $c<d$.
Together with the previously stated condition that $a<b$, this assumption ensures
that the analytic strip $a+c<\re\,s<b+d$ of $M[f h;s]$ is always nonempty.
Now, using Eqs.~(\ref{eqn:supp:MTM:MT}) \& (\ref{eqn:supp:MTM:IMT}) it is possible
to show that
\footnote{
To use the inverse transform of $M[h;s]$ the real number $r$ must be
within the analytic strip $a<\re\,s<b$.
On the other hand, the derivation could have been performed using the inverse
transform of $M[f;s]$ so that
\begin{align*}
   M[f h;s] &= \frac{1}{2\pi i}\int_{r'-i\infty}^{r'+i\infty}
                                         \!\!ds''\,M[h;s-s''] M[f;s''] 
    = \frac{1}{2\pi i}\int_{\sigma-r'-i\infty}^{\sigma-r'+i\infty}
                                         \!\!ds'\,M[h;s'] M[f;s-s'],
\end{align*}
where $r'$ lies in the analytic strip $c<\re\,s<d$ and the change of variable
$s'=s-s''$ has been used.
For this result to be equal to Eq.~(\ref{eqn:supp:MTM:M[fh;s]}),
$\sigma=\re\,s$ has to be such that $r=\re\,s-r'$. 
Since $a<r<b$ and $c<r'<d$, this is equivalent
to saying that $a+c<\re\,s<b+d$, which is the analytic strip of $M[f h;s]$.
That is, the two derivations are consistent as long as $s$ lies within the analytic
strip of $M[f h;s]$.
}
\begin{align}
   M[f h;s] &= \int_0^\infty\!\!dx\,x^{s-1} f(x) h(x) \nonumber \\
    &= \int_0^\infty\!\!dx\,x^{s-1} f(x)\lb\{\frac{1}{2\pi i}
        \int_{r-i\infty}^{r+i\infty}\!\!ds'\,x^{-s'} M[h;s']\rb\}
           \qquad (a<r<b) \nonumber \\
    &= \frac{1}{2\pi i}\int_{r-i\infty}^{r+i\infty}\!\!ds'\,M[h;s']
                        \int_0^\infty\!\!dx\,x^{s-s'-1} f(x) \nonumber \\
    &= \frac{1}{2\pi i}\int_{r-i\infty}^{r+i\infty}\!\!ds'\,M[h;s'] M[f;s-s'].
 \label{eqn:supp:MTM:M[fh;s]}
\end{align}
It is assumed here that the conditions necessary for the interchange of integration
order are satisfied.
The Parseval formula follows immediately from Eq.~(\ref{eqn:supp:MTM:M[fh;s]}) after
setting $s=1$:
\begin{equation}
   \int_0^\infty\!\!dx\,f(x) h(x) = 
            \frac{1}{2\pi i}\int_{r-i\infty}^{r+i\infty}\!\!ds\,M[h;s] M[f;1-s].
 \label{eqn:supp:MTM:Parseval}
\end{equation}

\begin{figure}[tb]
 \centering
 \begin{tabular}{c}
  \includegraphics[scale=0.52]{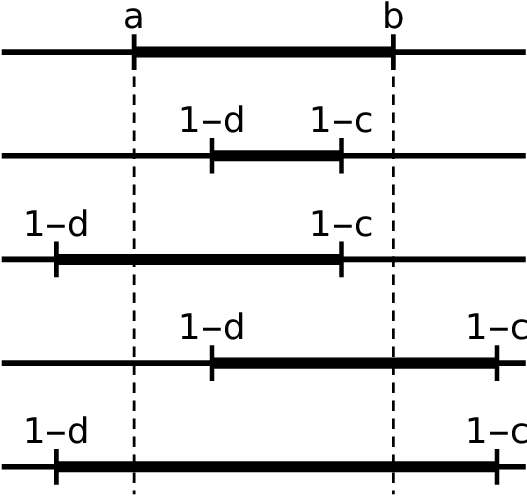}
 \end{tabular}
 \caption{
Intersection between the analytic strips of $M[f;1-s]$ and $M[h;s]$ for the case $a<b$
and $c<d$. In this case the Mellin transforms of both $f(x)$ and $h(x)$ are defined
in the ordinary sense. The upper line represents the analytic strip of $M[h;s]$,
while the lower four lines represent the four distinct cases for how the analytic
strip of $M[f;1-s]$ might overlap. All four cases can be summarised as the interval
$(\max\{a,1-d\},\min\{1-c,b\})$.
}
 \label{fig:supp:MTM:intersect1}
\end{figure}

\begin{figure}[tb]
 \centering
  \includegraphics[scale=0.52]{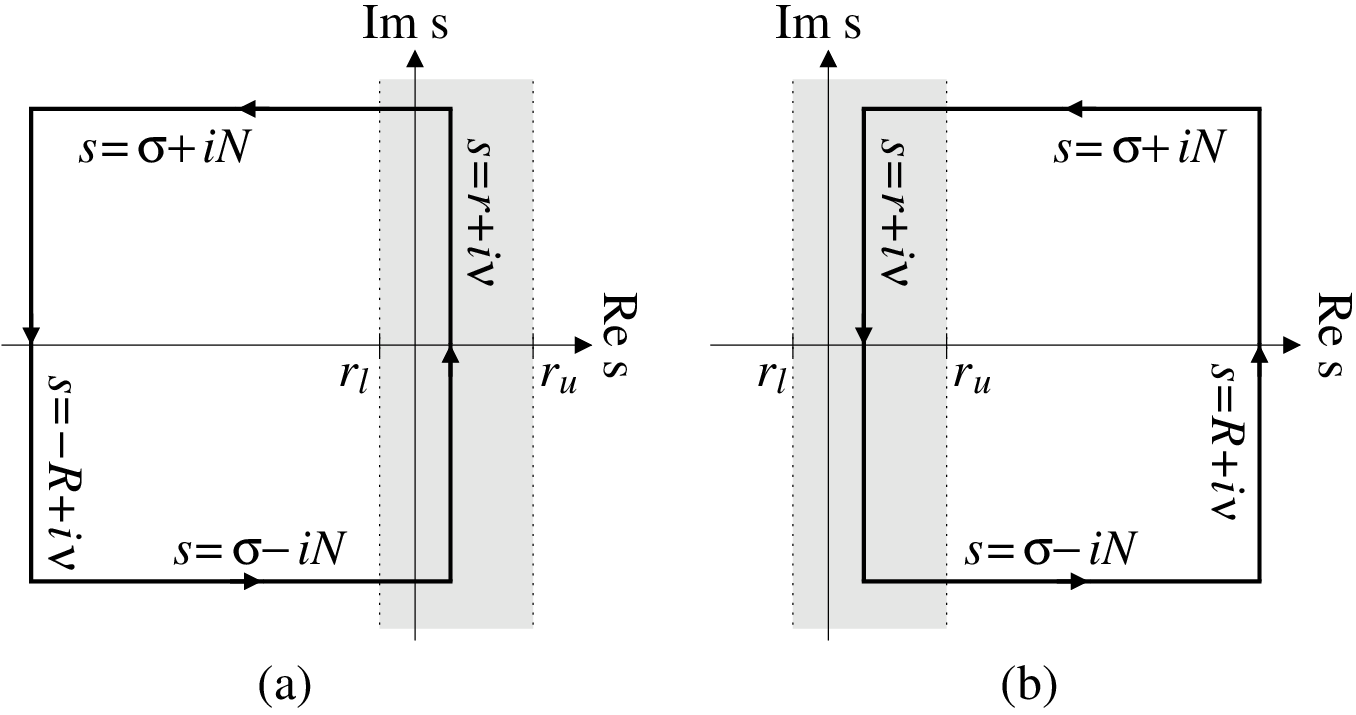}
 \caption{
Contour $C$ used to evaluate the complex integral in Eq.~(\ref{eqn:supp:MTM:G(s)})
when (a) $\lambda\rightarrow0$, and (b) $\lambda\rightarrow\infty$. Each contour
consists of four lines with the parametrisation for each line written next to it.
The shaded area represents the strip of analyticity $r_l<\re\,s<r_u$ of $G(s)$.
}
 \label{fig:supp:MTM:contour}
\end{figure}

Now, using the Parseval formula and Eq.~(\ref{eqn:supp:MTM:M[h(lambda x);s]}), it
is possible to show for $f(x)$ and $h(x)$ as above that
\begin{equation}
   I(\lambda) = \int_0^\infty\!\!dx\,f(x) h(\lambda x) =
     \frac{1}{2\pi i}\int_{r-i\infty}^{r+i\infty}\!\!ds\,\lambda^{-s}
                                                           M[f;1-s]M[h;s],
 \label{eqn:supp:MTM:I(lambda)}
\end{equation}
where $\lambda$ is a positive real number.
On this occasion it is assumed that $r$ lies within the intersection of the
strips of analyticity of $M[h;s]$ (i.e.~$a<\re\,s<b$) and $M[f;1-s]$
(i.e.~$1-d<\re\,s<1-c$).
The intersection will always be nonempty because the conditions for the left-hand side
of Eq.~(\ref{eqn:supp:MTM:I(lambda)}) to converge are
\begin{equation}
   a+c<1 \quad \mbox{\&} \quad b+d>1,
 \label{eqn:supp:MTM:converge}
\end{equation}
which means that $r$ can always be chosen so that $r_l<r<r_u$, with
$r_l=\max\{a,1-d\}$ and $r_u=\min\{b,1-c\}$ (Fig.~\ref{fig:supp:MTM:intersect1}).
%
The object from here is to construct an asymptotic expansion for $I(\lambda)$
as either $\lambda\rightarrow0$ or $\lambda\rightarrow\infty$.
To do this consider the contour integral
\begin{equation}
   \oint_{\mbox{\tiny$C$}} ds\,\lambda^{-s}G(s)
 \label{eqn:supp:MTM:G(s)}
\end{equation}
where $G(s)=M[f;1-s]M[h;s]$ and the contour $C$ is drawn depending on which
limit of $\lambda$ is chosen (Fig.~\ref{fig:supp:MTM:contour}).
If $\lambda\rightarrow0$ (or $\lambda\rightarrow\infty$)
then the contour must be closed in the left-hand (or right-hand) plane so as to
ensure that the term $\lambda^{-s}$ remains finite when $\sigma\rightarrow\infty$.
Treating the $\lambda\rightarrow0$ case first,
if $G(s)$ is such that $|G(\sigma\pm i N)|\rightarrow0$ as $N\rightarrow\infty$ then
the integrals along the horizontal sections will correspondingly
approach zero in the same limit. Furthermore, if $G(s)$ is absolutely integrable
then the integral along the vertical line at $\sigma=-R$ will be $O(\lambda^{-R})$
so that it becomes negligible as $R\rightarrow\infty$.
The contour integral therefore becomes
\begin{equation}
   \oint_C ds\,\lambda^{-s}G(s) = \int_{r-i\infty}^{r+i\infty}\!\!ds\,\lambda^{-s}G(s)
     = 2\pi i\!\!\sum_{\{s_n\leqslant r_l\}} \mbox{Res}\{\lambda^{-s} G(s)\}
\end{equation}
where $\{s_n\leqslant r_l\}$ is the set of all poles $s_n$ in the left-hand plane.
A similar process for the case where $\lambda\rightarrow\infty$ produces 
\begin{equation}
   \oint_C ds\,\lambda^{-s}G(s)
    = -\int_{r-i\infty}^{r+i\infty}\!\!ds\,\lambda^{-s}G(s)
     = 2\pi i\!\!\sum_{\{s_n\geqslant r_u\}} \mbox{Res}\{\lambda^{-s} G(s)\}
\end{equation}
with $\{s_n\geqslant r_u\}$ being the set of all poles in the right-hand plane.
On comparison with Eq.~(\ref{eqn:supp:MTM:I(lambda)}),
it is now clear that these two results provide asymptotic expansions for $I(\lambda)$
as sums over the residues at the poles of $G(s)$. That is,
\begin{equation}
   I(\lambda) \sim \lb\{\!\! \begin{aligned}
        &\sum_{\{s_n\leqslant r_l\}} \mbox{Res}\lb\{\lambda^{-s}
              M[f;1-s]\,M[h;s]\rb\}, & \qquad & \lambda\rightarrow0 \\
        -&\sum_{\{s_n\geqslant r_u\}} \mbox{Res}\lb\{\lambda^{-s}
              M[f;1-s]\,M[h;s]\rb\}, & \qquad &   \lambda\rightarrow\infty.
     \end{aligned} \rb.
 \label{eqn:supp:MTM:ResidueSeries}
\end{equation}
Evaluation of each term in the series is aided by analytic continuation of the
Mellin transform into the whole complex plane and, in particular,
Eqs.~(\ref{eqn:supp:MTM:analytCont0}) \& (\ref{eqn:supp:MTM:analytContInfty}).
As noted earlier, the poles of $M[h;s]$ in the left and right halves of the complex plane
correspond to the powers of $x$ in the asymptotic expansions of $h(x)$
about 0 and $\infty$, respectively.
In contrast, after recalling the asymptotic expansions of $f(x)$ in
Eq.~(\ref{eqn:supp:MTM:f}), then
\begin{equation}
   M[f;1-s] \sim \lb\{\!\! \begin{aligned}
        -&\sum_{m=0}^{M_n^0} \frac{f_{nm}^0 m!}{(s-c_n-1)^{m+1}},
                                          & \qquad & s\rightarrow c_n+1 \\
        &\sum_{m=0}^{M_n^\infty} \frac{f_{nm}^\infty m!}{(s+d_n-1)^{m+1}},
                                        & \qquad &   s\rightarrow -d_n+1.
     \end{aligned} \rb.
 \label{eqn:supp:MTM:M[f;1-s]}
\end{equation}
That is, due to the shift in argument from $s$ to $1-s$, the poles of $M[f;1-s]$
in the left and right halves of the complex plane correspond to 1 minus the powers of
$x$ in the expansions of $f(x)$ about $\infty$ and 0, respectively.


\begin{figure}[tb]
 \centering
  \includegraphics[scale=0.52]{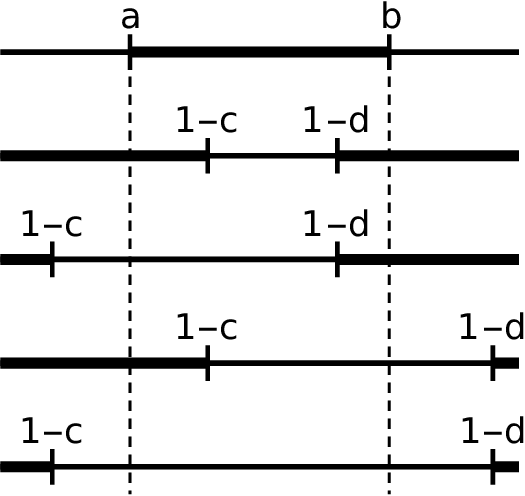}
 \caption{
Intersection between the analytic strips of $M[f;1-s]$ and $M[h;s]$ for the case $a<b$
and $c>d$. In this case the Mellin transform of $f(x)$ only exists in the generalised
sense. The upper line represents the analytic strip of $M[h;s]$,
while the lower four lines represent the four distinct cases for how it may overlap
the analytic strips of $M[f_1;1-s]$ \& $M[f_2;1-s]$. Only the second line is
consistent with the conditions for $I(\lambda)$ to converge
(re Eq.~(\ref{eqn:supp:MTM:converge})), in which case the
intersection of the analytic strips can be summarised as $(a,1-c)\cup(1-d,b)$.
The lower three lines respectively correspond to the cases where $I_1(\lambda)$,
$I_2(\lambda)$ and both $I_1(\lambda)$ \& $I_2(\lambda)$ do not converge.
}
 \label{fig:supp:genMTM:intersect}
\end{figure}

\subsection{Generalised methods for Mellin transforms}
\label{sub:supp:genMTM}

The Mellin transform of the function $f(x)$ was defined in Sec.~\ref{sub:supp:MTM}
under the assumption that $c<d$ so that the analytic strip is nonempty.
However, when $c>d$ the integral in Eq.~(\ref{eqn:supp:MTM:MT}) does not converge
so the Mellin transform in the ordinary sense does not exist.
Nevertheless, analytic continuation can be used to give the Mellin transform
meaning even for functions where $c>d$.

As in Sec.~\ref{sub:supp:MTM}, assume that $f(x)$ is
locally integrable on $(0,\infty)$ and has asymptotic expansions as defined in
Eq.~(\ref{eqn:supp:MTM:f}).
Now choose $x_0>0$ and define the functions
\begin{align}
   f_1(x) &= \lb\{\begin{array}{lll} f(x), & & 0\leqslant x < x_0 \\[1mm]
                        0, & & x_0\leqslant x <\infty \end{array}\rb. 
 \label{eqn:supp:genMTM:f1} \\[2mm]
   f_2(x) &= f(x) - f_1(x).
 \label{eqn:supp:genMTM:f2}
\end{align}
From these definitions and the contents of Sec.~\ref{sub:supp:MTM}
it is immediately true that $M[f_1;s]$ exists and has analytic
strip $c<\re\,s$. Furthermore, $M[f_1;s]$ can be
analytically continued to a meromorphic function in the left-hand plane $\re\,s<c$,
with the asymptotic behaviour of $f_1(x)$ as $x\rightarrow0$ dictating the behaviour
of $M[f_1;s]$ at the poles.
Similar characteristics hold for the analytic continuation of $M[f_2;s]$ into the
right-hand plane $\re\,s>d$.

Now, Eq.~(\ref{eqn:supp:genMTM:f2}) and the analytic continuation of $M[f_1;s]$ \&
$M[f_2;s]$ into the entire complex plane suggest a way to generalise the Mellin
transform of $f(x)$ when $c>d$.
First note that
\begin{equation}
   M[f;s] = M[f_1;s] + M[f_2; s]
 \label{eqn:supp:genMTM:genMTM}
\end{equation}
is clearly true for the Mellin transform of $f(x)$ in the ordinary sense, i.e.~when
$c<d$.  Moreover, the right-hand side is also well defined even when $c>d$ because both
$M[f_1;s]$ and $M[f_2;s]$ exist in the ordinary sense.
Using this fact, the left-hand side of Eq.~(\ref{eqn:supp:genMTM:genMTM}) is therefore
called the {\em Mellin transform of $f(x)$ in the generalised sense}.
As all of the properties of the Mellin transform discussed in Sec.~\ref{sub:supp:MTM}
apply for $M[f_1;s]$ and $M[f_2;s]$, Eq.~(\ref{eqn:supp:genMTM:genMTM}) also provides
a means to predict the behaviour of $M[f;s]$ throughout the whole complex plane when
required.

The asymptotic expansion of $I(\lambda)$ in Eq.~(\ref{eqn:supp:MTM:I(lambda)}) is now
extended for the case when the Mellin transform of $f(x)$ only
exists in the generalised sense.
With $f_1(x)$ \& $f_2(x)$ as defined in
Eqs.~(\ref{eqn:supp:genMTM:f1}) \& (\ref{eqn:supp:genMTM:f2}) and assuming the
conditions for convergence in Eq.~(\ref{eqn:supp:MTM:converge}) still apply,
it is possible to write
\begin{equation}
   I(\lambda) = \int_0^\infty dx\,f(x) h(\lambda x)
    =\int_0^\infty dx\,f_1(x) h(\lambda x)
     + \int_0^\infty dx\,f_2(x) h(\lambda x) = I_1(\lambda) + I_2(\lambda).
 \label{eqn:supp:genMTM:I(lambda)}
\end{equation}
Setting $G_1(s)=M[h;s] M[f_1;1-s]$, since $M[f_1;1-s]$ and $M[h;s]$ both exist in
the ordinary sense and have analytic strips that overlap for $a<\re\,s<1-c$
(Fig.~\ref{fig:supp:genMTM:intersect}), then
the results of Sec.~\ref{sub:supp:MTM} imply that
\begin{equation}
   I_1(\lambda)= 
    \frac{1}{2\pi i}\int_{r_1-i\infty}^{r_1+i\infty}\!\!ds\,\lambda^{-s} G_1(s)
    \sim \lb\{\begin{aligned}
       &\sum_{\{s_n\leqslant a\}} \mbox{Res}\{\lambda^{-s} G_1(s)\}, 
                                              & \qquad & \lambda\rightarrow0 \\
       -&\!\!\sum_{\{s_n\geqslant 1-c\}} \mbox{Res}\{\lambda^{-s} G_1(s)\}, 
                                              & \qquad & \lambda\rightarrow\infty,
     \end{aligned} \rb.
 \label{eqn:supp:genMTM:I1(lambda)}
\end{equation}
where $\{s_n\leqslant a\}$ \& $\{s_n\geqslant 1-c\}$ are the pole sets in
the left- and right-hand sides of the complex plane, respectively.
Similarly, if $G_2(s)=M[h;s] M[f_2;1-s]$ and realising that the analytic strips of
$M[f_2;1-s]$ and $M[h;s]$ intersect for $1-d<\re\,s<b$
(Fig.~\ref{fig:supp:genMTM:intersect}), then
\begin{equation}
   I_2(\lambda)= 
    \frac{1}{2\pi i}\int_{r_2-i\infty}^{r_2+i\infty}\!\!ds\,\lambda^{-s} G_2(s)
    \sim \lb\{\begin{aligned}
       &\!\!\!\sum_{\{s_n\leqslant 1-d\}} \mbox{Res}\{\lambda^{-s} G_2(s)\}, 
                                              & \qquad & \lambda\rightarrow0 \\
       -&\sum_{\{s_n\geqslant b\}} \mbox{Res}\{\lambda^{-s} G_2(s)\}, 
                                              & \qquad & \lambda\rightarrow\infty
     \end{aligned} \rb.
 \label{eqn:supp:genMTM:I2(lambda)}
\end{equation}
with $\{s_n\leqslant 1-d\}$ \& $\{s_n\geqslant b\}$ being the pole sets in
the left- and right-hand sides of the complex plane, respectively.

Placing the series of Eqs.~(\ref{eqn:supp:genMTM:I1(lambda)}) \&
(\ref{eqn:supp:genMTM:I2(lambda)}) into
Eq.~(\ref{eqn:supp:genMTM:I(lambda)}), the overall asymptotic expansion of
$I(\lambda)$ is therefore defined to be
\begin{equation}
   I(\lambda) \sim \lb\{\begin{aligned}
       & \sum_{\{s_n\leqslant a\}} \mbox{Res}\{\lambda^{-s} G_1(s)\} \quad
        + \sum_{\{s_n\leqslant 1-d\}} \mbox{Res}\{\lambda^{-s} G_2(s)\}, 
                                            & \qquad & \lambda\rightarrow0 \\
       -& \!\!\!\sum_{\{s_n\geqslant 1-c\}} \mbox{Res}\{\lambda^{-s} G_1(s)\} 
        \quad - \sum_{\{s_n\geqslant b\}} \mbox{Res}\{\lambda^{-s} G_2(s)\}, 
                                            & \qquad & \lambda\rightarrow\infty.
     \end{aligned} \rb.
 \label{eqn:supp:genMTM:ResidueSeries}
\end{equation}
The residues in each of the series can be evaluated using 
Eqs.~(\ref{eqn:supp:MTM:analytCont0}), (\ref{eqn:supp:MTM:analytContInfty}) \&
(\ref{eqn:supp:MTM:M[f;1-s]}) in a way that is analogous to that described in
Sec.~\ref{sub:supp:MTM}.

\subsection{Some details on the Mellin transforms of $h_{k0}(v)$ \& $h_{kk}(v)$}
\label{sub:supp:Mhkl}

It is helpful to compile a few facts about the Mellin transforms of $h_{k0}(v)$ and
$h_{kk}(v)$.
Starting with $h_{k0}(v)$, the singularities at $v=0$ and $\varpi_k$ are removable so
it is a continuous and locally integrable function on $(0,\infty)$.
It also has the asymptotic expansions
\begin{equation}
   h_{k0}(v) \sim \lb\{ \begin{aligned}
        & \sum_{n=0}^\infty h^{0}_{0n}\,v^{2n}, & & v\rightarrow 0 \\
        & (1-\cos v)\sum_{n=0}^\infty h^{\infty}_{0n}\,v^{-2n-4}, & &
                                            v\rightarrow\infty
    \end{aligned} \rb.
 \label{eqn:supp:Mhkl:hk0(v)_expan}
\end{equation}
with $h^{0}_{0n}=(-)^{n+1} \sum_{m=0}^n (-)^m/[{\varpi_k}^{2(m+1)}(2n-2m+2)!]$
and $h^{\infty}_{0n}= {\varpi_k}^{2n}$.
The powers of the leading terms in the expansions therefore stipulate that
$M[h_{k0};s]$ is analytic for the strip $0<\re\,s<4$.
Moreover, from the asymptotic behaviour as $v\rightarrow0$, $M[h_{k0};s]$ can
be analytically continued into the left-hand plane $\re\,s<0$ with poles at
$s=-2n$, and from the behaviour as $v\rightarrow\infty$, $M[h_{k0};s]$ can be
analytically continued into the right-hand plane $\re\,s>4$ with poles at $s=2n+4$.
Specifically,
\begin{equation}
   M[h_{k0};s] \sim \lb\{ \begin{aligned}
         &\frac{h^{0}_{0n}}{s+2n}, & &  s\rightarrow -2n \\
         -&\frac{h^{\infty}_{0n}}{s-2n-4}, & &  s\rightarrow 2n+4
    \end{aligned} \rb. 
 \label{eqn:supp:Mhkl:Mhk0_poles}
\end{equation}
for $n=0,1,2,\cdots$.

Next, $h_{kk}(v)$ has a removable singularity at $v=\varpi_k$ so it is also
continuous and locally integrable on the positive real line.
It has asymptotic expansions
\begin{equation}
   h_{kk}(v) \sim \lb\{ \begin{aligned}
        & \sum_{n=0}^\infty h^{0}_{kn}\,v^{2n+2}, & & v\rightarrow 0 \\
        & (1-\cos v)\sum_{n=0}^\infty h^{\infty}_{kn}\,v^{-2n-4}, & &
                                            v\rightarrow\infty
    \end{aligned} \rb.
 \label{eqn:supp:Mhkl:hkk(v)_expan}
\end{equation}
with $h^0_{kn}=(-)^n\sum_{m=0}^n (-)^m (m+1)/[{\varpi_k}^{2(m+2)}(2n-2m+2)!]$
and $h^\infty_{kn}=(n+1){\varpi_k}^{2n}$.
From the leading asymptotic behaviour it can be concluded that $M[h_{kk};s]$
is analytic in the strip $-2<\re\,s<4$. More generally, $M[h_{kk};s]$ can be
continued analytically into the whole complex plane such that
\begin{equation}
   M[h_{kk};s] \sim \lb\{ \begin{aligned}
         &\frac{h^{0}_{kn}}{s+2n+2}, & &  s\rightarrow -2n-2 \\
         -&\frac{h^{\infty}_{kn}}{s-2n-4}, & &  s\rightarrow 2n+4
    \end{aligned} \rb. 
 \label{eqn:supp:Mhkl:Mhkk_poles}
\end{equation}
for $n=0,1,2,\cdots$.

The exact values of $M[h_{k0};s]$ and $M[h_{kk};s]$ will also be required as
coefficients in $I_{kl}^{\mbox{\tiny $P$}}(\beta)$ and $I_{kl}^{\mbox{\tiny $Q$}}
(\beta)$ when $s$ is real and not equal to a pole.
In that case the Mellin transforms
can be evaluated using a combination of the results presented in
Sec.~\ref{sub:supp:Integrals}.
Using Eqs.~(\ref{eqn:supp:Integrals:CI2}) and (\ref{eqn:supp:Integrals:H(s;t,c)}),
for $s\in\mathbb{R}$ it is possible to identify the correspondence
\begin{equation}
   M[h_{k0};s] = \lim_{s'\rightarrow s}\cos(\pi s'/2)
                            \lb[H(s'-2;0,\varpi_k)-H(s'-2;1,\varpi_k)\rb],
 \label{eqn:supp:Mhkl:Mhk0}
\end{equation}
where the limit is required for the case when $s=2$ or an odd negative integer.
Similarly, combining Eq.~(\ref{eqn:supp:Integrals:CI2}) with
Eq.~(\ref{eqn:supp:Integrals:I(s;t,c)}) enables the correspondence
\begin{equation}
   M[h_{kk};s] = \frac{\pi{\varpi_k}^{s-3}}{4} +\lim_{s'\rightarrow s} \cos(\pi s'/2)
                                    \lb[ I(s';0,\varpi_k)-I(s';1,\varpi_k)\rb],
 \label{eqn:supp:Mhkl:Mhkk}
\end{equation}
with the limit again necessary for $s=0$, 2 and odd negative integer values.
Table~\ref{tab:supp:Mhkl:Mhkl} presents exact results for selected integer and
half-integer values of $s$.

It is useful to derive an asymptotic expansion for $M[h_{k0};s]$ in
the limit $k\rightarrow\infty$. This can be done by inserting
Eqs.~(\ref{eqn:supp:Integrals:H(s;0,c)}) \& (\ref{eqn:supp:Integrals:H(s;t,c)_rslt})
into Eq.~(\ref{eqn:supp:Mhkl:Mhk0}), and then using the identity
Eq.~(\ref{eqn:supp:Integrals:Gamma}) to write the result in terms of the
incomplete Gamma function $\Gamma(z,t)$. The incomplete Gamma function has the
well-known large $t$ expansion 
\begin{equation*}
   \Gamma(z,i t) \sim (i t)^{z-1} e^{-i t} \sum_{n=0}^\infty
                                   \frac{\Gamma(z)}{\Gamma(z-n)} (i t)^{-n},
\end{equation*}
which can be used to achieve the desired goal. After some manipulation to separate
out the real and imaginary parts, the result is
\begin{equation}
   M[h_{k0};s] \sim -\lim_{s'\rightarrow s}
    \lb\{\frac{\pi{\varpi_k}^{s'-4}}{2}\cot(\pi s'/2) + \frac{\pi}{2}\csc(\pi s'/2)
     \sum_{n=0}^\infty\frac{(-)^n {\varpi_k}^{-2n-2}}{\Gamma(3-s'-2n)} \rb\}.
 \label{eqn:supp:Mhkl:Mhk0_largek}
\end{equation}

A similar procedure can be used to obtain an expansion for $M[h_{kk};s]$.
Equations~(\ref{eqn:supp:Integrals:I(s;t,c)_rslt}) \& (\ref{eqn:supp:Integrals:I(s;0,c)})
are placed in Eq.~(\ref{eqn:supp:Mhkl:Mhkk}), and the expansion for the
incomplete Gamma function is used again to derive
\begin{align}
   M[h_{kk};s] &\sim \frac{\pi {\varpi_k}^{s-3}}{4} -\lim_{s'\rightarrow s}
    \lb\{\frac{\pi(s'-2){\varpi_k}^{s'-4}}{4}\cot(\pi s'/2)  \rb. \nonumber \\
     & \hspace{40mm} \lb. +\frac{\pi}{2}\csc(\pi s'/2)
     \sum_{n=0}^\infty\frac{(-)^n (n+1) {\varpi_k}^{-2n-4}}{\Gamma(1-s'-2n)} \rb\}.
 \label{eqn:supp:Mhkl:Mhkk_largek}
\end{align}

%
%

\begin{table}
 \centering
 \captionsetup{width=10.5cm}
 \caption{Mellin transform of $h_{k0}(v)$ \& $h_{kk}(v)$ for selected values of $s$.}
 \begin{tabular}{@{\vrule height 17.0pt depth 10pt  width0pt}c|cc}
     \hline
      $s$ & $M[h_{k0};s]$ & $M[h_{kk};s]$ \\ \hline
   $2n+4$, $n=0,1,2,\cdots$ & {\it pole} & {\it pole} \\
   $2n+3$, $n=0,1,2,\cdots$ & 0 & $\frac{\pi{\varpi_k}^{2n}}{4}$ \\
   $5/2$ & $-\frac{\pi\aSk}{2{\varpi_k}^{3/2}}$ 
          & $\frac{\pi\aCk}{4{\varpi_k}^{1/2}}-\frac{\pi\aSk}{8{\varpi_k}^{3/2}}$ \\
   2 & $-\frac{\gamma_e+\ln\varpi_k-\Cismall(\varpi_k)}{{\varpi_k}^2}$
                                           & $\frac{\Sismall(\varpi_k)}{2\varpi_k}$ \\
   $3/2$ & $-\frac{\sqrt{2\pi}}{{\varpi_k}^2}+\frac{\pi\aCk}{2{\varpi_k}^{5/2}}$
          & $\frac{\sqrt{2\pi}}{4{\varpi_k}^2}-\frac{\pi\aCk}{8{\varpi_k}^{5/2}}
                                           + \frac{\pi\aSk}{4{\varpi_k}^{3/2}}$ \\
   1 & $-\frac{\pi}{2{\varpi_k}^2}$ & $\frac{\pi}{4{\varpi_k}^2}$ \\
   $1/2$ & $-\frac{2\sqrt{2\pi}}{3{\varpi_k}^2}-\frac{\pi\aSk}{2{\varpi_k}^{7/2}}$ 
          & $\frac{\pi\aCk}{4{\varpi_k}^{5/2}}+\frac{3\pi\aSk}{8{\varpi_k}^{7/2}}$ \\
   0 & {\it pole} & $\frac{\gamma_e+\ln\varpi_k-\Cismall(\varpi_k)}{{\varpi_k}^4} 
                                         + \frac{\Sismall(\varpi_k)}{2{\varpi_k}^3}$ \\
   $-1/2$ & $-\frac{\sqrt{2\pi}}{{\varpi_k}^4}+\frac{4\sqrt{2\pi}}{15{\varpi_k}^2}
          +\frac{\pi\aCk}{2{\varpi_k}^{9/2}}$ & $\frac{5\sqrt{2\pi}}{4{\varpi_k}^4}
             +\frac{\pi\aSk}{4{\varpi_k}^{7/2}}-\frac{5\pi\aCk}{8{\varpi_k}^{9/2}}$ \\
   $-1$ & $\frac{\pi({\varpi_k}^2-6)}{12{\varpi_k}^4}$ & $\frac{3\pi}{4{\varpi_k}^4}$ \\
   $-3$ & $-\frac{\pi({\varpi_k}^4-20{\varpi_k}^2+120)}{240{\varpi_k}^6}$ &
         $-\frac{\pi({\varpi_k}^2-15)}{12{\varpi_k}^6}$ \\
   $-5$ & $\frac{\pi({\varpi_k}^6-42{\varpi_k}^4+840{\varpi_k}^2-5040)}{10080{\varpi_k}^8}$
    & $\frac{\pi({\varpi_k}^4-40{\varpi_k}^2+420)}{240{\varpi_k}^8}$ \\
   $-2n-2$, $n=0,1,2,\cdots$ & {\it pole} & {\it pole} \\
   \hline
 \end{tabular}
 \label{tab:supp:Mhkl:Mhkl}
\end{table}

\subsection{Some details on the asymptotic expansion of $U_{k0}$ \& $U_{kk}$ for
Models~1 \& 2}
\label{sub:supp:Models1&2}


Due to the high level of repetition, only the important intermediate details of the MTM
procedure have been included for Models~1 \& 2.
A more detailed presentation is made for Model~3~in Sec.~\ref{sub:supp:Model3}.
It should also be noted that the notation is the same as that used in
Sec.~\ref{sec:asymptotics} of the main text.

\vspace{2mm}
\noindent
{\it Model~1}.
\begin{itemize}
\item
Spectral density:
\begin{equation*}
   u_2(\omega) = 2\Dzero\lb[1-\frac{\eta}{1+(\omega\tau)^2}\rb].
\end{equation*}

\item
Expansions of $f(v)$:
\begin{equation*}
   f(v) = \frac{u_2(v/\tau)}{2\Dzero} = 1-\frac{\eta}{1+v^2}
    \sim \lb\{\begin{aligned}
       & 1-\eta\sum_{n=0}^\infty (-)^n v^{2n}, & \qquad & v\rightarrow0 \\
       & 1+\eta\sum_{n=1}^\infty (-)^n v^{-2n}, & \qquad & v\rightarrow\infty
     \end{aligned} \rb.
\end{equation*}
\begin{align*}
   \implies & \theta_{1n}=\theta_{2n}=2n, & & f_n^0=\delta_{n,0}-(-)^n\eta,
    & & f_n^\infty=\delta_{n,0}+(1-\delta_{n,0})(-)^n \eta.
\end{align*}

\item
Poles of $M[f_j;1-s]$:

Leading terms in the expansions imply that $M[f_1;1-s]$ and $M[f_2;1-s]$ have analytic
strips $\re\,s<1$ and $\re\,s>1$, respectively. Furthermore,
\begin{align*}
   \begin{aligned}
     & M[f_1;1-s] \sim -\frac{f_n^0}{s-2n-1}, & \quad & s\rightarrow 2n+1 \\[2mm]
     & M[f_2;1-s] \sim \frac{f_n^\infty}{s+2n-1}, & & s\rightarrow -2n+1
   \end{aligned} & & (n=0,1,2,\cdots).
\end{align*}


\item
Pole sets of $G_{lj}(s)$:

The analytic strips of $M[f_1;1-s]$ and $M[h_{kl};s]$ intersect for
$-2\delta_{k,l}<\re\,s<1$ so that $a_{l1}=-2\delta_{k,l}$ and $b_{l1}=1$.
Hence,
\begin{align*}
   \{s_n\leqslant -2\delta_{k,l}\}&=\{-2(n+\delta_{k,l})\}
       =\{-2\delta_{k,l},-2-2\delta_{k,l},-4-2\delta_{k,l},\cdots\} \\[2mm]
   \{s_n\geqslant1\}&=\{2(n+2)\}\cup\{2n+1\}
        = \{4,6,8,\cdots\}\cup\{1,3,5,\cdots\}
\end{align*}
\begin{align*}
   \implies & \quad Q_{l1}=\emptyset, \quad P_{l1}=R_{l1}=\{0,1,2,\cdots\}.
\end{align*}

The analytic strips of $M[f_2;1-s]$ and $M[h_{kl};s]$ intersect for $1<\re\,s<4$ so that
$a_{l2}=1$ and $b_{l2}=4$. It follows that
\begin{align*}
   \{s_n\leqslant1\} &
       =\{-2(n+\delta_{k,l}\}\cup\{-2n+1\} \\
    & \qquad\qquad = \{-2\delta_{k,l},-2-2\delta_{k,l},-4-2\delta_{k,l},\cdots\}\cup
                                     \{1,-1,-3,-5,\cdots\} \\[2mm]
   \{s_n\geqslant4\}&=\{2(n+2)\}=\{4,6,8,\cdots\}
\end{align*}
\begin{align*}
   \implies & \quad Q_{l2}=\emptyset, \quad P_{l2}=R_{l2}=\{0,1,2,\cdots\}.
\end{align*}

\item
Exact Mellin transform of $f(v)$:

Using Eqs.~(\ref{eqn:supp:Integrals:H(s;t,c)}), (\ref{eqn:supp:Integrals:H(s;0,c)}) \&
(\ref{eqn:supp:Integrals:M[v^alpha;s]}),
\begin{align*}
   M[f;s] = M[1;s]-\eta H(s;0,1)=-\frac{\pi\eta}{2\sin(\pi s/2)}.
\end{align*}

\item
Combining the results:
\begin{equation*}
   I_{kl}^{\mbox{\tiny $P$}}(\beta) \sim \lb\{ \begin{aligned}
    & M[h_{kl};1]\,\beta^{-1} -\eta M[h_{kl};-1]\,\beta +\eta M[h_{kl};-3]\,\beta^3
              + O\lb(\beta^5\rb), & &   \beta\rightarrow0 \\[2mm]
    & (1-\eta) M[h_{kl};1]\beta^{-1} +\eta M[h_{kl};3] \beta^{-3}
      -\eta M[h_{kl};5]\beta^{-5} + O\lb(\beta^{-7}\rb), & &    \beta\rightarrow\infty
     \end{aligned} \rb.
\end{equation*}

\begin{equation*}
   I_{kl}^{\mbox{\tiny $R$}}(\beta) \sim \lb\{ \begin{aligned}
    & -\frac{\pi\eta}{4{\varpi_k}^{2+2\delta_{k,l}}}\,\beta^{2\delta_{k,l}}
     + \frac{\pi\eta[{\varpi_k}^2-12(1+\delta_{k,l})]}
      {48{\varpi_k}^{4+2\delta_{k,l}}}\,\beta^{2+2\delta_{k,l}}
       + O\lb(\beta^{4+2\delta_{k,l}}\rb) , & & \beta\rightarrow0 \\[2mm]
    & -\frac{\pi\eta}{2}\,\beta^{-4} + O\lb(\beta^{-6}\rb), & &    \beta\rightarrow\infty
     \end{aligned} \rb.
\end{equation*}


%
%


\end{itemize}



\noindent
{\it Model~2}.
\begin{itemize}
\item
Spectral density:
\begin{equation*}
   u_2(\omega) = 2\Dzero\lb[1-\zi\,\frac{1-\sqrt{|\omega|\tau/2}
                          +|\omega|\tau\sqrt{|\omega|\tau/2}}{1+(\omega\tau)^2}\rb]
\end{equation*}

\item
Expansions of $f(v)$:
\begin{align*}
   f(v) = \frac{u_2(v/\tau)}{2\Dzero} &= 1-\zeta\,\frac{1-\sqrt{v/2}+v \sqrt{v/2}}{1+v^2} \\
    &\qquad \sim \lb\{\begin{aligned}
       & 1-\zeta\lb(1-\frac{v^{1/2}}{\sqrt{2}}+\frac{v^{3/2}}{\sqrt{2}}\rb)
                     \sum_{n=0}^\infty (-)^n v^{2n}, & \quad & v\rightarrow0 \\
       & 1+\zeta\lb(1-\frac{v^{1/2}}{\sqrt{2}}+\frac{v^{3/2}}{\sqrt{2}}\rb)
                     \sum_{n=1}^\infty (-)^n v^{-2n}, & \quad & v\rightarrow\infty
     \end{aligned} \rb.
\end{align*}
\begin{align*}
   \implies & \theta_{1n}=\theta_{2n}=\vphi_n/2, 
    & & f_n^0=\delta_{n,0}-\zeta\,\re\,{\alpha_2}^{\vphi_n}
    & & f_n^\infty=\delta_{n,0}+(1-\delta_{n,0})\,\zeta\,\re\,(-\alpha_1)^{\vphi_n}
\end{align*}
where $\vphi_n=n+\lfloor(n+1)/3\rfloor$, $\alpha_1=\sqrt{i}$ and $\alpha_2=i\sqrt{i}$.

\item
Poles of $M[f_j;1-s]$:

$M[f_1;1-s]$ and $M[f_2;1-s]$ have analytic strips
$\re\,s<1$ and $\re\,s>1$, respectively. Also,
\begin{align*}
   \begin{aligned}
     & M[f_1;1-s] \sim -\frac{f_n^0}{s-\vphi_n/2-1}, 
                                           & \quad & s\rightarrow \vphi_n/2+1 \\[2mm]
     & M[f_2;1-s] \sim \frac{f_n^\infty}{s+\vphi_n/2-1}, & & s\rightarrow -\vphi_n/2+1
   \end{aligned} & & (n=0,1,2,\cdots).
\end{align*}

\item
Pole sets of $G_{lj}(s)$:

The analytic strips of $M[f_1;1-s]$ and $M[h_{kl};s]$ intersect for
$-2\delta_{k,l}<\re\,s<1$ so that $a_{l1}=-2\delta_{k,l}$ and $b_{l1}=1$.
Hence,
\begin{align*}
   \{s_n\leqslant -2\delta_{k,l}\}&=\{-2(n+\delta_{k,l})\}
       =\{-2\delta_{k,l},-2-2\delta_{k,l},-4-2\delta_{k,l},\cdots\} \\[2mm]
   \{s_n\geqslant1\}&=\{2(n+2)\}\cup\{\vphi_n/2+1\}
        = \{4,6,8,\cdots\}\cup\{1,3/2,5/2,3,7/2,\cdots\}
\end{align*}
\begin{align*}
   \implies & \quad Q_{l1}=\emptyset, \quad P_{l1}=R_{l1}=\{0,1,2,\cdots\}.
\end{align*}

The analytic strips of $M[f_2;1-s]$ and $M[h_{kl};s]$ intersect for $1<\re\,s<4$ so that
$a_{l2}=1$ and $b_{l2}=4$. It follows that
\begin{align*}
   \{s_n\leqslant1\} &
       =\{-2(n+\delta_{k,l}\}\cup\{-\vphi_n/2+1\} \\
    & \qquad = \{-2\delta_{k,l},-2-2\delta_{k,l},-4-2\delta_{k,l},\cdots\}\cup
                                     \{1,1/2,-1/2,-1,\cdots\} \\[2mm]
   \{s_n\geqslant4\}&=\{2(n+2)\}=\{4,6,8,\cdots\}
\end{align*}
\begin{align*}
   \implies & \quad Q_{l2}=\emptyset, \quad P_{l2}=R_{l2}=\{0,1,2,\cdots\}.
\end{align*}

\item
Exact Mellin transform of $f(v)$:

From Eqs.~(\ref{eqn:supp:Integrals:H(s;t,c)}), (\ref{eqn:supp:Integrals:H(s;0,c)}) \&
(\ref{eqn:supp:Integrals:M[v^alpha;s]}),
\begin{align*}
   M[f;s] = M[1;s]-\zeta\lb[H(s;0,1)-\frac{H(s+1/2;0,1)}{\sqrt{2}}
      +\frac{H(s+3/2;0,1)}{\sqrt{2}}\rb]=-\frac{2\pi\zeta\cos(\pi s/2)}{\sin(2\pi s)}.
\end{align*}

\item
Keeping only a few of the most dominant terms:
\begin{equation*}
   I_{kl}^{\mbox{\tiny $P$}}(\beta) \sim \lb\{ \begin{aligned}
    & M[h_{kl};1]\,\beta^{-1} -\frac{\zeta}{\sqrt{2}} M[h_{kl};1/2]\,\beta^{-1/2} \\
    & \hspace{30mm} +\frac{\zeta}{\sqrt{2}} M[h_{kl};-1/2]\,\beta^{1/2}
              + O\lb(\beta\rb), & &   \beta\rightarrow0 \\[2mm]
    & (1-\zeta) M[h_{kl};1]\beta^{-1} +\frac{\zeta}{\sqrt{2}} M[h_{kl};3/2] \beta^{-3/2} \\
    & \hspace{30mm} -\frac{\zeta}{\sqrt{2}} M[h_{kl};5/2]\beta^{-5/2} 
                  + O\lb(\beta^{-3}\rb), & \qquad &    \beta\rightarrow\infty
     \end{aligned} \rb.
\end{equation*}

\begin{equation*}
   I_{kl}^{\mbox{\tiny $R$}}(\beta) \sim \lb\{ \begin{aligned}
    & -\frac{\pi\zeta}{4{\varpi_k}^{2+2\delta_{k,l}}}\,\beta^{2\delta_{k,l}}
     - \frac{\pi\zeta[{\varpi_k}^2-12(1+\delta_{k,l})]}{48{\varpi_k}^{4+2\delta_{k,l}}}
        \,\beta^{2+2\delta_{k,l}}
       + O\lb(\beta^{4+2\delta_{k,l}}\rb) , & \quad & \beta\rightarrow0 \\[2mm]
    & \frac{\pi\zeta}{2}\,\beta^{-4} + O\lb(\beta^{-6}\rb), & &    \beta\rightarrow\infty
     \end{aligned} \rb.
\end{equation*}


\end{itemize}

\subsection{Asymptotic expansion of $U_{k0}$ \& $U_{kk}$ for Model~3}
\label{sub:supp:Model3}

The dispersive diffusivity of this model is
\begin{equation*}
   \cD(\omega) = \frac{\Dzero}{1+\zeta+2 z_\omega\lb[\sqrt{\zeta+(1-z_\omega)^2}-
                             (1-z_\omega)\rb]},
 \label{eqn:supp::Model3:D(w)}
\end{equation*}
where $z_\omega=i\sqrt{i\omega\tau}$.
Although a closed form for $u_2(\omega)=2\re\,\cD(\omega)$ can be obtained, the 
derivation is tedious and contains awkward nested square roots.
A simpler way to obtain the asymptotic
behaviour of $u_2(\omega)$ is to use the expansions for $\cD(\omega)$
\begin{equation}
   \frac{\cD(\omega)}{\Dzero} \sim \lb\{ \begin{aligned}
        & \sum_{n=0}^\infty d_n^0\,{\alpha_2}^n (\omega\tau)^{n/2},
                     \quad             & & \omega\tau\rightarrow 0 \\
        & \sum_{n=0}^\infty (-)^n d^\infty_n\,{\alpha_1}^n(\omega\tau)^{-n/2},
                                  \quad     & & \omega\tau\rightarrow\infty
    \end{aligned} \rb.
 \label{eqn:supp:Model3:D(w)expansion}
\end{equation}
with
\begin{align*}
   d^0_0&=1/(1+\zi) & & d^\infty_0=1 \\
   d^0_1&=2[1-\sqrt{1+\zi}]/(1+\zi)^2 & & d^\infty_1=\zi \\
   d^0_2&=2[3+\zi-3\sqrt{1+\zi}]/(1+\zi)^3, & & d^\infty_2=\zi[4+3\zi]/4 \\
   d^0_3&=[16+8\zi-(16+\zi)\sqrt{1+\zi}]/(1+\zi)^4  & & 
                                   d^\infty_3=\zi[4+5\zi+2\zi^2]/4\\
   d^0_4&=[40+24\zi-(40+5\zi)\sqrt{1+\zi}]/(1+\zi)^5 & & 
                               d^\infty_4=\zi[16+24\zi+18\zi^2+5\zi^3]/16
\end{align*}
being the first five coefficients in both expansions.
The coefficients $d_n^0$ and $d^\infty_n$ are all real numbers so that the complex
nature of the function is completely retained in the numbers $\alpha_1=\sqrt{i}
=(1+i)/\sqrt{2}$ and $\alpha_2=i\sqrt{i}=(-1+i)/\sqrt{2}$, which are both 8th complex
roots of unity and hence generate the cyclic group of order 8 under multiplication
(see Table~\ref{tab:supp:Model3:cyc_gps}).
This means that expansions for $u_2(\omega)$ can be constructed by simply
omitting the terms in Eq.~(\ref{eqn:supp:Model3:D(w)expansion}) with $n=4l+2$
($l=0,1,2,\cdots$) because $\re\,{\alpha_1}^n=\re\,{\alpha_2}^n=0$ for
those values of $n$.
Explicitly,
\begin{equation}
   \frac{u_2(\omega)}{2\Dzero} \sim \lb\{ \begin{aligned}
     & d_0^0 - \frac{d_1^0}{\sqrt{2}}(\omega\tau)^{\frac{1}{2}}
         +\frac{d^0_3}{\sqrt{2}}(\omega\tau)^{\frac{3}{2}}
         -d^0_4(\omega\tau)^2 + O(\omega\tau)^{\frac{5}{2}}, 
          \quad & & \omega\tau\rightarrow 0 \\
     & d^\infty_0 - \frac{d^\infty_1}{\sqrt{2}}(\omega\tau)^{-\frac{1}{2}}
         + \frac{d_3^\infty}{\sqrt{2}}(\omega\tau)^{-\frac{3}{2}}
          -d_4^\infty(\omega\tau)^{-2}
         +O(\omega\tau)^{-\frac{5}{2}},
          \quad & & \omega\tau\rightarrow\infty. 
    \end{aligned} \rb.
 \label{eqn:supp:Model3:u2(w_k)expansion}
\end{equation}

\begin{table}
 \centering
 \captionsetup{width=8.2cm}
 \caption{$\alpha_1$ and $\alpha_2$ generate the cyclic group of order 8.}
 \begin{tabular}{@{\vrule height 12.0pt depth0pt  width0pt}c|ccccccccc}
     \hline
      $n$ & 0 & 1 & 2 & 3 & 4 & 5 & 6 & 7 & 8 \\ \hline
   ${\alpha_1}^n$ & 1 & $\sqrt{i}$ & $i$ & $\alpha_2$ & $-1$
                                  & $-\alpha_1$ & $-i$ & $-\alpha_2$ & 1 \\
   ${\alpha_2}^n$ & 1 & $i\sqrt{i}$ & $-i$ & $\alpha_1$ & $-1$
                                    & $-\alpha_2$ & $i$ & $-\alpha_1$ & 1 \\
   \hline
 \end{tabular}
 \label{tab:supp:Model3:cyc_gps}
\end{table}

Now to consider $f(v)=u(v/\tau)/2\Dzero$, the expansions in the previous paragraph imply
that 
\begin{equation}
   f(v) \sim \lb\{ \begin{aligned}
        & \sum_{n=0}^\infty f^0_{n}\,v^{\vphi_n/2},  & & v\rightarrow 0 \\
        & \sum_{n=0}^\infty f^\infty_{n}\,v^{-\vphi_n/2}, \quad & & v\rightarrow\infty
    \end{aligned} \rb.
 \label{eqn:supp:Model3:f(v)_expan}
\end{equation}
where $\vphi_n=n+\lfloor(n+1)/3\rfloor$,
$f_n^0=d_{\vphi_n}^0\,\re\,{\alpha_2}^{\vphi_n}$ and
$f_n^\infty= d_{\vphi_n}^\infty \re\,(-\alpha_1)^{\vphi_n}$.
It is evident from the powers of the leading terms that the analytic strip of $M[f;s]$
is empty,
which means that the generalised methods of Sec.~\ref{sub:supp:genMTM} must be used
to evaluate the Mellin transform of $f(v)$.
Choose $v_0\in(0,\infty)$ and then define $f_1(v)=f(v)$ for $0\leqslant v<v_0$,
$f_1(v)=0$ for $v_0\leqslant v$ and $f_2(v)=f(v)-f_1(v)$.
From the behaviour of $f_1(v)$ as $v\rightarrow0$, $M[f_1;s]$ is analytic for
$\re\,s>0$ and can be analytically continued to $\re\,s<0$ with
$M[f_1;s]\sim f^0_n/(s+\vphi_n/2)$ as $s\rightarrow-\vphi_n/2$.
Similarly, from the behaviour of $f_2(v)$ as $v\rightarrow\infty$, $M[f_2;s]$ is analytic
for $\re\,s<0$ and can be analytically continued to $\re\,s>0$ with
$M[f_2;s]\sim-f^\infty_n/(s-\vphi_n/2)$ as $s\rightarrow\vphi_n/2$. After shifting
$s$ to $1-s$ it is clear that $M[f_1;1-s]$ and $M[f_2;1-s]$ have analytic strips
$\re\,s<1$ and $\re\,s>1$, respectively, and that
\begin{align}
   & M[f_1;1-s] \sim -\frac{f_n^0}{s-\vphi_n/2-1},
                                    & & \hspace{-2.5cm} s\rightarrow \vphi_n/2+1
 \label{eqn:supp:Model3:M[f1;s]} \\[2mm]
   & M[f_2;1-s] \sim \frac{f_n^\infty}{s+\vphi_n/2-1}, 
                                    & & \hspace{-2.5cm} s\rightarrow -\vphi_n/2+1.
 \label{eqn:supp:Model3:M[f2;s]}
\end{align}

The Mellin transform technique will be applied for $U_{k0}$ first.
It is clear from Sec.~\ref{sub:supp:Models1&2} and the results above that the analytic
strips of $M[f_1;1-s]$ and
$M[h_{k0};s]$ intersect for $0<\re\,s<1$, and the analytic strips of $M[f_2;1-s]$ and
$M[h_{k0};s]$ intersect for $1<\re\,s<4$.
With reference to Eqs.~(\ref{eqn:supp:MTM:I(lambda)}) \&
(\ref{eqn:supp:genMTM:I(lambda)}), this means that it is valid to write
\begin{align*}
   I_{k0}(\beta) 
    &= \int_0^\infty\!\!dv\,\lb[f_1(v)+f_2(v)\rb] h_{k0}(\beta v) \nonumber \\
     & \hspace{2cm}= \frac{1}{2\pi i}\int_{c_1-i\infty}^{c_1+i\infty}
                                                       \!\!ds\,\beta^{-s} G_{01}(s)
      +\frac{1}{2\pi i}\int_{c_2-i\infty}^{c_2+i\infty}\!\!ds\,\beta^{-s} G_{02}(s)
 \label{eqn:supp:Model3:Ik0(beta)}
\end{align*}
for $c_1\in\lb(0,1\rb)$, $c_2\in\lb(1,4\rb)$ and $G_{0j}(s)=M[f_j;1-s]M[h_{k0};s]$,
$j=1,2$.
The analog of Eq.~(\ref{eqn:supp:genMTM:ResidueSeries}) therefore applies
\begin{equation}
   I_{k0}(\beta) \sim \lb\{\begin{aligned}
       & \sum_{\{s_n\leqslant 0\}} \mbox{Res}\{\beta^{-s} G_{01}(s)\} \quad
        + \sum_{\{s_n\leqslant 1\}} \mbox{Res}\{\beta^{-s} G_{02}(s)\}, 
                                            & & \qquad \beta\rightarrow0 \\
       & -\!\!\!\!\sum_{\{s_n\geqslant 1\}} \mbox{Res}\{\beta^{-s} G_{01}(s)\} 
        \quad - \sum_{\{s_n\geqslant 4\}} \mbox{Res}\{\beta^{-s} G_{02}(s)\}, 
                                            & & \qquad \beta\rightarrow\infty
     \end{aligned} \rb.
 \label{eqn:supp:Model3:ResidueSeries_Uk0}
\end{equation}
with $n=0,1,2,\cdots$.
The set of poles of $G_{01}(s)$ is a union of the poles of $M[f_1;1-s]$ and
$M[h_{k0};s]$:
\begin{align*}
   & \mbox{left-hand plane:} & & \{s_n\leqslant 0\}=\{-2n\}=\{0,-2,-4,\cdots\} \\
   & \mbox{right-hand plane:} & & \{s_n\geqslant1\}=\{2n+4\}\cup\{\vphi_n/2+1\}
                                                         = \{1,3/2,5/2,\cdots\},
\end{align*}
and the set of poles of $G_{02}(s)$ is a union of the poles of $M[f_2;1-s]$ and
$M[h_{k0};s]$:
\begin{align*}
   & \mbox{left-hand plane:} & & \{s_n\leqslant1\}=\{-2n\}\cup\{-\vphi_n/2+1\}
                                                         =\{1,1/2,0,-1/2,\cdots\} \\
   & \mbox{right-hand plane:} & & \{s_n\geqslant4\}=\{2n+4\} = \{4,6,\cdots\}.
\end{align*}
The poles have been ordered on the far right according to the significance of the
corresponding power of $\beta$ as it approaches 0 or $\infty$.
As there are no repetitions in any of the sets, it follows that
\begin{align*}
   & Q_{01}=Q_{02}=\emptyset, & & P_{01}=R_{01}=P_{02}=R_{02}=\{0,1,2,\cdots\}.
\end{align*}
The poles are all simple so it is relatively easy to calculate the residues to obtain
\begin{equation}
   I_{k0}^{\mbox{\tiny $P$}}(\beta) \sim \lb\{ \begin{aligned}
       & \sum_{n=0}^\infty \beta^{\vphi_n/2-1} f_n^\infty M[h_{k0};-\vphi_n/2+1],
                                                     & \qquad &   \beta\rightarrow0 \\
       & \sum_{n=0}^\infty \beta^{-\vphi_n/2-1} f_n^0 M[h_{k0};\vphi_n/2+1],
                                                       & &    \beta\rightarrow\infty
     \end{aligned} \rb.
 \label{eqn:supp:Model3:Ik0P}
\end{equation}
\begin{equation}
   I_{k0}^{\mbox{\tiny $R$}}(\beta) \sim \lb\{ \begin{aligned}
       & \sum_{n=0}^\infty \beta^{2n} h^{0}_{0n} M[f;2n+1],
                                                         & &   \beta\rightarrow0 \\
       & \sum_{n=0}^\infty \beta^{-2n-4} h^{\infty}_{0n} M[f;-2n-3],
                                   & \qquad &    \beta\rightarrow\infty
     \end{aligned} \rb.
 \label{eqn:supp:Model3:Ik0R}
\end{equation}
where $M[f;2n+1]$ and $M[f;-2n-3]$ are used in the generalised sense of
Eq.~(\ref{eqn:supp:genMTM:genMTM}).

Following a similar process for $U_{kk}$ now, the strips of analyticity
of $M[f_1;1-s]$ and $M[h_{kk};s]$ intersect for $-2<\re\,s<1$, and those for
$M[f_2;1-s]$ and $M[h_{kk};s]$ intersect for $1<\re\,s<4$. Using
Eqs.~(\ref{eqn:supp:MTM:I(lambda)}) \& (\ref{eqn:supp:genMTM:I(lambda)}) again
\begin{align*}
   I_{kk}(\beta) &= \int_0^\infty\!\!dv\,\lb[f_1(v)+f_2(v)\rb] h_{kk}(\beta v) \\
    &\hspace{2cm}= \frac{1}{2\pi i}\int_{c_1-i\infty}^{c_1+i\infty}
                                                       \!\!ds\,\beta^{-s} G_{k1}(s)
     + \frac{1}{2\pi i}\int_{c_2-i\infty}^{c_2+i\infty}\!\!ds\,\beta^{-s} G_{k2}(s)
 \label{eqn:supp:Model3:Ikk(beta)}
\end{align*}
with $c_1\in\lb(-2,1\rb)$, $c_2\in\lb(1,4\rb)$ and $G_{kj}(s)=M[f_j;1-s]M[h_{kk};s]$,
$j=1,2$.
Subsequently, Eq.~(\ref{eqn:supp:genMTM:ResidueSeries}) implies that
\begin{equation}
   I_{kk}(\beta) \sim \lb\{\begin{aligned}
       & \sum_{\{s_n\leqslant -2\}} \mbox{Res}\{\beta^{-s} G_{k1}(s)\} \quad
        + \sum_{\{s_n\leqslant 1\}} \mbox{Res}\{\beta^{-s} G_{k2}(s)\}, 
                                            & & \qquad \beta\rightarrow0 \\
       & -\!\!\!\!\sum_{\{s_n\geqslant 1\}} \mbox{Res}\{\beta^{-s} G_{k1}(s)\} 
        \quad - \sum_{\{s_n\geqslant 4\}} \mbox{Res}\{\beta^{-s} G_{k2}(s)\}, 
                                            & & \qquad \beta\rightarrow\infty
     \end{aligned} \rb.
 \label{eqn:supp:Model3:ResidueSeries_Ukk}
\end{equation}
for $n=0,1,2,\cdots$.
Combining the poles of $M[f_1;1-s]$ and $M[h_{kk};s]$ to obtain the poles of
$G_{k1}(s)$:
\begin{align*}
   & \mbox{left-hand plane:} & & \{s_n\leqslant -2\}=\{-2n-2\}=\{-2,-4,\cdots\} \\
   & \mbox{right-hand plane:} & & \{s_n\geqslant1\}=\{2n+4\}\cup\{\vphi_n/2+1\}
                                                         = \{1,3/2,5/2,\cdots\},
\end{align*}
and similarly merging the poles of $M[f_2;1-s]$ and $M[h_{kk};s]$ as the set of
poles of $G_{k2}(s)$:
\begin{align*}
   & \mbox{left-hand plane:} & & \{s_n\leqslant1\}=\{-2n-2\}\cup\{-\vphi_n/2+1\}
                                                         =\{1,1/2,-1/2,\cdots\} \\
   & \mbox{right-hand plane:} & & \{s_n\geqslant4\}=\{2n+4\} = \{4,6,\cdots\}.
\end{align*}
The poles are all simple with no repetitions so
\begin{align*}
   & Q_{01}=Q_{02}=\emptyset, & & P_{01}=R_{01}=P_{02}=R_{02}=\{0,1,2,\cdots\}.
\end{align*}
After calculating the residues
\begin{equation}
   I_{kk}^{\mbox{\tiny $P$}}(\beta) \sim \lb\{ \begin{aligned}
       & \sum_{n=0}^\infty \beta^{\vphi_n/2-1} f_n^\infty M[h_{kk};-\vphi_n/2+1],
                                                     & \qquad &   \beta\rightarrow0 \\
       & \sum_{n=0}^\infty \beta^{-\vphi_n/2-1} f_n^0 M[h_{kk};\vphi_n/2+1],
                                                     & &    \beta\rightarrow\infty
     \end{aligned} \rb.
 \label{eqn:supp:Model3:IkkP}
\end{equation}
\begin{equation}
   I_{kk}^{\mbox{\tiny $R$}}(\beta) \sim \lb\{ \begin{aligned}
       & \sum_{n=0}^\infty \beta^{2n+2} h^{0}_{kn} M[f;2n+3],
                                                & \qquad &   \beta\rightarrow0 \\
       & \sum_{n=0}^\infty \beta^{-2n-4} h^{\infty}_{kn} M[f;-2n-3], 
                                                & &    \beta\rightarrow\infty
     \end{aligned} \rb.
 \label{eqn:supp:Model3:IkkR}
\end{equation}
with $M[f;2n+3]$ and $M[f;-2n-3]$ used to represent the generalised Mellin transform
in the sense of Eq.~(\ref{eqn:supp:genMTM:genMTM}).

The Mellin transforms of $h_{k0}(v)$ and $h_{kk}(v)$ that appear as coefficients
in Eqs.~(\ref{eqn:supp:Model3:Ik0P}) \& (\ref{eqn:supp:Model3:IkkP}) can be inserted from
Table~\ref{tab:supp:Mhkl:Mhkl}.
Unfortunately, a closed form for the coefficients involving the Mellin
transform of $f(v)$ is not available. However, as it is only the
value that is required rather than the functional form of the result,
given specific $s$ and $\zi$ those coefficients can be obtained with numerical
techniques.
The method outlined on page~117 of \cite{BleisteinHandelsmanBook} must be used to do
this because the analytic strip of $M[f;1-s]$ is empty.


Inserting the information from Eqs.~(\ref{eqn:supp:Model3:Ik0P}),
(\ref{eqn:supp:Model3:Ik0R}), (\ref{eqn:supp:Model3:IkkP}) \&
(\ref{eqn:supp:Model3:IkkR}) into Eq.~(\ref{eqn:asymptotics:MTM:Ikl}),
and recalling that $\beta=\varpi_k/\omega_k\tau$, the final results are
\begin{equation}
   \frac{U_{k0}}{2\Dzero} \sim \lb\{ \begin{aligned}
     & d_0^0 - A_{01}^\infty\frac{d_1^0}{\sqrt{2}}
                                            (\omega_k\tau)^{\frac{1}{2}}
         +A_{03}^\infty\frac{d^0_3}{\sqrt{2}}
                                            (\omega_k\tau)^{\frac{3}{2}} 
         + O(\omega_k\tau)^{2}, \quad & & \omega_k\tau\rightarrow 0 \\
     & d_0^\infty -A_{01}^0\frac{d^\infty_1}{\sqrt{2}}
                                            (\omega_k\tau)^{-\frac{1}{2}}
        + k M[f;1] (\omega_k\tau)^{-1}
         +O(\omega_k\tau)^{-\frac{3}{2}},
          \quad & & \omega_k\tau\rightarrow\infty
    \end{aligned} \rb.
 \label{eqn:supp:Model3:Uk0expansion}
\end{equation}
\begin{equation*}
   A_{01}^\infty = \frac{2}{\pi k^{1/2}}-\frac{\aCk}{\varpi_k}, \qquad
    A_{03}^\infty = \frac{\aSk}{\varpi_k}, \qquad
   A_{01}^0 =\frac{8k^{1/2}}{3}+\frac{\aSk}{\varpi_k}.
\end{equation*}
\vspace{2mm}
\begin{equation}
   \frac{U_{kk}}{2\Dzero} \sim \lb\{ \begin{aligned}
     & d_0^0 - A_{k1}^\infty \frac{d_1^0}{\sqrt{2}}
                                            (\omega_k\tau)^{\frac{1}{2}}
         +A_{k3}^\infty \frac{d^0_3}{\sqrt{2}} 
                                            (\omega_k\tau)^{\frac{3}{2}}
         + O(\omega_k\tau)^{2}, & \quad & \omega_k\tau\rightarrow 0 \\
     & d_0^\infty 
        -A_{k1}^0 \frac{d^\infty_1}{\sqrt{2}}
                                                (\omega_k\tau)^{-\frac{1}{2}}
        +A_{k3}^0 \frac{d_3^\infty}{\sqrt{2}} 
                                                (\omega_k\tau)^{-\frac{3}{2}}
         +O(\omega_k\tau)^{-2}, & & \omega_k\tau\rightarrow\infty
    \end{aligned} \rb.
 \label{eqn:supp:Model3:Ukkexpansion}
\end{equation}
\begin{align*}
   & A_{k1}^\infty=\frac{1}{\pi k^{1/2}} -\frac{\aCk}{2\varpi_k}+\aSk \quad & &
    A_{k3}^\infty=-\frac{\aSk}{2\varpi_k}+\aCk \quad \\
   & A_{k1}^0=\frac{3\aSk}{2\varpi_k}+\aCk \quad & &
    A_{k3}^0=\frac{5}{\pi k^{1/2}} -\frac{5\aCk}{2\varpi_k}+\aSk.
\end{align*}
Here $\aSk=2S(2k^{1/2})$ and $\aCk=2C(2k^{1/2})$ where $S(z)$ and $C(z)$ are the
Fresnel integrals.
The coefficients $A_{0n}^\infty$, $A_{0n}^0$, $A_{kn}^\infty$ and $A_{kn}^0$ have
been used to
help identify differences in the asymptotic behaviour of $U_{k0}$ \& $U_{kk}$ in
comparison to that of $u_2(\omega_k)$
(re Eq.~(\ref{eqn:supp:Model3:u2(w_k)expansion})).

\subsection{Structural universality}
\label{sub:supp:Universality}

The goal here is to derive the asymptotic behaviour of $U_{k0}$ and $U_{kk}$
given only the information in Eq.~(\ref{eqn:intro:u2}).
It shall simply be assumed that all of the conditions hold for the MTM to be applied.
It shall also be assumed that there is a time-scale with respect to which the
asymptotic frequency-dependence of the model may be investigated.
As there is no obvious choice for such a time-scale in the given information,
an unspecified parameter $\tau$ shall be introduced in the following so that the
problem can be written in the dimensionless form of
Eq.~(\ref{eqn:asymptotics:MTM:Ikl}).


%

The first step in the procedure is to note that, after making the variable change
$v=\omega\tau$ in Eq.~(\ref{eqn:intro:u2}),
\begin{equation}
   2\Dzero f(v)=u_2(v/\tau) \sim \lb\{ \begin{array}{ll}
      2\Dinf + (\cinf/\tau^\vartheta)\, v^\vartheta, 
                                & \quad v\rightarrow0 \vspace{2mm}\\
      2\Dzero - (\czero\tau^{1/2})\,v^{-1/2}, & \quad  v\rightarrow\infty.
   \end{array}\rb.
 \label{eqn:supp:Universality:f(v)}
\end{equation}
With respect to the notation in Eq.~(\ref{eqn:asymptotics:poles:f}) it follows that
\begin{align}
   \{\theta_{1n}\}&=\{0,\vartheta,\cdots\},
              &     \{f_n^0\}&=\{\Dinf/\Dzero,\cinf/2\tau^\vartheta \Dzero,\cdots\}
 \label{eqn:supp:Universality:rho1n} \\
   \{\theta_{2n}\}&=\{0,1/2,\cdots\},
              &     \{f_n^\infty\}&=\{1,-\czero\tau^{1/2}/2 \Dzero,\cdots\}.
 \label{eqn:supp:Universality:rho2n}
\end{align}
The values of $\theta_{10}$ \& $\theta_{20}$ imply that the analytic strip
of $M[f;s]$ is empty so the generalised Mellin transform must be used.
Defining $f_1(v)$ \& $f_2(v)$ as in Eqs.~(\ref{eqn:asymptotics:MTM:f1}) \&
(\ref{eqn:asymptotics:MTM:f2}),
it can be said that
$M[f_1;1-s]$ is analytic for $\re\,s<1$ while $M[f_2;1-s]$ is analytic for $\re\,s>1$.
Furthermore, analytic continuation into the entire complex plane implies that
\begin{align}
   M[f_1;1-s] &\sim \lb\{ \begin{aligned}
     & -\frac{f_0^0}{s-1}, & & s\rightarrow1 \\
     & -\frac{f_1^0}{s-\vartheta-1}, & & s\rightarrow\vartheta+1
    \end{aligned} \rb.
 \label{eqn:supp:Universality:M[f1;1-s]_theta>0} \\[2mm]
   M[f_2;1-s] &\sim \lb\{ \begin{aligned}
     & \frac{f_0^\infty}{s-1}, & & s\rightarrow1 \\
     & \frac{f_1^\infty}{s-1/2}, & & s\rightarrow1/2.
    \end{aligned} \rb.
 \label{eqn:supp:Universality:M[f2;1-s]}
\end{align}

\begin{figure}[tb]
 \centering
    \includegraphics[scale=0.52]{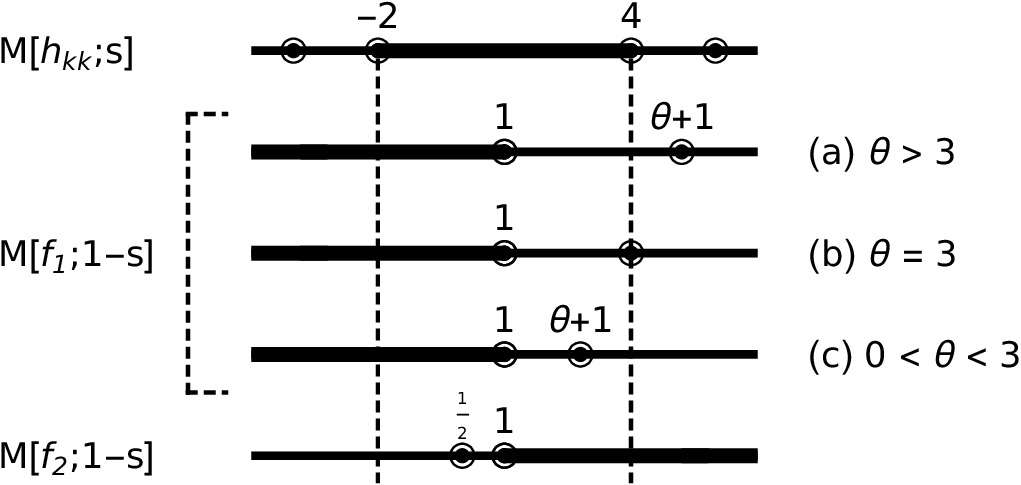}
 \caption{
Illustration demonstrating how the analytic strips and known poles of $M[f_1;1-s]$ \&
$M[f_2;1-s]$ are positioned along the real axis of the complex plane relative to
the analytic strip and poles of $M[h_{kk};s]$. There are 3 separate cases depending
on the value of $\vartheta$.
}
 \label{fig:supp:Universality:kkStrips}
\end{figure}


\begin{figure}[tb]
 \centering
    \includegraphics[scale=0.52]{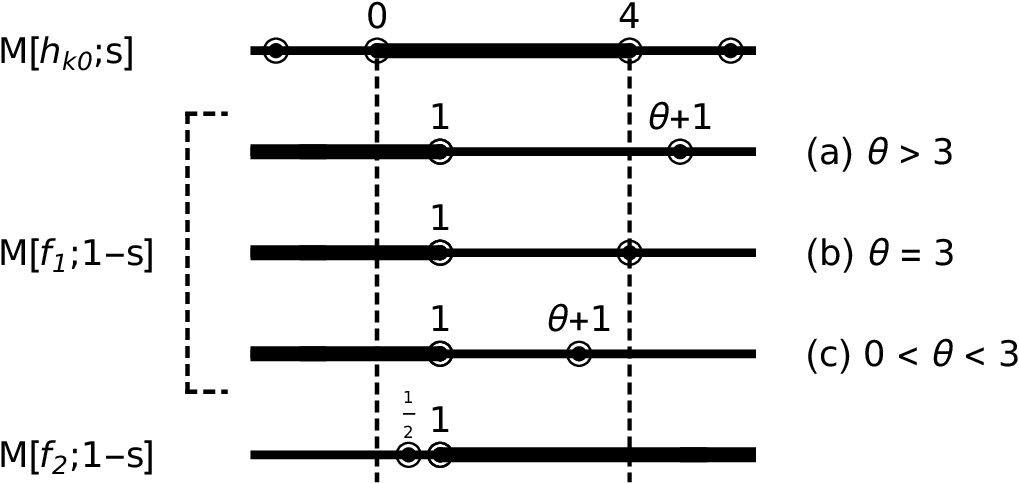}
 \caption{
Illustration demonstrating how the analytic strips and known poles of $M[f_1;1-s]$ \&
$M[f_2;1-s]$ are positioned along the real axis of the complex plane relative to
the analytic strip and poles of $M[h_{k0};s]$. There are 3 separate cases depending
on the value of $\vartheta$.
}
 \label{fig:supp:Universality:k0Strips}
\end{figure}


If required the Mellin transform of $f(v)$ can be evaluated using
$M[f;1-s]=M[f_1;1-s]+M[f_2;1-s]$.
Note that, due to the limited number of terms in the asymptotic expansions given
for $u_2(\omega)$, nothing is known about the behaviour of $M[f;1-s]$ beyond the
poles at $\vartheta+1$, $1/2$ and $1$.

The next step is to consider how the exponent $\vartheta$ affects the problem.
With reference to the conditions in Eq.~(\ref{eqn:supp:MTM:converge}), $I_{kl}(\beta)$
is convergent for $\vartheta>-1-2\delta_{k,l}$, which is always satisfied because
$\vartheta>0$. 
Therefore, the value of $\vartheta$ only affects the problem through the
positions of the known poles of $M[f_1;1-s]$ relative to the poles of $M[h_{kl};s]$
(Figs.~\ref{fig:supp:Universality:kkStrips} \& \ref{fig:supp:Universality:k0Strips}).
There are three separate cases:
\begin{enumerate}
\item[(a)]
$0<\vartheta<3$.
The analytic strip of $M[f_1;1-s]$ intersects with that of $M[h_{kl};s]$ for
$-2\delta_{k,l}<\re\,s<1$, which means that the known poles of $G_{l1}(s)$ are
\begin{align*}
   & \mbox{left-hand plane:} & & \{s_n\leqslant -2\delta_{k,l}\}
                 =\{-2n-2\delta_{k,l}\}=\{-2\delta_{k,l},-2-2\delta_{k,l},\cdots\} \\
   & \mbox{right-hand plane:} & & \{s_n\geqslant1\}=\{2n+4\}\cup\{1,\vartheta+1\}
                                                         = \{1,\vartheta+1,4\cdots\}.
\end{align*}
Similarly, the analytic regions of $M[f_2;1-s]$ and $M[h_{kl};s]$ overlap for
$1<\re\,s<4$ so the known poles of $G_{l2}(s)$ are
\begin{align*}
   & \mbox{left-hand plane:} & & \{s_n\leqslant 1\}=
         \{-2n-2\delta_{k,l}\}\cup\{1,1/2\}=\{1,1/2,\cdots,-2\delta_{k,l},\cdots\} \\
   & \mbox{right-hand plane:} & & \{s_n\geqslant4\}=\{2n+4\} = \{4,\cdots\}.
\end{align*}
There are no repetitions so
\begin{align}
   & Q_{l1}=Q_{l2}=\emptyset, & & P_{l1}=P_{l2}=\{0,1\}, 
                                           & & R_{l1}=R_{l2}=\{0,1,2,\cdots\}.
 \label{eqn:supp:Universality:Ql1a}
\end{align}
The residues are easy to calculate because the poles are simple, and so
\begin{equation}
   I_{kl}^{\mbox{\tiny $P$}}(\beta) \sim \lb\{ \begin{aligned}
       & f_0^\infty M[h_{kl};1]\beta^{-1}+f_1^\infty 
               M[h_{kl};1/2]\beta^{-1/2}, & \qquad &   \beta\rightarrow0 \\[1mm]
       & f_0^0 M[h_{kl};1]\beta^{-1}+f_1^0 
          M[h_{kl};\vartheta+1]\beta^{-\vartheta-1}, & &    \beta\rightarrow\infty
     \end{aligned} \rb.
 \label{eqn:supp:Universality:IklPa}
\end{equation}
\begin{equation}
   I_{kl}^{\mbox{\tiny $R$}}(\beta) \sim \lb\{ \begin{aligned}
       & h_{l0}^0 M[f;1+2\delta_{k,l}]\beta^{2\delta_{k,l}}, 
                                       & \qquad &   \beta\rightarrow0 \\[1mm]
       & h_{l0}^\infty M[f;-3]\beta^{-4}, & &    \beta\rightarrow\infty
     \end{aligned} \rb.
 \label{eqn:supp:Universality:IklRa}
\end{equation}

\item[(b)]
$\vartheta=3$.
Even though the intersection of the analytic regions of $M[f_1;1-s]$ and $M[h_{kl};s]$
is the same as for case (a), in this special case the pole $\vartheta+1=4$ of
$M[f_1;1-s]$ coincides with the first pole of $M[h_{kl};s]$ so that $G_{l1}(s)$ has
a second order pole. Accordingly,
\begin{align*}
    \{s_n\geqslant1\}=\{2n+4\}\cup\{1,\vartheta+1\} = \{1,(4,4),\cdots\}
\end{align*}
and $Q_{l1}=\{(1,0)\}$, $P_{l1}=\{0\}$ \& $R_{l1}=\{1,2,3,\cdots\}$.
It follows that for $\beta\rightarrow\infty$,
\begin{align*}
    I_{kl}^{\mbox{\tiny $P$}}(\beta) &\sim f_0^0 M[h_{kl};1]\beta^{-1} \\[2mm]
    I_{kl}^{\mbox{\tiny $R$}}(\beta) &\sim O(\beta^{-6}) \\[2mm]
    I_{kl}^{\mbox{\tiny $Q$}}(\beta) &\sim
      h_{l0}^\infty f_{1}^0 \beta^{-4}\ln\beta + K_{l10}^1\,\beta^{-4}
\end{align*}
with
\begin{equation*}
   K_{l10}^1 = h_{l0}^\infty M[f_2;-3] - \lim_{s\rightarrow4}
     \frac{d}{ds}\lb\{(s-4)^2 G_{l1}(s)\rb\}.
\end{equation*}
As $\vartheta$ does not affect the expansion as $\beta\rightarrow0$, from here on the
details for that limit are the same as for case (a) above.

\item[(c)]
$\vartheta>3$.
The only difference between this case and case (a) is the order of the poles of
$G_{l1}(s)$ in the right-hand plane. That is,
\begin{align*}
    \{s_n\geqslant1\}=\{2n+4\}\cup\{1,\vartheta+1\} = \{1,4,\vartheta+1,\cdots\}.
\end{align*}
Otherwise, the sets $Q_{l1}$, $P_{l1}$ \& $R_{l1}$ are unaltered from those in
Eq.~(\ref{eqn:supp:Universality:Ql1a}) so $I_{kl}^{\mbox{\tiny $P$}}(\beta)$ \&
$I_{kl}^{\mbox{\tiny $R$}}(\beta)$ are the same as in
Eqs.~(\ref{eqn:supp:Universality:IklPa}) \& (\ref{eqn:supp:Universality:IklRa}),
respectively. It must be remembered though that, as $\beta\rightarrow\infty$, the
leading $O(\beta^{-4})$ term in $I_{kl}^{\mbox{\tiny $R$}}(\beta)$ now dominates the
$O(\beta^{-\vartheta-1})$ term in $I_{kl}^{\mbox{\tiny $P$}}(\beta)$ when the two
quantities are summed to form $I_{kl}(\beta)$.

\end{enumerate}

Recalling Eq.~(\ref{eqn:asymptotics:MTM:Ikl}) and the fact that
$\beta=\varpi_k/\omega_k\tau$ and $f(v)=u_2(v/\tau)/2\Dzero$, the preceding information
enables the known terms in the expansions of $U_{kk}$ to be summarised.
For the high-frequency limit,
\begin{align*}
   U_{kk} \sim 2\Dzero - \Czero_{k1}\, {\omega_k}^{-1/2}
         + \cdots + \Czero_{k2}\,{\omega_k}^{-3}, & & \omega_k\tau\rightarrow\infty
\end{align*}
where
\begin{equation*}
   \Czero_{k1}/c_0=\aCk+3\aSk/2\varpi_k=A_{k1}^0 \qquad\qquad
       \Czero_{k2} = 4\varpi_k \Dzero M[f;3]/\pi\tau^3. 
\end{equation*}
For the low-frequency limit, $\omega_k\tau\rightarrow0$, the qualitative behaviour
of $U_{kk}$ depends on $\vartheta$,
\begin{equation}
   U_{kk} \sim \lb\{ \begin{aligned}
    & 2\Dinf + \Cinf_{k1}(\vartheta)\,{\omega_k}^\vartheta + \cdots
                               + \Cinf_{k3}\,{\omega_k}^3, & & 0<\vartheta<3 \\
    & 2\Dinf - \Cinf_{k2}\,{\omega_k}^3\ln\omega_k\tau
                                 + \Cinf_{k4}\,{\omega_k}^3, & & \vartheta=3 \\
    & 2\Dinf + \Cinf_{k3}\,{\omega_k}^3 + \cdots
                   + \Cinf_{k1}(\vartheta)\,{\omega_k}^\vartheta, & & 3<\vartheta
    \end{aligned} \rb.
\end{equation}
where
\begin{align*}
   \Cinf_{k1}(\vartheta) &= (4{\varpi_k}^{2-\vartheta}M[h_{kk};\vartheta+1]/\pi) \cinf  &
   \Cinf_{k3} &= 8\tau^3 \Dzero M[f;-3]/\pi\varpi_k \\
   \Cinf_{k2} &= (4/\pi\varpi_k) \cinf & 
   \Cinf_{k4} &= \Cinf_{k2} \lb[\ln\varpi_k + 2\Dzero K^1_{k10}\tau^3/\cinf\rb].
\end{align*}

Similarly, the expansion of $U_{k0}$ in the high-frequency limit is
\begin{align}
   U_{k0} \sim 2\Dzero - \Czero_{01}\, {\omega_k}^{-1/2} + \cdots
         + \Czero_{02} {\omega_k}^{-1}, & & \omega_k\tau\rightarrow\infty
\end{align}
with
\begin{equation*}
   \Czero_{01}/c_0=8k^{1/2}/3+\aSk/\varpi_k=A_{01}^0 \qquad\qquad
     \Czero_{02}=2\varpi_k \Dzero M[f;1]/\pi\tau.
\end{equation*}
When $\omega_k\tau\rightarrow0$,
\begin{equation}
   U_{k0} \sim \lb\{ \begin{aligned}
    & 2\Dinf + \Cinf_{01}(\vartheta)\,{\omega_k}^\vartheta + \cdots
                               + \Cinf_{03}\,{\omega_k}^3, & & 0<\vartheta<3 \\
    & 2\Dinf - \Cinf_{02}\,{\omega_k}^3\ln\omega_k\tau
                                 + \Cinf_{04}\,{\omega_k}^3, & & \vartheta=3 \\
    & 2\Dinf + \Cinf_{03}\,{\omega_k}^3 + \cdots
                + \Cinf_{01}(\vartheta)\,{\omega_k}^\vartheta, & & 3<\vartheta
    \end{aligned} \rb.
\end{equation}
where
\begin{align*}
   \Cinf_{01}(\vartheta) &= -(2{\varpi_k}^{2-\vartheta}M[h_{k0};\vartheta+1]/\pi)
                                                               c_\infty  &
   \Cinf_{03} &= -4\tau^3 \Dzero M[f;-3]/\pi\varpi_k \\
   \Cinf_{02} &= -(2/\pi\varpi_k) c_\infty & 
   \Cinf_{04} &= \Cinf_{02}\lb[\ln\varpi_k + 2\Dzero K^1_{010}\tau^3/\cinf\rb].
\end{align*}

\subsection{Behaviour of $\Cinf_{k1}(k,\vartheta)$ \& $\Cinf_{01}(k,\vartheta)$}
\label{sub:supp:Cl1inf}


It is of interest to consider how the coefficients $\Cinf_{k1}(k,\vartheta)$ \&
$\Cinf_{01}(k,\vartheta)$ behave with respect to $\vartheta$ and $k$.
The coefficients may be evaluated exactly using Eqs.~(\ref{eqn:supp:Mhkl:Mhk0})
\& (\ref{eqn:supp:Mhkl:Mhkk}) together with some of the results for $H(s;t,c)$ and
$I(s;t,c)$ in Sec.~\ref{sub:supp:Integrals}.

\begin{figure}[tbp]
 \centering
 \begin{tabular}{cc}
  \subfloat[]{\setcounter{subfigure}{1}
          \includegraphics[scale=0.72]{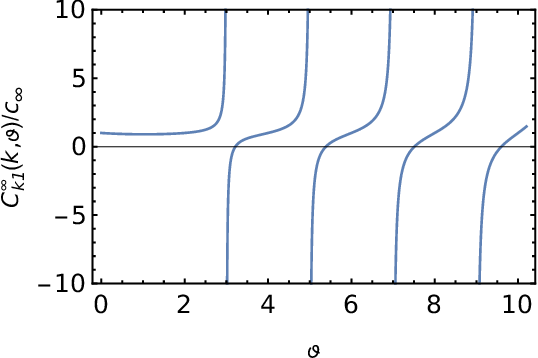}} \quad  & \quad
  \subfloat[]{\setcounter{subfigure}{2}
                       \includegraphics[scale=0.72]{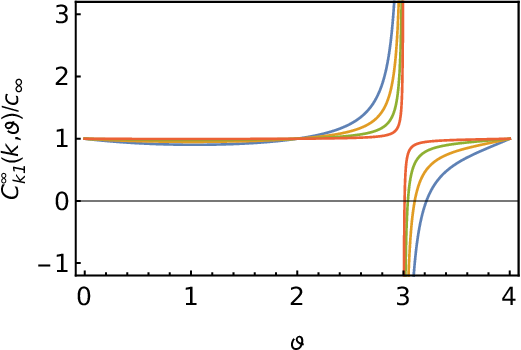}}
 \end{tabular}
 \caption{
Plots illustrating the characteristics of $\Cinf_{k1}(k,\vartheta)/\cinf$ as a
function of $\vartheta$ and $k$.
(a) Plot of $\Cinf_{k1}(k,\vartheta)/\cinf$ for $k=1$.
The singularities at $\vartheta=3,5,7\cdots$ correspond to the poles of
$M[h_{kk};\vartheta+1]$, while the region $0<\vartheta<3$ lies within the analytic
strip of the same function.
$\Cinf_{k1}(k,\vartheta)/\cinf$ resembles $1+\tan(\pi\vartheta/2)$ for $\vartheta>2$.
(b) Magnification of the region $0<\vartheta<4$ to demonstrate how
$\Cinf_{k1}(k,\vartheta)/\cinf$ behaves as a function of $k$.
There are curves for $k=1$ (blue), 2 (orange), 5 (green) and 20 (dark orange).
Apart from the singularities, $\Cinf_{k1}(k,\vartheta)/\cinf$ always goes to 1.
There are some special cases at $\vartheta=0,2,4\cdots$ where
$\Cinf_{k1}(k,\vartheta)/\cinf=1$ independent of $k$, but otherwise
$\Cinf_{k1}(k,\vartheta)/\cinf$ approaches the limit as $k^{-1}$ for increasing $k$.
}
 \label{fig:supp:Cl1inf:Ck1inf}
\end{figure}

Figure~\ref{fig:supp:Cl1inf:Ck1inf} showcases the dependence of
$\Cinf_{k1}(k,\vartheta)/\cinf$ on $\vartheta$ and $k$.
The singularities at $\vartheta=3,5,7\cdots$ correspond to the poles of
$M[h_{kk};\vartheta+1]$, while the the region $0<\vartheta<3$ falls within the
analytic strip of the same function.
Many of the characteristics of the figure can be understood
by considering the large $k$ behaviour of the coefficient.
After setting $s=\vartheta+1$ and multiplying Eq.~(\ref{eqn:supp:Mhkl:Mhkk_largek}) by
$B_{kk}{\varpi_k}^{-\vartheta}$,
\begin{equation}
   \frac{\Cinf_{k1}(k,\vartheta)}{\cinf} \sim 1+\frac{(\vartheta-1)
    \tan(\pi\vartheta/2)} {\varpi_k} - \frac{2\sec(\pi\vartheta/2)}
     {\Gamma(-\vartheta)}\,{\varpi_k}^{-\vartheta-2} \lb\{1+O({\varpi_k}^{-2})\rb\}.
 \label{eqn:supp:Cl1inf:Ck1inf}
\end{equation}

For $\vartheta>0$, the $O({\varpi_k}^{-1})$ term always dominates the
$O({\varpi_k}^{-\vartheta-2})$ term.
$\Cinf_{k1}(k,\vartheta)$ is always exactly 1 for
$\vartheta=0,2,4\cdots$, which is evidenced by the plots at different $k$ converging
to the same point in Fig.~\ref{fig:supp:Cl1inf:Ck1inf}b, but otherwise the
coefficient approaches 1 as $1/k$ for increasing $k$. There is also a special case
where Eq.~(\ref{eqn:supp:Cl1inf:Ck1inf}) must be evaluated in the limit
$\vartheta\rightarrow1$ so that
\begin{equation}
   \frac{\Cinf_{k1}(k,1)}{\cinf} \sim
     1 - \frac{2}{\pi\varpi_k} + \frac{4}{\pi{\varpi_k}^3} + O({\varpi_k}^{-5}).
 \label{eqn:supp:Cl1inf:Ck1inf_theta=1}
\end{equation}
The dominance of the $O({\varpi_k}^{-1})$ term is why $\Cinf_{k1}(k,\vartheta)$
resembles plots of $1+\tan(\pi\vartheta/2)$ for $\vartheta>2$
(Fig.~\ref{fig:supp:Cl1inf:Ck1inf}a).

\begin{figure}[tbp]
 \centering
 \begin{tabular}{cc}
  \subfloat[]{\setcounter{subfigure}{1}
        \includegraphics[scale=0.72]{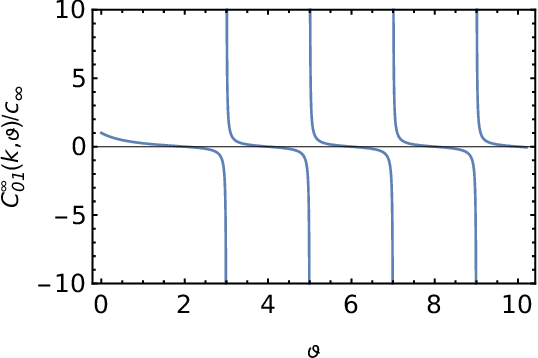}} \quad  & \quad
  \subfloat[]{\setcounter{subfigure}{2}
        \includegraphics[scale=0.72]{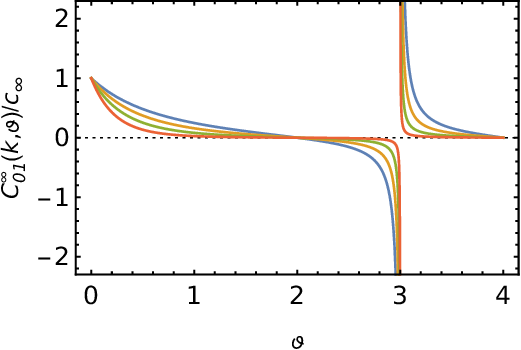}}
 \end{tabular}
 \caption{
Plots showcasing the features of $\Cinf_{01}(k,\vartheta)/\cinf$ as a function
of $\vartheta$ and $k$.
(a) Plot of $\Cinf_{01}(k,\vartheta)/\cinf$ for $k=1$.
The singularities at $\vartheta=3,5,7\cdots$ correspond to the poles of
$M[h_{k0};\vartheta+1]$, while the region $0<\vartheta<3$ falls within the analytic
strip of the same function.
$\Cinf_{01}(k,\vartheta)/\cinf$ looks like $-\tan(\pi\vartheta/2)$ for $\vartheta>2$.
(b) Replot of the region $0<\vartheta<4$ to demonstrate how
$\Cinf_{01}(k,\vartheta)/\cinf$ behaves as a function of $k$.
Curves for $k=1$ (blue), 2 (orange), 5 (green) and 20 (dark orange) are shown.
There are some special cases that are independent of $k$; that is,
$\Cinf_{01}(k,\vartheta)/\cinf$ is equal to $1$ for $\vartheta=0$,
and equal to 0 for $\vartheta=2,4,6\cdots$.
Otherwise, apart from the singularities, $\Cinf_{01}(k,\vartheta)/\cinf$ always
goes to 0 for increasing $k$ but the  way it does so depends on the value of
$\vartheta$. The limit is approached as $k^{-\vartheta}$ for $0<\vartheta<1$, as
$k^{-1}\ln k$ for $\vartheta=1$, and as $k^{-1}$ for $\vartheta>1$.
}
 \label{fig:supp:Cl1inf:C01inf}
\end{figure}

Figure~\ref{fig:supp:Cl1inf:C01inf} presents $\Cinf_{01}(k,\vartheta)/\cinf$
as a function of $\vartheta$ and $k$.
The singularities at $\vartheta=3,5,7\cdots$ correspond to the poles
of $M[h_{k0};\vartheta+1]$, while the region $0<\vartheta<3$ lies within the
analytic strip of the same function.
Setting $s=\vartheta+1$ and multiplying Eq.~(\ref{eqn:supp:Mhkl:Mhk0_largek}) by
$B_{k0}{\varpi_k}^{-\vartheta}$, the following large $k$ expansion can be derived
\begin{equation}
   \frac{\Cinf_{01}(k,\vartheta)}{\cinf} \sim -\frac{\tan(\pi\vartheta/2)}{\varpi_k}
    + \frac{\sec(\pi\vartheta/2)}{\Gamma(2-\vartheta)}\,{\varpi_k}^{-\vartheta}
                                                 \lb\{1+O({\varpi_k}^{-2})\rb\}.
 \label{eqn:supp:Cl1inf:C01inf}
\end{equation}
This expression resembles Eq.~(\ref{eqn:supp:Cl1inf:Ck1inf}) in several ways, but
on this occasion the large $k$ behaviour depends on the value of $\vartheta$:
\begin{itemize}
\item
$\vartheta=0$.
Although this case is not relevant to the asymptotic behaviour of $U_{k0}$, for
completeness it is mentioned that $\Cinf_{01}(k,0)=1$ independent of $k$.

\item
$0<\vartheta<1$.
The $O({\varpi_k}^{-\vartheta})$ term is dominant here so $\Cinf_{01}(k,\vartheta)$
goes to zero as $k^{-\vartheta}$.

\item
$\vartheta=1$.
Taking the limit of Eq.~(\ref{eqn:supp:Cl1inf:C01inf}) as $\vartheta\rightarrow1$
produces the finite result
\begin{equation}
   \frac{\Cinf_{01}(k,1)}{\cinf} \sim 
    \frac{2\ln\varpi_k}{\pi\varpi_k}+\frac{2\gamma_e}{\pi\varpi_k}
                                          + O({\varpi_k}^{-3}),
 \label{eqn:supp:Cl1inf:C01inf_theta=1}
\end{equation}
which aligns with the fact that $M[h_{k0};\vartheta+1]$ has no pole at $\vartheta=1$.
The coefficient approaches zero as $\ln k/k$ for large $k$.

\item
$\vartheta>1$.
The $O({\varpi_k}^{-1})$ term dominates for this range of $\vartheta$,
and that is why the plot in Fig.~\ref{fig:supp:Cl1inf:C01inf}a closely
reflects the characteristics of $-\tan(\pi\vartheta/2)$.
$\Cinf_{01}(k,\vartheta)$ always approaches 0 as $1/k$ except when
$\vartheta=2,4,6\cdots$, in which case the coefficient equals 0 independent of $k$.

\end{itemize}

\subsection[Some comments on two previous derivations of the ADC]{Some comments on
two previous derivations of the ADC in the high-frequency limit}
\label{sub:supp:ADCcomments}

\vspace{2mm}
\noindent
{I. Reference~\cite{Sukstanskii2013}.}
\vspace{1mm}

The derivation in \cite{Sukstanskii2013} is based on equating the ADC signal model
with the second order term of the cumulant expansion framework for describing DWI.
After some manipulation the relationship was written in terms of the cumulative
diffusion coefficient, $D(t)$, like so
\begin{equation*}
   b \tilde{D}_G(t) = \frac{\gamma^2}{2} \int_0^t\!\!d\tau_1\int_0^{\tau_1}\!\!d\tau_2\,
     g(\tau_1) g(\tau_2) (\tau_1-\tau_2) D(\tau_1-\tau_2).
\end{equation*}
The well-known expression for $D(t)$ in the short-time limit
\begin{equation}
   D(t) = \Dzero\lb[1-\frac{4}{3\sqrt{\pi}}\lb(\frac{t}{t_D}\rb)^{1/2}\rb],
 \label{eqn:supp:ADCcomments:D(t)}
\end{equation}
which is said to be valid under the condition that the time $t$
is much smaller than the characteristic time $t_D=d^2/(S/V)^2 \Dzero$,
was then directly inserted under the integral. With $g(\tau)=g_0\cos(\omega\tau-\vphi)$,
the integrals were evaluated to finally obtain
\begin{equation}
   \tilde{D}_G(t=2\pi N/\omega) = \Dzero\lb[1-\frac{c'(\vphi,N)}{(\omega t_D)^{1/2}}\rb]
 \label{eqn:supp:ADCcomments:DG}
\end{equation}
where
\begin{equation*}
   c'(\vphi,N) = \frac{32\pi N^{3/2}\sin^2\vphi + 12\pi N\, C(2N^{1/2})
           + 3(3+4\sin^2\vphi)\, S(2N^{1/2})}{6\sqrt{2}\pi N(1+2\sin^2\vphi)},
\end{equation*}
and $C(x)$ and $S(x)$ are the Fresnel integrals.

The approach taken in the present manuscript starts with the same assumed equivalence
between the ADC and cumulant expansion signal models. However, rather than working
with a time-dependent diffusion metric, the cumulant expansion term is transformed
so that the ADC is written as an integral involving the frequency-dependent metric
$u_2(\omega)$ (i.e.~Eq.~(27) in \cite{Kershaw2021} with $U_{kk}$ \& $U_{k0}$ replaced
by their integral definitions).
The high-frequency asymptotic behaviour of $u_2(\omega)$ was previously known from
its derivation in \cite{Novikov2011}, and with that information the MTM was applied
under the condition $\beta=T/\tau\rightarrow0$ to derive asymptotic expansions
for $U_{kk}$ \& $U_{k0}$ and eventually
Eq.~(\ref{eqn:Discussion:ADC:HiLim}) in Sec.~\ref{sub:Discussion:ADC}.

\begin{table}
 \centering
 \captionsetup{width=12.2cm}
 \caption{Translating between the notations and conditions used in
References~\cite{Sukstanskii2013} \& \cite{Novikov2019} and the present manuscript.}
 \begin{tabular}{@{\vrule height 12.0pt depth0pt  width0pt}lccc}
     \hline
   $\;$Quantity & Ref.~\cite{Sukstanskii2013} & Ref.~\cite{Novikov2019}
                                                         & Present work \\ \hline
   $\;$number of oscillations & $N$ & $N$ & $k$ \\
   $\;$MPG duration & $t$ & $T$ & $T$ \\
   $\;$MPG frequency & $\omega$ & $\omega_0$ & $\omega_k$ \\
   $\;$generic time variables & $\tau$, $\tau_1$, $\tau_2$ & $t$
                                                            & $t$, $t_1$, $t_2$ \\
   $\;$MPG amplitude & $g_0$ & $g_0/\gamma$ & $G$ \\
   $\;$MPG phase & $\vphi$ & $\vphi$ & $\phi$ \\
   $\;$cumulative diffusion coefficient & $D(t)$ & $D(T)$ & $D(T)$ \\
   $\;$characteristic diffusion-time & $t_D$ &  & $\tau$ \\
   $\;$finite-$N$ correction factor & $c'(\vphi,N)$ & $\tilde{c}(\vphi,N)/\sqrt{2}$
                                                  & $\sqrt{2}\,\Czerol(k,\phi)/\czero$ \\
   $\;$apparent diffusion coefficient & $\tilde{D}_G(t)$ & $-\frac{\ln s}{b}$
                                                           & $\ADC_k(\phi)$ \\[1mm]
   \hline
   $\;$strict condition & $t\ll t_D$ & $(S/V)\sqrt{\Dzero T}\ll 1$
                                                  & $\beta=T/\tau\rightarrow0$ \\
   $\;$weak condition & $t/N\ll t_D$ & $(S/V)\sqrt{\Dzero/\omega_0}\ll1$
                                             & $k\rightarrow\infty$ \\[1mm]
   \hline
 \end{tabular}
 \label{tab:supp:ADCcomments:notation}
\end{table}

To compare the two derivations it is helpful to first translate the separate notations.
Table~\ref{tab:supp:ADCcomments:notation} provides a map between the notations and
conditions used in each study.
Also, although $\tau$ has been used in this work as a nonspecific time-scale
characterising the response of a system, it could be taken as equivalent to
the quantity $t_D$ in the short-time expansion of $D(t)$.
All other quantities, such as $\Dzero$, $S/V$ and $d$, have the
same meaning in both studies.
After applying the translation it is found that the expression for
$\tilde{D}_G(t=2\pi N/\omega)$ in Eq.~(\ref{eqn:supp:ADCcomments:DG}) is equivalent
to the result derived for $\ADC_k(\phi)$ in Eq.~(\ref{eqn:Discussion:ADC:HiLim}).
In particular, the quantity $c'(\vphi,N)$ is equivalent to
$\sqrt{2} \Czerol(k,\phi)/\czero$.

Even though the results of the two studies are equivalent, 
several additional comments on differences between the works and other issues are
included below:

\begin{enumerate}

\item
The author of \cite{Sukstanskii2013} did not anticipate any practical
restrictions on the parameters of an oscillating MPG.
It also seems that oscillating-gradient DWI was thought of as an experiment
where $N$ is set first, and then the duration of the MPG is varied to
increase or decrease the frequency $\omega$.
Within that conceptual set-up the observer would be able to arbitrarily set the
observation frequency without limitation to observe the
high-frequency regime for any system.

In contrast, the present study began with the realisation that the number of
oscillations and duration of an MPG are limited in practice, which means that the
resolution and range of the frequency $\omega_k$ is restricted.
To maximise the spectral range and resolution it was suggested in
\cite{Kershaw2021} that experiments are conducted by setting the duration
to be as long as possible, and then the frequency is altered by varying the number
of oscillations.
It is possible to set the number of oscillations and then alter the MPG duration to
change the frequency as conceived in \cite{Sukstanskii2013}, but the window of
accessible frequencies will be unaltered.
Overall, the practical restrictions on $\omega_k$ mean that the system response can
only be observed within a limited immoveable window of frequencies, and whether the
observed response is in the high-frequency regime or otherwise depends on the
nature of the system.

\item
It was stated after Eq.~(17) in \cite{Sukstanskii2013} that
the condition $c'(\vphi,N)(\omega t_D)^{-1/2}\ll1$ must be satisfied for
Eq.~(\ref{eqn:supp:ADCcomments:DG}) to be a valid asymptotic expansion.
This condition and the large $N$ behaviour of $c'(\vphi,N)$ led the author of
\cite{Sukstanskii2013} to conclude that the critical time-scale depends
on the waveform selected for the applied MPG. More explicitly, after substituting
$\omega=2\pi N/t$ the condition becomes $[c'(\vphi,N)/(2\pi N)^{1/2}](t/t_D)^{1/2}\ll1$.
Since $c'(\vphi,N)$ is $O(N^{1/2})$ for large $N$ and $\vphi\neq0$, the condition
therefore implies that the high-frequency expansion of the ADC is valid if $t\ll t_D$.
On the other hand, $c'(\vphi,N)$ is $O(1)$ for the special case $\vphi=0$, which
means that the expansion is valid for the weaker condition $t/N\ll t_D$.
That is, non-$\cos$-type MPGs require the duration of the whole MPG to
be dominated by the characteristic diffusion-time, whereas for $\cos$-type MPGs
only the period of the oscillations needs to be much smaller than the characteristic
diffusion-time.

In comparison, the derivations in this work assume from the beginning that
$T/\tau\rightarrow0$, which is equivalent to the stricter condition of
\cite{Sukstanskii2013}. Nevertheless, given
the equivalence between $c'(\vphi,N)$ and $\sqrt{2} \Czerol(k,\phi)/\czero$, if $\phi$
is set to zero in Eq.~(\ref{eqn:Discussion:ADC:HiLim}) then the same
reasoning as used in \cite{Sukstanskii2013} can be applied to obtain the weak condition
$T/k\ll\tau$ for $\cos$-type MPGs.

\item
There is a small error in Eq.~(17) of \cite{Sukstanskii2013}. It was
stated that $c'(0,N)=(1/\sqrt{2})(1-1/\pi^2 N^{3/2})$ for $N\gg1$, whereas the
expression should actually be $c'(0,N)=(1/\sqrt{2})(1+3/4\sqrt{2}\pi N)$.
The factor of $-1/\pi^2 N^{3/2}$ actually appears in the third term of the large
$N$ expansion of $c'(0,N)$.

\item
The quantity $k$ in Eqs.~(19), (21) \& (22) of \cite{Sukstanskii2013} was never
defined.

\end{enumerate}

\vspace{2mm}
\noindent
{II. Reference~\cite{Novikov2019}.}
\vspace{1mm}

Even though the ADC is not explicitly mentioned in \cite{Novikov2019}, the
high-frequency behaviour of the signal observed using an MPG with a finite number
of oscillations is discussed in Appendix C.
The discussion is based on a signal equation derived in the preceding appendix,
Appendix~B.
That equation, Eq.~(B5), was written in terms of the time-dependent diffusion
metrics $\cD(t)$, the retarded velocity autocorrelation function, and $D(T)$, the
cumulative diffusion coefficient.\footnote{The equivalence of Eq.~(B5) and
Eq.~(\ref{eqn:intro:lnSk}) was previously addressed in the Supporting Material
(Sec.~S-9) of \cite{Kershaw2021}.} Expressions for $\cD(t)$ and $D(T)$ in the
short-time limit were inserted into Eq.~(B5) and the integrals were evaluated.
The final result, Eq.~(C6), is equivalent to the result obtained in
Ref. \cite{Sukstanskii2013} (i.e.~Eq.~(\ref{eqn:supp:ADCcomments:DG}) above), and
hence it is also equivalent to the result presented in this work.

Some additional remarks on \cite{Novikov2019} are listed below:
\begin{enumerate}

\item
In the paragraph after Eq.~(C6) it is mentioned
that the finite-$N$ correction factor, $\tilde{c}(\vphi,N)$ (see
Table~\ref{tab:supp:ADCcomments:notation} for the translation between studies),
diverges as $N^{1/2}$ for large $N$.
This is the same issue that was highlighted by the author of \cite{Sukstanskii2013}.
The authors of \cite{Novikov2019} point out that the problem of divergence only
arises when the strict condition, $(S/V)\sqrt{\Dzero T}\ll1$, is violated.
It is also noted that the $N^{1/2}$ dependence is due to
the factor of $T^{1/2}$ in the short-time expansion of $D(T)$ and subsequently
``forcing Eq.~(C6) to mimic the form of Eq.~(32).''\footnote{Eq.~(32) of
\cite{Novikov2019} is the short-time expansion of $D(t)$. The authors probably meant
Eq.~(33), which is the high-frequency expansion of $\re\,\cD(\omega)$.}
A similar situation occurs in the current manuscript. The $O(k^{1/2})$ term in
$A^0_{01}$ originates from writing the high-frequency asymptotic behaviour of $U_{k0}$
(re~Eq.~(\ref{eqn:Universality:HiLim:Ukl})) in terms of $\omega_k\tau$
instead of the true asymptotic variable $\beta$.
The present authors agree that the problem of divergence is spurious and will not
occur as long as the strict condition, $\beta\rightarrow0$, holds.

\item
A physical interpretation for the log signal as the sum of a pure oscillating gradient
(OG) component and a pulsed gradient (PG) component is discussed in \cite{Novikov2019}.
The OG component corresponds to the signal when $\varphi=0$, while the PG
component is the additional signal that appears when $\varphi$ is nonzero.
The authors of \cite{Novikov2019} propose that the decomposition into components
allows Eq.~(C4) to be used to probe the $S/V$ limit for any value of $\varphi$ as
long as the weak condition, $(S/V)\sqrt{\Dzero/\omega_0}\ll1$, is met.
This is in contrast to the conclusion in \cite{Sukstanskii2013} where the weak condition
only applies when $\varphi=0$.

In the notation of the present manuscript the OG \& PG components correspond to
$U_{kk}$ \& $U_{k0}$, respectively.
Recalling that $A^0_{k1}\sim O(1)$, the same reasoning as used in \cite{Novikov2019}
with respect to Eq.~(C4) could be applied here to argue that under suitable conditions
the $U_{kk}$ contribution to Eq.~(\ref{eqn:intro:lnSk}) may probe the high-frequency
limit regardless of the behaviour of the $U_{k0}$ contribution.
This is an interesting idea, but its application is hindered by the practical
limitations on MPG duration and frequency.
It is also unclear how the OG contribution to the signal could be reliably isolated
when $\phi\neq0$.

\item
There is some potential for confusion in \cite{Novikov2019} with regards to how the
asymptotic results derived from Eq.~(B5) in Appendix~B and Appendix~C are obtained.

The derivation in Appendix~C proceeds by directly inserting
Eq.~(C1) into Eq.~(B5) without indicating what the asymptotic variable is.
However, as the result is equivalent to that obtained here using the MTM, 
 it can be concluded that the asymptotic variable
in the derivation of Appendix~C must be $T$ (with $N$ kept constant).\footnote{Even
though $\beta=T/\tau$ was the asymptotic variable used in this work,
the effective asymptotic quantity is $T$ because $\tau$, which is a property of
the sample, is constant.}
That is, similar to what has been done in this manuscript, Eq.~(C6) is an asymptotic
expansion of the signal as $T\rightarrow0$ that is written in terms of the frequency
$\omega_0$ (using the fact $T=2\pi N/\omega_0$).

On the other hand, it was previously noted in \cite{Kershaw2021}
(Supplementary Material, Sec.~S-9) that
Eq.~(B6) was derived from Eq.~(B5) under the asumption that $\omega_0$ remains
constant. Since $\omega_0=2\pi N/T$, the assumption means that $N$ and $T$ are
a coupled set of asymptotic variables.
That is, the asymptotic behaviour is obtained by replacing every occurrence of $N$ 
with $\omega_0 T/2\pi$ (or $T$ with $2\pi N/\omega_0$) and expanding as $T$ goes to 0
(or $N$ goes to $\infty$).
Deriving the expansion in this way produces a different result to when $T$ and $N$
are uncoupled. For example, using the exact $U_{k0}$ \& $U_{kk}$ calculated for
Model~2\ in \cite{Kershaw2021}, the asymptotic behaviour with $T$ and $k$ coupled
is:
%
%
\begin{align*}
   \frac{U_{k0}}{2\Dzero} &\sim 1-\zi+\frac{2\zi (\omega_k\tau)^{1/2}}{\pi^{1/2}}
     \,{\varpi_k}^{-1/2} -\frac{\zi (\omega_k\tau)^{1/2}}{\sqrt{2}}\lb[\frac{1+
      \omega_k\tau+\sqrt{2}(\omega_k\tau)^{5/2}}{1+(\omega_k\tau)^2}\rb]{\varpi_k}^{-1}
       +O(\varpi_k)^{-3/2} \\
   \frac{U_{kk}}{2\Dzero} &\sim u_2(\omega_k) -\frac{\zi (\omega_k\tau)^{1/2}}
     {\sqrt{2}}\lb[\frac{1-\omega_k\tau+5(\omega_k\tau)^2-8(\omega_k\tau)^{5/2}
      +3(\omega_k\tau)^3} {2[1+2(\omega_k\tau)^2]^2}\rb] {\varpi_k}^{-1}
       +O(\varpi_k)^{-5/2}.
\end{align*}
Comparison of these expressions with the $\omega_k\tau\rightarrow\infty$ parts of
Eqs.~(\ref{eqn:asymptotics:apply:Model2:Uk0}) \&
(\ref{eqn:asymptotics:apply:Model2:Ukk}), which were derived with $\beta$
as the sole asymptotic variable, finds that the results are quite different.
In particular, notice that the lowest order term in $U_{kk}$ is $u_2(\omega_k)$, which
is not the case in Eq.~(\ref{eqn:asymptotics:apply:Model2:Ukk}).

\item
The present authors are also aware that the asymptotic behaviour of the signal might
be investigated in a third way.
In Sec.~\ref{sub:asymptotics:meth} it was noted that $\omega_k$ may be manipulated
by independently varying either $k$ or $T$, but practical limitations restrict the
range of both quantities. It was therefore argued that it is the ratio of $T$ and
the characteristic time-scale $\tau$ that determines the regime for OGSE-DWI
observations. Varying $k$ provides only a small perturbation to the
observed response because the range of $k$ is limited. However, if the
practical limitation on $k$ could be ignored, increasing $k$ (with constant $T$)
would concurrently decrease the period of the oscillation so that the condition
for the high-frequency limit, $\omega_k\tau=2\pi k\tau/T\rightarrow\infty$, is
eventually satisified.
That is, the high-frequency limit of the sample is probed under the condition
$k\rightarrow\infty$, which is a form of the weak condition because it implies that
the period of the oscillation, $T/k$, is the critical time-scale.
In general, the asymptotic expansions of $U_{k0}$, $U_{kk}$ and the ADC obtained in
this way differ from the results obtained under the strict condition. The exact
$U_{k0}$ \& $U_{kk}$ for Model~2 of \cite{Kershaw2021} are used to demonstrate this:
\begin{align*}
   \frac{U_{k0}}{2\Dzero} &\sim 1-\zi-\zi\,
     \frac{1-2\sqrt{\beta/\pi}-e^{\beta}\erfc(\sqrt{\beta})}{\beta}
     -\frac{\zi}{\sqrt{2}\beta}\lb(\omega_k\tau\rb)^{-3/2}
                                             + O\lb(\omega_k\tau\rb)^{-2} \\
   \frac{U_{kk}}{2\Dzero} &\sim 1-\frac{\zi}{\sqrt{2}}\lb(\omega_k\tau\rb)^{-1/2}
    -\frac{\zi(3-2\beta)}{2\sqrt{2}\beta}\lb(\omega_k\tau\rb)^{-3/2}
                                       + O\lb(\omega_k\tau\rb)^{-2}. 
\end{align*}

These expansions have been written in terms of powers of $\omega_k\tau$ to aid
comparison with the high-frequency results of
Eqs.~(\ref{eqn:asymptotics:apply:Model2:Uk0})--(\ref{eqn:asymptotics:apply:Model2:Ukk})
in Sec.~\ref{sec:apply}.
One obvious difference is that the coefficients to the powers of $\omega_k\tau$
here are functions of $\beta$, whereas the coefficients in
Eqs.~(\ref{eqn:asymptotics:apply:Model1:Uk0}), (\ref{eqn:asymptotics:apply:Model1:Ukk}),
(\ref{eqn:asymptotics:apply:Model2:Uk0}) \& (\ref{eqn:asymptotics:apply:Model2:Ukk})
feature $k$ (or $\varpi_k$) but not $\beta$.
This is consistent with the fact that the asymptotic variable here is $k$, while the
actual asymptotic variable used for the results in Sec.~\ref{sec:apply} is $\beta$.
With future improvements in gradient hardware it might become possible to implement
MPGs with very high values of $k$. Even so, there are other issues (e.g.~high SAR,
insufficient diffusion-weighting or signal attenuation) that might prevent the use
of large $k$. So, although it might be possible in principle, probing the
high-frequency/short-time limit in this way does not seem to be practical.

\item
There is a typographical error in the upper part of Eq.~(C2) in \cite{Novikov2019}.
Rather than $\sqrt{2}/\omega_0[\cdots]$, the right-hand side should be
$\sqrt{2/\omega_0}[\cdots]$. The subsequent results have not been affected by this
error.

\end{enumerate}

\subsection[Low-frequency asymptotic behaviour of the ADC]{Low-frequency asymptotic
behaviour of the ADC for a single-harmonic MPG}
\label{sub:supp:ADCLoLim}

For a single-harmonic MPG the ADC can be written in terms of $U_{kk}$ \& $U_{k0}$
as \cite{Kershaw2021}
\begin{equation}
   \ADC_k(\phi) = \frac{U_{kk}+2U_{k0}\sin^2\phi}{2+4\sin^2\phi}.
 \label{eqn:supp:ADCLoLim:ADC}
\end{equation}
One way to determine the low-frequency behaviour of the ADC is to insert the
appropriate asymptotic expressions for $U_{k0}$ \& $U_{k0}$ from
Sec.~\ref{sub:Universality:LoLim}. However, that approach is problematic due to
the dependence of the results on $\vartheta$.
For example, as $\vartheta$ approaches 3 from below $\Cinf_{k1}(k,\vartheta)\rightarrow
+\infty$ while $\Cinf_{01}(k,\vartheta)\rightarrow-\infty$, so it is not immediately
clear which behaviour will dominate in the asymptotic behaviour of the ADC. This
problem could be solved with some careful analysis, but a more straightforward
method is to rewrite Eq.~(\ref{eqn:supp:ADCLoLim:ADC}) as an integral and
then apply the MTM.

An integral expression for the ADC is constructed
by introducing the integral definitions of $U_{k0}$ \& $U_{kk}$
into Eq.~(\ref{eqn:supp:ADCLoLim:ADC}) so that
\begin{equation}
   \ADC_k(\phi) = \frac{1}{2}\int_0^\infty\!\!d\omega\,u_2(\omega)H_{k}(\omega;T,\phi)
 \label{eqn:supp:ADCLoLim:ADC2}
\end{equation}
with
\begin{equation*}
   H_{k}(\omega;T,\phi) = \frac{H_{kk}(\omega;T)+2H_{k0}(\omega;T)\sin^2\phi}
                                                        {1+2\sin^2\phi}.
\end{equation*}
Using the properties of the $H_{kl}(\omega;T)$ it can be shown that
$\int_0^\infty d\omega\,H_k(\omega;T,\phi)=1$ and 
$H_k(\omega;T,\phi)=B_k(\phi) T\,h_k(\omega T;\phi)$, where
\begin{equation*}
   h_k(v;\phi) = \frac{(1-\cos v)[v^2\cos^2\phi+{\varpi_k}^2\sin^2\phi]}
                                         {v^2[v^2-{\varpi_k}^2]^2}
\end{equation*}
and $B_k(\phi)=4{\varpi_k}^2/(1+2\sin^2\phi)\pi$.
It is therefore possible to rewrite Eq.~(\ref{eqn:supp:ADCLoLim:ADC2}) in the
dimensionless form
\begin{equation}
   I_k(\beta;\phi) =\frac{\ADC_k(\phi)}{\Dzero \beta B_k(\phi)} 
                             = \int_0^\infty\!\!dv\,f(v) h_k(\beta v;\phi)
 \label{eqn:supp:ADCLoLim:Ik}
\end{equation}
with $f(v)=u_2(v/\tau)/2\Dzero$ as in Sec.~\ref{sub:asymptotics:MTM}.

From here the procedure is very similar to that used in
Sec.~\ref{sub:supp:Universality} but with some of the details removed.
The first step is to record the asymptotic behaviour of $h_k(v;\phi)$:
\begin{equation*}
   h_k(v;\phi) \sim \lb\{ \begin{aligned}
        & \sum_{n=0}^\infty h^{0}_{n}(\phi)\,v^{2n}, & & v\rightarrow 0 \\
        & (1-\cos v)\sum_{n=0}^\infty h^{\infty}_{n}(\phi)\,v^{-2n-4}, & &
                                            v\rightarrow\infty.
    \end{aligned} \rb.
\end{equation*}
Since $h_k(v;\phi)=h_{kk}(v)-h_{k0}(v)\sin^2\phi$ it is possible to write the
coefficients in these expansions as (re Secs.~\ref{sub:asymptotics:poles} \&
\ref{sub:supp:Mhkl})
\begin{align*}
   h_n^0(\phi) &= \lb\{ \begin{aligned}
       & -h_{00}^0\sin^2\phi, & \quad & n=0 \\
       & h_{kn-1}^0-h_{0n}^0\sin^2\phi, & & n\geqslant1
    \end{aligned} \rb. \\[2mm]
   h_n^\infty(\phi) &= \lb\{ \begin{aligned}
       & \cos^2\phi, & \quad & n=0 \\
       & h_{kn}^\infty-h_{0n}^\infty\sin^2\phi, & & n\geqslant1.
    \end{aligned} \rb. 
\end{align*}
Notice that $h_0^0(0)=0$ and $h_0^\infty(\pi/2)=0$, which means that the analytic
strip and poles of $M[h_k;s]$ depend on the value of $\phi$.
In fact, the analytic strip may be summarised as
$-2\delta_{0,\phi}<\re\,s<4+2\delta_{\pi/2,\phi}$
and the behaviour of $M[h_k;s]$ at its poles is
\begin{equation*}
   M[h_k;s] \sim \lb\{ \begin{aligned}
     & \frac{h_{n+\delta_{0,\phi}}^0(\phi)}{s+2n+2\delta_{0,\phi}}, 
                                      & & s\rightarrow-2n-2\delta_{0,\phi} \\
     & -\frac{h_{n+\delta_{\pi/2,\phi}}^\infty(\phi)}{s-2n-4-2\delta_{\pi/2,\phi}},
                                 & & s\rightarrow2n+4+2\delta_{\pi/2,\phi}
    \end{aligned} \rb.
\end{equation*}
for $n=0,1,2,\cdots$.

Assuming the same behaviour for $f(v)$ and its Mellin transform as in
Sec.~\ref{sub:asymptotics:poles}, the convergence of $I_k(\beta;\phi)$ requires
that $\vartheta>-1-2\delta_{0,\phi}$. The expansion of $I_k(\beta;\phi)$ as
$\beta\rightarrow\infty$ can therefore be broken down into the following cases:
\begin{enumerate}
\item[(a)]
$0<\vartheta<3+2\delta_{\pi/2,\phi}$.
The analytic strips of $M[f_1;1-s]$ and $M[f_2;1-s]$ intersect with that of
$M[h_k;s]$ for $-1-2\delta_{0,\phi}<\re\,s<1$ and $1<\re\,s<4+2\delta_{\pi/2,\phi}$,
respectively, so the corresponding known poles in the right-hand plane are
\begin{align*}
   & \{s_n\geqslant1\}=\{2n+4+2\delta_{\pi/2,\phi}\}\cup\{1,\vartheta+1\}
                         = \{1,\vartheta+1,\cdots,4+2\delta_{\pi/2,\phi},\cdots\} \\
   & \{s_n\geqslant4+2\delta_{\pi/2,\phi}\}=\{2n+4+2\delta_{\pi/2,\phi}\}
                                 = \{4+2\delta_{\pi/2,\phi},\cdots\}.
\end{align*}
There are no second order poles so the residues are easily calculated and
\begin{equation*}
   I_k(\beta;\phi) \sim \frac{\Dinf}{\Dzero}M[h_k;1]\beta^{-1}
     + \frac{\cinf}{2\Dzero\tau^\vartheta}M[h_k;\vartheta+1]\beta^{-\vartheta-1}.
\end{equation*}

\item[(b)]
$\vartheta=3+2\delta_{\pi/2,\phi}$.
In this case the pole at $\vartheta+1=4+2\delta_{\pi/2,\phi}$ of $M[f_1;1-s]$ coincides
with the first pole of $M[h_{kl};s]$. As such,
\begin{align*}
   & \{s_n\geqslant1\}=\{2n+4+2\delta_{\pi/2,\phi}\}\cup\{1,\vartheta+1\}
                       = \{1,(4+2\delta_{\pi/2,\phi},4+2\delta_{\pi/2,\phi}),\cdots\}
\end{align*}
and the expansion is
\begin{equation*}
   I_k(\beta;\phi) \sim \frac{\Dinf}{\Dzero}M[h_k;1]\beta^{-1}
    + \frac{\cinf h_{\delta_{\pi/2,\phi}}^\infty(\phi)}
      {2\Dzero\tau^{3+2\delta_{\pi/2,\phi}}} \beta^{-4-2\delta_{\pi/2,\phi}} \ln\beta.
\end{equation*}

\item[(c)]
$\vartheta>3+2\delta_{\pi/2,\phi}$.
The order of the poles switches from that in case (a):
\begin{align*}
   & \{s_n\geqslant1\}=\{2n+4+2\delta_{\pi/2,\phi}\}\cup\{1,\vartheta+1\}
                                 = \{1,4+2\delta_{\pi/2,\phi},\vartheta+1,\cdots\}
\end{align*}
so that now
\begin{equation*}
   I_k(\beta;\phi) \sim \frac{\Dinf}{\Dzero}M[h_k;1]\beta^{-1}
     + h_{\delta_{\pi/2,\phi}}^\infty(\phi) M[f;-3-2\delta_{\pi/2,\phi}]
                                            \beta^{-4-2\delta_{\pi/2,\phi}}.
\end{equation*}

\end{enumerate}

Now recalling Eq.~(\ref{eqn:supp:ADCLoLim:Ik}) and using the facts that
$\beta=\varpi_k/\omega_k\tau$ and $M[h_k;s]=M[h_{kk};s]-M[h_{k0};s]\sin^2\phi$,
the expansion of the ADC in the low-frequency limit can be summarised as
\begin{equation*}
   \ADC_k(\phi) \sim \lb\{ \begin{aligned}
   & \Dinf + \Cinf_1(k,\phi,\vartheta)\,{\omega_k}^{\vartheta},
                              & & 0<\vartheta<3+2\delta_{\pi/2,\phi} \\
   & \Dinf - \Cinf_2(k,\phi)\,{\omega_k}^{3+2\delta_{\pi/2,\phi}}
                \ln\omega_k\tau, & & \vartheta=3+2\delta_{\pi/2,\phi} \\
   & \Dinf + \Cinf_3(k,\phi)\,{\omega_k}^{3+2\delta_{\pi/2,\phi}},
                                   & & \vartheta>3+2\delta_{\pi/2,\phi}
     \end{aligned} \rb.
\end{equation*}
where
\begin{align*}
   \frac{\Cinf_1(k,\phi,\vartheta)}{\cinf} &= \frac{2{\varpi_k}^{2-\vartheta}}
                                    {(1+2\sin^2\phi)\pi} M[h_k;\vartheta+1] \\
   \frac{\Cinf_2(k,\phi)}{\cinf} &= \frac{2{\varpi_k}^{-1-2\delta_{\pi/2,\phi}}}
                {(1+2\sin^2\phi)\pi} h_{\delta_{\pi/2,\phi}}^\infty(\phi) \\
   \frac{\Cinf_3(k,\phi)}{\Dzero} &= \frac{2{\varpi_k}^{-1-2\delta_{\pi/2,\phi}}}
     {(1+2\sin^2\phi)\pi} h_{\delta_{\pi/2,\phi}}^\infty(\phi)
        M[f;-3-2\delta_{\pi/2,\phi}] \tau^{3+2\delta_{\pi/2,\phi}}.
\end{align*}

For $\phi=0$, the result is equivalent to the expansion of $U_{kk}/2$ as expected.
In particular, $\Cinf_1(k,0,\vartheta)=\Cinf_{k1}(k,\vartheta)/2$,
$\Cinf_2(k,0)=\Cinf_{k2}(k)/2$ and $\Cinf_3(k,0)=\Cinf_{k3}(k)/2$.

For $0<\phi<\pi/2$, the analytic strip of $M[h_k;\vartheta+1]$ narrows to
$-1<\vartheta<3$.
That is, a pole that did not exist for $\phi=0$ appears at $\vartheta=-1$, which
corresponds to the fact that $h_0^0(\phi)\neq0$ when $\phi\neq0$.

\begin{figure}[tbp]
 \centering
 \begin{tabular}{cc}
  \subfloat[]{\setcounter{subfigure}{1}
          \includegraphics[scale=0.72]{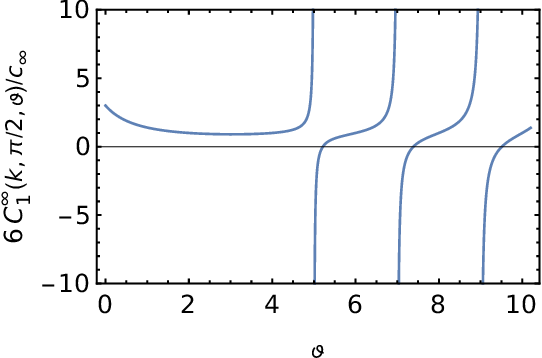}} \quad  & \quad
  \subfloat[]{\setcounter{subfigure}{2}
                       \includegraphics[scale=0.72]{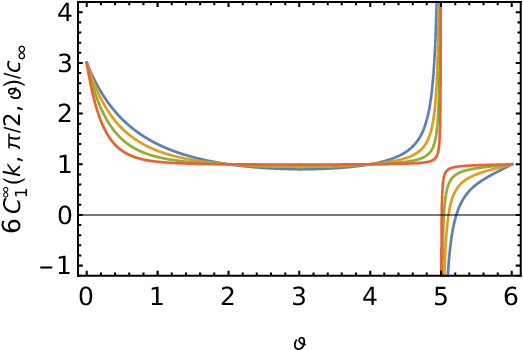}}
 \end{tabular}
 \caption{
Plots illustrating the characteristics of $\Cinf_1(k,\pi/2,\vartheta)/\cinf$ as a
function of $\vartheta$ and $k$ for $\phi=\pi/2$.
(a) Plot of $\Cinf_1(k,\pi/2,\vartheta)/\cinf$ for $k=1$.
The singularities at $\vartheta=5,7,9\cdots$ correspond to the poles of
$M[h_{k};\vartheta+1]$, and the singularity-free region $0<\vartheta<5$ is within
the analytic strip of the same function.
$\Cinf_1(k,\pi/2,\vartheta)/\cinf$ resembles $\Cinf_{k1}(\vartheta,\pi/2)/\cinf$
for $\vartheta>5$ (re Fig.~\ref{fig:supp:Cl1inf:Ck1inf}).
(b) Magnification of the region $0<\vartheta<6$ with curves for
$k=1$ (blue), 2 (orange), 5 (green) and 20 (dark orange).
For $\vartheta>0$, $6\Cinf_1(k,\pi/2,\vartheta)/\cinf$ always goes to 1 for
increasing $k$ but the way it does so depends on the value of $\vartheta$.
The limit is approached as $k^{-\vartheta}$ for $0<\vartheta<1$, as
$k^{-1}\ln k$ for $\vartheta=1$, and as $k^{-1}$ for $1<\vartheta$.
There are also some special cases that are independent of $k$; that is,
$6\Cinf_1(k,\pi/2,\vartheta)/\cinf$ is equal to $3$ for $\vartheta=0$,
and equal to 1 for $\vartheta=2,4,6\cdots$.
}
 \label{fig:supp:ADCLoLim:C1inf}
\end{figure}

For $\phi=\pi/2$, the pole of $M[h_k;\vartheta+1]$ at $\vartheta=3$ disappears because
$h_0^\infty(\pi/2)=0$. This broadens the analytic strip of $M[h_k;\vartheta+1]$ to
$-1<\vartheta<5$, and
also indicates that the divergences in $\Cinf_{k1}(k,\vartheta)$ and
$\Cinf_{01}(k,\vartheta)$ cancel each other when $\phi=\pi/2$.
Figure~\ref{fig:supp:ADCLoLim:C1inf} illustrates the behaviour of
$\Cinf_1(k,\pi/2,\vartheta)$ with respect to $\vartheta$ and $k$. Many features of
the figure can be qualitatively understood using the large $k$ expansion
\begin{equation*}
   \frac{6 \Cinf_1(k,\pi/2,\vartheta)}{\cinf} \sim 1+
    \lim_{\vartheta'\rightarrow\vartheta}
    \lb\{\frac{(\vartheta'-3)\tan(\pi\vartheta'/2)}{\varpi_k} 
    + \frac{2\sec(\pi\vartheta'/2)}{\Gamma(2-\vartheta')}{\varpi_k}^{-\vartheta'}
     + O({\varpi_k}^{-\vartheta'-2}) \rb\}.
\end{equation*}
Overall, for $\vartheta>0$ it is found that
\begin{equation*}
   \Cinf_1(k,\pi/2,\vartheta)\,{\omega_k}^\vartheta \sim O({\varpi_k}^\vartheta),
\end{equation*}
which means that plots of the ADC against $\omega_k$ will tend
towards the apparent dependence on frequency.

\subsection{Various integrals}
\label{sub:supp:Integrals}

\begin{figure}[tb]
 \centering
  \includegraphics[scale=0.52]{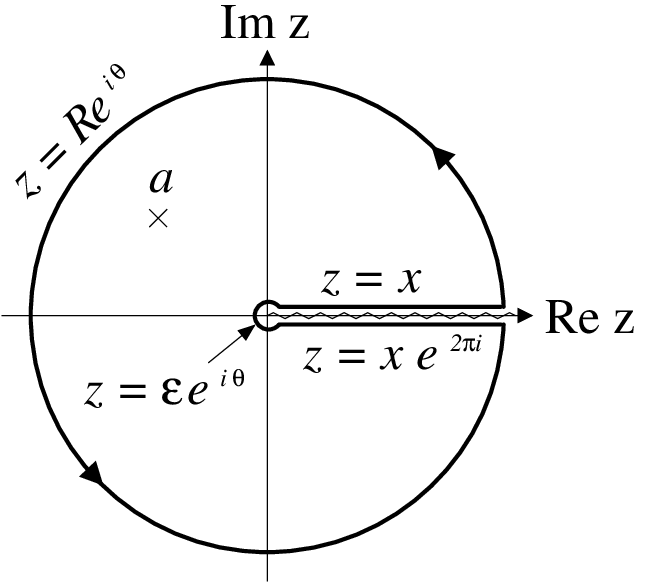}
 \caption{
Contour $C$ used to evaluate Eq.~(\ref{eqn:supp:Integrals:CI}).
The contour runs either side of a branch cut drawn along the positive real axis,
and $a$ is an arbitrary complex number within the contour.
}
 \label{fig:supp:Integrals:C}
\end{figure}


A number of integrals used in the preceding sections are evaluated here.

\begin{enumerate}
\item
Letting $s$ be a complex number such that $0<\re\,s<1$, and $a\neq0$ be any complex
number that does not lie on the positive real axis, consider the integral
\begin{equation}
   J(s;a) = \ointctrclockwise_C dz\,\frac{z^{s-1}}{z+a}
 \label{eqn:supp:Integrals:CI}
\end{equation}
around the contour $C$ shown in Fig.~\ref{fig:supp:Integrals:C}.
The condition on the real part of $s$ ensures that the integrals around the two
circular segments will both be zero in the limits $\epsilon\rightarrow0$ and
$R\rightarrow\infty$. The remaining contributions give
\begin{equation*}
   J(s;a) = \int_0^\infty\!\!dx\,\frac{x^{s-1}}{x+a} + 
     \int_\infty^0\!\!dx\,\frac{x^{s-1}e^{2\pi i(s-1)}}{x+a}
       = 2\pi i\,\mbox{Res}\lb\{\frac{z^{s-1}}{z+a}\rb\}_{z=-a}
         = 2\pi i (-a)^{s-1}.
\end{equation*}
After some simple algebra it is found that
\begin{equation}
   \int_0^\infty\!\!dx\,\frac{x^{s-1}}{x+a} = \frac{\pi a^{s-1}}{\sin\pi s}.
 \label{eqn:supp:Integrals:J(s;a)}
\end{equation}
The left-hand side of this result defines the Mellin transform of $1/(x+a)$
(i.e.~$M[1/(x+a);s]$), which means that even though the integral was evaluated
with the restrictions on $\re\,s$, analytic continuation of the Mellin transform
guarantees that the result holds for all $s\in\mathbb{C}$.

\item
Given real constants $t\geqslant0$ and $c>0$, another integral of interest is
\begin{equation}
   H(s;t,c) = \int_0^\infty\!\!dv\,\frac{v^{s-1}e^{-v t}}{v^2+c^2}
 \label{eqn:supp:Integrals:H(s;t,c)}
\end{equation}
with $1<\re\,s<4$.
Partial fraction decomposition of the denominator is first applied so that
\begin{equation*}
   H(s;t,c)
    =\frac{1}{2ic}\int_0^\infty\!\!dv\,v^{s-1}e^{-v t}\lb[\frac{1}{v-ic}
     -\frac{1}{v+ic}\rb] 
       = \frac{\bar{J} - J}{2ic}
\end{equation*}
with $\bar{J}$ representing the complex conjugate of $J$.
For $t=0$, Eq.~(\ref{eqn:supp:Integrals:J(s;a)}) with $a=i c$ may be used to
evaluate $J$ so that the overall result simplifies to
\begin{equation}
    H(s;0,c) = \frac{\pi c^{s-2}}{2\sin\pi s/2}.
 \label{eqn:supp:Integrals:H(s;0,c)}
\end{equation}
On the other hand, if $t>0$ then
\begin{align*}
   J &= \int_0^\infty\!\!dv\,\frac{v^{s-1} e^{-v t}}{v+ic}
    = e^{i c t}\int_0^\infty\!\!dv\,v^{s-1} \int_t^\infty\!\!db\,e^{-b (v+i c)}
    = e^{i c t} \Gamma(s)\int_t^\infty\!\!db\,b^{-s} e^{-i c b}.
\end{align*}
After setting $w=c b$,
\begin{align*}
    J = c^{s-1} e^{i c t} \Gamma(s)\int_{ct}^\infty\!\!dw\,
                      \frac{\cos w - i\sin w}{w^s}
    = c^{s-1} e^{i c t} \Gamma(s)\lb[\cint(1-s,c t)- i\sint(1-s,ct)\rb].
\end{align*}
where $\cint(z,t)$ and $\sint(z,t)$ are the upper generalised cosine and sine
integrals, respectively
(Sec.~8.21, Digital Library of Mathematical Functions).
Some straightforward algebra subsequently finds that
\begin{align}
   H(s;t,c) = c^{s-2}\Gamma(s)\lb[\sint(1-s,c t)\cos c t 
                                         -\cint(1-s,c t)\sin c t\rb].
 \label{eqn:supp:Integrals:H(s;t,c)_rslt}
\end{align}
As $H(s;t,c)$ is the Mellin transform of $e^{-v t}/(v^2+c^2)$, it can be analytically
continued into the whole complex plane to be a meromorphic function with poles at
$s=-n$, $n=0,1,2,\cdots$.


\hspace{0.5em}
When $s$ takes on positive integer values (i.e.~$s=n$, $n=1,2,\cdots$) or half-integer
values (i.e.~$s=n+1/2$, $n\in\mathbb{Z}$), the generalised sine and cosine integrals
may be written in terms of simpler functions.
For example, for nonzero $t$ and $s=1$
\begin{align*}
   \cint(0,ct)=\int_{ct}^\infty\!\!dw\,\frac{\cos w}{w} = -\Ci(ct)
\end{align*}
\begin{align*}
   \sint(0,ct)=\int_{ct}^\infty\!\!dw\,\frac{\cos w}{w} 
    = \lb\{\int_0^\infty-\int_0^{ct}\rb\}dw\,\frac{\sin w}{w} = \frac{\pi}{2}-\Si(ct)
\end{align*}
where $\Ci(x)$ and $\Si(x)$ are the standard cosine and sine integrals, respectively,
and $\Si(\infty)=\pi/2$ has been used. Subsequently,
\begin{equation}
   H(1;t,c) = \frac{1}{c}\lb\{
     \lb[\frac{\pi}{2}-\Si(ct)\rb]\cos c t + \Ci(ct)\sin c t\rb\}.
 \label{eqn:supp:Integrals:H(1;t,c)}
\end{equation}
Similarly, when $s=1/2$
the variable change $w=\pi u^2/2$ can be applied to show that
\begin{align*}
   \cint(1/2,ct)=\int_{ct}^\infty\!\!dw\,\frac{\cos w}{\sqrt{w\,}}
     = \sqrt{2\pi}\lb\{\int_0^\infty-
               \int_0^{\sqrt{2c t/\pi}}\rb\}du\,\cos\frac{\pi u^2}{2}
    = \sqrt{\frac{\pi}{2}}\lb[1-2C(\sqrt{2c t/\pi})\rb]
\end{align*}
\begin{align*}
   \sint(1/2,ct)=\int_{ct}^\infty\!\!dw\,\frac{\sin w\,}{\sqrt{w\,}}
    = \sqrt{\frac{\pi}{2}}\lb[1-2S(\sqrt{2c t/\pi})\rb]
\end{align*}
where $C(x)$ and $S(x)$ are the Fresnel cosine and sine integrals, respectively,
and $C(\infty)=S(\infty)=1/2$ has been used.
This means that
\begin{equation}
   H(1/2;t,c) = \frac{\pi}{\sqrt{2}\,c^{3/2}}\lb\{
     \lb[1-2S(\sqrt{2ct/\pi})\rb]\cos c t -\lb[1-2C(\sqrt{2ct/\pi})\rb]\sin c t\rb\}.
 \label{eqn:supp:Integrals:H(1/2;t,c)}
\end{equation}

\hspace{0.5em}
It is also useful to know that differentiating $H(s;t,c)$ with respect to $t$ finds
that
\begin{equation}
   H(s+1;t,c) = -\frac{d}{dt}H(s;t,c).
 \label{eqn:supp:Integrals:dHdt}
\end{equation}
As examples of applying this relationship,
\begin{align}
   H(3/2;t,c) &= -\frac{d}{dt}H(1/2;t,c) = \frac{\pi}{\sqrt{2}\,c^{1/2}}\lb\{
     \lb[1-2C(\sqrt{2ct/\pi})\rb]\cos c t\rb. \nonumber \\
      & \hspace{6cm}\lb.+\lb[1-2S(\sqrt{2ct/\pi})\rb]\sin c t\rb\}
 \label{eqn:supp:Integrals:H(3/2;t,c)}
\end{align}
\begin{equation}
   H(2;t,c) = -\frac{d}{dt}H(1;,t,c) = \lb[\frac{\pi}{2}-\Si(ct)\rb]\sin ct
      - \Ci(ct)\cos ct.
 \label{eqn:supp:Integrals:H(2;t,c)}
\end{equation}
On the other hand, $H(s;t,c)$ may be evaluated for noninteger values of $\re\,s<0$ by
integrating Eq.~(\ref{eqn:supp:Integrals:dHdt}) so that
\begin{equation}
   H(s-1;t,c) = H(s-1;0,c) - \int_0^t\!\!db\,H(s;b,c),
 \label{eqn:supp:Integrals:intHdt}
\end{equation}
and then using Eq.~(\ref{eqn:supp:Integrals:H(s;0,c)}) for $H(s-1;0,c)$.

\hspace{0.5em}
Unfortunately, repeated differentiation and integration can be arduous.
To make evaluation of $H(s;t,c)$ simpler a recursion relation can be
developed by noting that
\begin{equation}
   i^{-z}\Gamma(z,i t)=\cint(z,t)-i\sint(z,t),
 \label{eqn:supp:Integrals:Gamma}
\end{equation}
and then using
the well-known identity $\Gamma(z+1,t)=t^z e^{-t}+z\Gamma(z,t)$ to show that
\begin{align*}
   \cint(z+1,t)&=-t^{z}\sin t-z\sint(z,t) \\
   \sint(z+1,t)&=t^{z}\cos t+z\cint(z,t).
\end{align*}
It follows that
\begin{equation}
   H(s+2;t,c)= t^{-s}\Gamma(s)-c^2 H(s;t,c).
 \label{eqn:supp:Integrals:HRecursionRelation}
\end{equation}

\item
Consider
\begin{equation}
   I(s;t,c) = \int_0^\infty\!\!dv\,\frac{v^{s-1}e^{-v t}}{(v^2+c^2)^2}
 \label{eqn:supp:Integrals:I(s;t,c)}
\end{equation}
for real constants $t\geqslant0$ and $c>0$, and $-1<\re\,s<4$.
This integral can be evaluated by noting that
\begin{equation*}
   \frac{1}{(v^2+c^2)^2} = -\frac{1}{2c}\frac{d}{dc}\frac{1}{\lb(v^2+c^2\rb)},
\end{equation*}
so that it is possible to write
\begin{equation}
   I(s;t,c) 
    = -\frac{1}{2c}\frac{d}{dc}\int_0^\infty\!\!dv\,\frac{v^{s-1}e^{-v t}}{v^2+c^2}
    = -\frac{1}{2c}\frac{d}{dc} H(s;t,c).
 \label{eqn:supp:Integrals:I=dH/dc}
\end{equation}
Using the fact that
\begin{align*}
   \frac{d}{dc}\sint(1-s,c t)&=-t^{1-s}c^{-s}\sin c t \\
     \frac{d}{dc}\cint(1-s,c t)&=-t^{1-s}c^{-s}\cos c t,
\end{align*}
it follows from Eq.~(\ref{eqn:supp:Integrals:H(s;t,c)_rslt}) that for $t>0$
\begin{align}
   I(s;t,c) &= -\frac{c^{s-4}\Gamma(s)}{2}\lb\{
                \sint(1-s,c t)\lb[(s-2)\cos c t - c t \sin c t\rb]\rb. \nonumber \\
    & \hspace{3cm}  \lb.  - \cint(1-s,c t)\lb[(s-2)\sin c t + c t \cos c t\rb] \rb\}.
 \label{eqn:supp:Integrals:I(s;t,c)_rslt}
\end{align}
Since $I(s;t,c)$ is the Mellin transform of $e^{-v t}/(v^2+c^2)^2$, analytic
continuation into the whole complex plane finds that it is a meromorphic function
with poles at $s=-n$, $n=0,1,2,\cdots$.
For the special case when $t=0$, differentiation of
Eq.~(\ref{eqn:supp:Integrals:H(s;0,c)}) gives
\begin{equation}
   I(s;0,c) = -\frac{\pi (s-2) c^{s-4}}{4\sin\pi s/2}.
 \label{eqn:supp:Integrals:I(s;0,c)}
\end{equation}

\hspace{0.5em}
It is also worthwhile mentioning that, similar to the case for item 2 above, a
recursion relation could be derived for $I(s;t,c)$.
However, a separate recursion relation for this quantity is
unnecessary as it is simpler to use Eq.~(\ref{eqn:supp:Integrals:HRecursionRelation})
followed by Eq.~(\ref{eqn:supp:Integrals:I=dH/dc}).

\item
Mellin transform of $v^\alpha (\ln v)^p$ for $\alpha\in\mathbb{C}$ and $p$ a finite
non-negative integer. If $v_o$ is a finite positive real number let $f_1(v)$ be equal
to $v^\alpha (\ln v)^p$ for $0<v<v_o$ and 0 otherwise.
Also, let $f_2(v)=v^\alpha (\ln v)^p-f_1(v)$.
Even though the final result is true for arbitrary $v_0$, for demonstration purposes
it is convenient to choose $v_0=1$. In that case repeated integration-by-parts gives
\begin{align*}
   & \int_0^\infty\!\!dv\,v^{s-1} f_1(v)
       = \frac{(-)^p\,p!}{(s+\alpha)^{p+1}}, & & \re\,s>-\re\,\alpha \\
   & \int_0^\infty\!\!dv\,v^{s-1} f_2(v)
       = -\frac{(-)^p\,p!}{(s+\alpha)^{p+1}}, & & \re\,s<-\re\,\alpha.
\end{align*}
Analytic continuation of these results into the whole complex plane defines
$M[f_1;s]$ and $M[f_2;s]$. Therefore,
\begin{align}
   M[v^\alpha (\ln v)^p;s] = M[f_1;s] + M[f_2;s] = 0, & & (s\neq-\alpha)
 \label{eqn:supp:Integrals:M[v^alpha;s]}
\end{align}
with the point $s=-\alpha$ being a pole.

\item
The following integral identity is an edited version of the contour integral
evaluated in Sec.~S-1 of the Supporting Material from \cite{Kershaw2021}.
It is assumed that $F(z)$ is a complex function obeying the conditions:
\begin{enumerate}
 \item[($i$)]
There are no poles along the positive real and imaginary axes, nor within the
first quadrant of the complex plane.
 \item[($ii$)]
For every $R>0$ there is a real number $c(R)$ such that
$|F(R e^{i\theta})|\leqslant c(R)$ and $\lim_{R\rightarrow\infty}c(R)/R$ $=0$.
\end{enumerate}
Then, if $l$ is either $0$ or $k$,
\begin{align}
   P\int_{0}^\infty\!\!dx\, \frac{F(x)(1-e^{i x})}
                              {(x^2-{\varpi_k}^2)(x^2-{\varpi_l}^2)} 
   &=\frac{\pi F(\varpi_k \delta_{k,l})}{2(3\delta_{k,l}-1){\varpi_k}^2} \nonumber \\
    &\hspace{20mm} + i P\int_0^\infty\!\!dy\,\frac{F(i y)(1-e^{-y})}
                      {(y^2+{\varpi_k}^2)(y^2+{\varpi_l}^2)}.
 \label{eqn:supp:Integrals:CI2}
\end{align}

\end{enumerate}

\end{document}